\documentclass[draftclsnofoot,onecolumn,12pt,twoside]{IEEEtran} % for WCOM
\def\psfancypar#1#2{\begingroup\def\par{\endgraf\endgroup\lineskiplimit=0pt}
               \setbox2=\hbox{\large\sc #2}
               \newdimen\tmpht \tmpht \ht2 \advance\tmpht by \baselineskip
               \font\hhuge=Times-Bold at \tmpht
               \setbox1=\hbox{{\hhuge #1}}
               \count7=\tmpht \count8=\ht1
               \divide\count8 by 1000 \divide\count7 by \count8 
               \tmpht=.001\tmpht\multiply\tmpht by \count7 
               \font\hhuge=Times-Bold at \tmpht
               \setbox1=\hbox{{\hhuge #1}}
               \noindent
                \hangindent1.05\wd1
               \hangafter=-2 {\hskip-\hangindent
               \lower1\ht1\hbox{\raise1.0\ht2\copy1}%
                \kern-0\wd1}\copy2\lineskiplimit=-1000pt}

\def\thetabf{{\mbox{\boldmath$\theta$\unboldmath}}}

\newcommand{\E}{\mbox{{\rm E}}}

\newcommand{\abf}{\mbox{${\bf a}$}}

\def\boxit#1{\vbox{\hrule\hbox{\vrule\kern3pt
        \vbox{\kern3pt#1\kern3pt}\kern3pt\vrule}\hrule}}

\def\reals{ { {\rm  I \kern-0.15em R }  } }
\def\complex{ {\,{{\rm C} \kern-0.50em \raise0.20ex {  |}}\, }}
\def\alphabf{\hbox{\boldmath$\alpha$\unboldmath}}

\def\mubf{\hbox{\boldmath$\mu$\unboldmath}}

\def\xibf{\hbox{\boldmath$\xi$\unboldmath}}

\def\psibf{\hbox{\boldmath$\psi$\unboldmath}}

\def\Sigmabf{\hbox{$\bf \Sigma$}}

\def\abf{{\bf a}}

\def\ebf{{\bf e}}
\def\fbf{{\bf f}}

\def\hbf{{\bf h}}

\def\mbf{{\bf m}}
\def\nbf{{\bf n}}

\def\rbf{{\bf r}}

\def\ubf{{\bf u}}

\def\wbf{{\bf w}}
\def\xbf{{\bf x}}
\def\ybf{{\bf y}}

\def\rbf{{\bf r}}
\def\xbf{{\bf x}}
\def\ybf{{\bf y}}
\def\Abf{{\bf A}}
\def\Bbf{{\bf B}}
\def\Cbf{{\bf C}}

\def\Ebf{{\bf E}}
\def\Fbf{{\bf F}}

\def\Ibf{{\bf I}}
\def\Jbf{{\bf J}}

\def\Mbf{{\bf M}}

\def\Qbf{{\bf Q}}
\def\Rbf{{\bf R}}
\def\Sbf{{\bf S}}

\def\Ubf{{\bf U}}

\def\Wbf{{\bf W}}
\def\Xbf{{\bf X}}

\def\Cc{{\cal C}}

\def\Nc{{\cal N}}

\def\be{\vskip .3cm \begin{equation}}
\def\ee{\end{equation} \vskip .4cm \noindent}
\def\defeq{{\stackrel{\Delta}{=}}}
\newcommand{\R}{\mbox{$\hat {\bf R}_{N}$}}

\def\Rxx{\Rbf_{\ssstyle X\kern-.1em X}}

\let\ssstyle=\scriptscriptstyle

\def\Kout{\setbox1=\hbox{\Huge\bf K}\hbox to
1.05\wd1{\hspace{.05\wd1}
}}
\def\Sout{\setbox1=\hbox{\Huge\bf S}\hbox to 1.05\wd1{\hspace{.05\wd1}
}}

  \ifx\LabelFigloaded\MYundefined\relax
  \else
   \fi

  \def\LabelFigloaded{\relax}% now loaded

  \chardef\LabelFigCatAt\the\catcode`\@
  \catcode`\@=11

 \let\LabelFigwlog@ld\wlog
 \def\wlog#1{\relax}

 \ifx\\\MYundefined@
    \let\\\relax
 \fi

  \def\ms@g{\immediate\write16}

 \def\N@wif{\csname newif\endcsname }
 \def\Temp@ {\N@wif\ifIN@}
 \ifx\INN@\MYundefined@
    \else \let\Temp@\relax
 \fi
 \Temp@

  \def\IN@{\expandafter\INN@\expandafter}
  \long\def\INN@0#1@#2@{\long\def\NI@##1#1##2##3\ENDNI@
    {\ifx\m@rker##2\IN@false\else\IN@true\fi}%
     \expandafter\NI@#2@@#1\m@rker\ENDNI@}
  \def\m@rker{\m@@rker}
 
  \newtoks\Initialtoks@  \newtoks\Terminaltoks@
  \def\SPLIT@{\expandafter\SPLITT@\expandafter}
  \def\SPLITT@0#1@#2@{\def\TTILPS@##1#1##2@{%
     \Initialtoks@{##1}\Terminaltoks@{##2}}\expandafter\TTILPS@#2@}

 \def\Shifted@@#1#2#3{\setbox0=\hbox{#3}%
   \raise -\dp0\vbox {\kern-#2%
       \hbox {\kern#1\unhbox0\kern-#1}%
           \kern#2}}

 \newcount\gridcount
 \newbox\auxGridbox@ \newbox\hGridbox@ \newbox\vGridbox@
 \newbox\Labelbox@ \newbox\auxLabelbox@
 \newbox\Coordinatebox@
 \newtoks\Labeltoks@
 \newdimen\Wdd@ \newdimen\Htt@
 \newdimen\Wddd@ \newdimen\Httt@
 
 \def\Wr@{\immediate\write16}

 \newdimen\GL@wd%% grid-line width
 \def\GridLineWidth#1{\GL@wd=#1}

 \def\gobble#1{}
 \def\EdgeErr@{\Wr@{}%
      \Wr@{\string\Edges\space argument
      1, 10, 100 or 1000 please\string!}%
      }

 \newcount\Edgect@

 \def\Sweepup#1\endSweepup{}

 \def\SetEdges@{%
    \edef\Zr@@s{\expandafter\gobble\number\Edgect@\empty}%
        \count255=0\Zr@@s\relax
        \ifnum\count255=\z@\else\EdgeErr@\show\tailtest\fi
        \count255=1\Zr@@s\relax%\showthe\count255
        \ifnum\count255=\Edgect@\relax\else\EdgeErr@\show\leadtest\fi
    \EdgGl@b\edef\Zr@s{\expandafter\gobble\Zr@@s\empty}%\show\Zr@s
    \ifnum\Edgect@>\@ne\relax\EdgGl@b\let\L@Dc\empty
        \else\EdgGl@b\edef\L@Dc{\string.}\fi
    \ifnum\Edgect@>\@ne\relax
        \EdgGl@b\edef\Edgescale@##1{\divide##1 by \Edgect@}%
        \else\EdgGl@b\edef\Edgescale@##1{}\fi
    }

 \def\Edges#1{\Edgect@=#1\relax
     \let\EdgGl@b\global \SetEdges@}

 \Edges{1}%% default

 \def\hhrule{\hrule height \GL@wd\vskip-.\GL@wd}

 \def\hRule@{%
   \advance\gridcount -2%
   \vfil\hhrule\vfil
   \llap{\smash{\raise -2.5pt
     \hbox{\L@Dc\number\gridcount\Zr@s\kern2pt}}}%
   \hhrule
   }

\def\vvrule{\vrule width \GL@wd \kern-\GL@wd}

 \def\vRule@{\advance\gridcount 2%
   \hfil\vvrule\hfil
   \setbox\auxGridbox@=\vbox to 0pt
      {\vskip \Htt@\vskip 2pt
        \hbox to 0pt{\hss\L@Dc\number\gridcount\Zr@s\hss}\vss}%
      \wd\auxGridbox@=0pt \box\auxGridbox@
   \vvrule
   }

 \def\PlaceGrid@@{\gridcount=10 
  \setbox\hGridbox@=\hbox{%
        \hbox{%
             \hskip-.4pt\vrule
             \vbox to \Htt@{%
               \offinterlineskip\parindent=\z@\relax
               \hbox to \Wdd@{\hfil}
               \hRule@\hRule@\hRule@\hRule@
               \vfil\hhrule\vfil}%
             \vrule\hskip-.4pt}
    }%
  \gridcount=0%
  \setbox\vGridbox@=\hbox{%
      \vbox{\offinterlineskip\parindent=0pt\hsize=0pt
         \vskip-.4pt\hrule%
         \hbox to \Wdd@{%
                 \vtop to \Htt@{\vfil}%
                 \vRule@\vRule@\vRule@\vRule@
                 \hfil\vvrule\hfil}%
         \hrule\vskip-.4pt}}%
  \wd\hGridbox@=0pt\ht\hGridbox@=0pt
  \wd\vGridbox@=0pt\ht\vGridbox@=0pt
  \hbox{\box\hGridbox@\box\vGridbox@}%
  }

 \def\LabelsGlobal{\def\LabGl@b{\global}}
 \def\LabelsLocal{\def\LabGl@b{}}
 \LabelsGlobal %% default

 \def\SetLabels#1\endSetLabels{%
   \LabGl@b\Labeltoks@={#1()\\}%
   }

 \LabGl@b\Labeltoks@={()\\}

 \def\ShowGrid{\LabGl@b\let\PlaceGrid@\PlaceGrid@@}
 \def\HideGrid{\LabGl@b\let\PlaceGrid@\relax}
 \def\Grids{\ShowGrid\LabGl@b\let\GridSwitch@\ShowGrid}
 \def\noGrids{\HideGrid\LabGl@b\let\GridSwitch@\HideGrid}

 \noGrids

 \def\bAdjust@@{%
     \setbox\auxLabelbox@=\hbox{\raise \dp\auxLabelbox@
            \box\auxLabelbox@}}
 \def\bAdjust@{\let\vAdjust@\bAdjust@@}

 \def\eAdjust@@{\dimen0=-.5\ht\auxLabelbox@
     \advance\dimen0 by .5\dp\auxLabelbox@
     \setbox\auxLabelbox@=
            \hbox{\raise\dimen0\box\auxLabelbox@}}
 \def\eAdjust@{\let\vAdjust@\eAdjust@@}

 \def\tAdjust@@{%
     \setbox\auxLabelbox@=\hbox{\raise-\ht\auxLabelbox@
            \box\auxLabelbox@}}
 \def\tAdjust@{\let\vAdjust@\tAdjust@@}

 \let\vAdjust@\relax

 \def\lAdjust@{\let\hAdjust@\rlap}
 \def\rAdjust@{\let\hAdjust@\llap}

 \let\hAdjust@\relax\let\vAdjust@\relax

 \def\FetchLabel@#1(#2)#3\\{%
     \IN@0#2@@\ifIN@
        \setbox0=\hbox{\ignorespaces#1#3\unskip}%
        \ifdim\wd0>0pt
           \ms@g{}%
           \ms@g{ !!! Bad label(s)? !!!}%
           \message{ #1(#2)#3}%
        \fi
        \def\LabelMole@##1\endFetchLabel@{%
            \IN@0()\\@##1@%
            \ifIN@\def\Temp@{\FetchLabel@##1\endFetchLabel@}%
            \else\def\Temp@{}%
            \fi
            \Temp@
           }%
     \else
       \ignorespaces#1\unskip
       \setbox\auxLabelbox@=%
         \hbox to 0pt{\hss\ignorespaces\hAdjust@
          {\ignorespaces#3\unskip}\hss}%
       \vAdjust@
       \let\hAdjust@\relax\let\vAdjust@\relax
       \AugmentLabelBox@@{#2}%
       \ht\Labelbox@=0pt\dp\Labelbox@=0pt
       \let\LabelMole@\FetchLabel@%
     \fi\LabelMole@}

 \newtoks\XYSep@ %\XYSep@{*}
 \def\SetXYSeparator#1{%
     \IN@0#1@@\ifIN@\XYSep@{*}%
     \else
     \XYSep@{#1}%
     \fi
     }

 \SetXYSeparator*

 \def\AugmentLabelBox@@#1{%
     \IN@0\the\XYSep@ @#1@\ifIN@
       \SPLIT@0\the\XYSep@ @#1@%
       \setbox\Labelbox@=\hbox to 0pt{%
         \unhbox\Labelbox@
         \Shifted@@{\the\Initialtoks@\Wddd@}%
         {\the\Terminaltoks@\Httt@}%
         {\box\auxLabelbox@}}%
     \else
         \ms@g{}%
         \ms@g{ !!! Bad insertion point. !!!}%
         \message{ (#1\ this point was rejected.)}%
     \fi
    }

 \def\FetchOption@#1[#2]#3\endFetchOption@{%
    \def\temp{#1}%\show\temp
    \ifx\temp\empty
       \Edgect@=#2\relax%\showthe\Edgect@
       \let\EdgGl@b\relax
       \SetEdges@%\def\Edgescale@##1{\divide##1 by \Edgect@\relax}%
       \Cleaner@#3%
    \fi}

 \def\Cleaner@#1[@]{\Labeltoks@{#1}}
     
 \def\PlaceLabels@@{\mathsurround=0pt%\bgroup
     \def\Cr@{\\}%
     \let\L\lAdjust@\let\R\rAdjust@
     \let\B\bAdjust@\let\E\eAdjust@\let\T\tAdjust@
     \expandafter\FetchOption@\the\Labeltoks@[@]\endFetchOption@
     \Wddd@=\Wdd@ \Edgescale@\Wddd@ %\showthe\Edgect@
     \Httt@=\Htt@ \Edgescale@\Httt@
     \expandafter\FetchLabel@\the\Labeltoks@\endFetchLabel@
     \box\Labelbox@%\egroup
     }%

 \let \PlaceLabels@\PlaceLabels@@

 \def\AffixLabels#1{\setbox\Coordinatebox@=\hbox{#1}%
      \Wdd@=\wd\Coordinatebox@ \Htt@=\ht\Coordinatebox@
      \advance\Htt@ \dp\Coordinatebox@
      \hbox{\copy\Coordinatebox@\kern-\Wdd@ 
           \Shifted@@{0pt}{-\dp\Coordinatebox@}%
           {\PlaceLabels@\PlaceGrid@}%
           \kern\Wdd@}%
      \GridSwitch@ %% next grid hidden
      \LabGl@b\Labeltoks@{()\\}%
      }
 
   \let\wlog\LabelFigwlog@ld   %%restore logging
   \catcode`\@=\LabelFigCatAt  %%12 or 13

                                By

              Raymond S\'eroul <A18645@FRCCSC21.BITNET>
                                and 
              Laurent Siebenmann <lcs@topo.math.u-psud.fr>
    
              VERSIONS: July 1991, Oct 1991, Jan 1992, July 1992

INTRODUCTION

      This labelling package is intended for TeX users who
rely on non-TeX sources for for their graphics inserts.  It
provides means for adding TeX labels to such inserts with a
minimum of fuss. 

       For most labels, TeX users have in the past found it
reasonably convenient to rely on non-TeX sources. Typical
occasions when an inescapable need for TeX labels seemed to
arise are

 (a) when the graphics program lacks certain exotic or complex
mathematical symbols

 (b) when the very highest typographical quality is wanted for the
labels

 (c) when labels included with the graphics fail to print, 
 and you cannot figure out why (cf. boxedeps.doc).  The labels
 provided by labelfig.tex are 100% portable.

       Since this package first appeared, many users, who in the
past scarcely dreamed of using TeX labels, have come to use
nothing but.  So it is now appropriate to add

Intoxication Warning:  TeX labels may be addictive and expensive. 

     If you have a fast preview you may disagree, and even find
that this package provides an agreeable paste-up environment; see
extra applications at end.

     Note to publishers: It is possible and convenient to ultimately
export the TeX labels produced by labelfig.tex to become an integral
part of the EPS file. This is often desired by a publisher who typically
uses an "upmarket" graphics or page layout program, with which the
staff is skilled in perfecting figures.  See Appendix I for
a recipe.

     The authors are grateful to Patrick Ion of Math Reviews for
helpful comments and encouragement.

BASIC INSTRUCTIONS

    After reading in the macro file using

preview or proof your figure with a coordinate grid printed on
top, by typing the following:

    \ShowGrid  % shows grid  for next figure only
    \AffixLabels{<the graphics insertion>}

Here <the graphics insertion> is what you would type to insert
the graphics object alone without the grid.  This must provide
for the space around it. For example <the graphics insertion>
might well be \BoxedEPSF{MyFigure scaled 700} using the
boxedeps.tex macro package (from same source); this provides a
TeX box containing the encapsulated PostScript insert specified by
the file MyFigure. \AffixLabels{...} provides the grid (supposing
\ShowGrid is present) and later, once you have specified labels
using the grid, it will "tack on" the labels.

     The grid is a sort of (usually elongated) checkerboard of
ten rows and ten columns and its (internal) partitions are by
default numbered  .1, ... ,.9  both horizontally (X-coordinate
running left to right) and vertically (Y-coordinate running bottom
to top).  Thus the points enclosed by the grid correspond to the
points of the unit square in the cartesian "X-Y" plane, the lower
left corner corresponding to the origin (0,0).  By extrapolation,
the full page corresponds to a larger rectangle in the plane.

     These coordinates serve to position labels as follows.
Before the \AffixLabels{...} command type label specifications:

  \SetLabels
   (<X-coordinate>*<Y-coordinate>) <first label> \\
   .
   .
   .
   (<X-coordinate>*<Y-coordinate>)  <last label> \\
  \endSetLabels

Each row specifies one label and is terminated by \\.  In each
row, the position indicator comes first; it is written as a
standard cartesian point except that the X- and Y- coordinates
are separated by * rather than a comma because TeX allows a
comma as decimal point. There are no dimension units to specify
as the unit is the grid itself.

     By default, this cartesian point specifies where the middle
of the baseline of the label will be located.  However if you precede
the point by \L [or \R] the left [or right] edge of the baseline will
be located there. Similarly you may also precede the point by \T, \E,
or \B to vertically align the top equator or bottom of the label box
at the specified point.  This gives nine standard positions of
the label with respect to the insertion point --- corresponding to
the eight principle points of the compas and the center

                     \L\T     \T      \R\T

                     \L\E     \E      \R\E

                     \L\B     \B      \R\B

But this neglects the default "baseline" level of TeX,
giving potentially three more positions

                     \L    <no tag>   \R

For text, the baseline level is often the preferred. Its relation to
the others is variable. It will often coincide with the bottom level,
as happens for "X".  But it is often distinct, as for "g", in which
case you have in all 12 distinct positions rather than 9.

     It is convenient to think of this specification of label
position as attaching the label by a thumb-tack to the coordinate
grid. There are up to twelve positions of the thumb-tack on the
label, while the position of the thumb-tack on the coordinate grid is
arbitrary.  Normally, one choses the position of the thumb-tack on
the label to be the one that is the closest to the item being
labeled.  There are good reasons for this "rule of thumb":

   (a)  It facilitates correct positioning at first try.

   (b)  If the scale of the figure must be altered after labels
have been affixed, the labels have a good chance of remaining well
positioned.

   (c)  The visible grid need not extend beyond the "bounding box"
for the figure, because the best preferred position is always
(at least almost) within the bounding box .

The second reason is particularly important. Indeed it often
happens that scale has to be altered after labelling begins, in
order to either provide space for the labels, or to adjust
proportions between the labels and the figure.  (The size of labels
is unaffected by scaling.)

     Here is an artificial but self-contained test which uses
TeX rules to make a graphics object.

TEST

    Do not skip this!

 \def\FrameIt#1{\hbox{\vrule$\vcenter {\hrule\kern3pt%
             \hbox {\kern3pt #1\kern3pt}%
               \kern3pt\hrule}$\relax\vrule}}

 \def\Caption#1#2{\FrameIt{%
       \vtop {\hsize=#1\relax \parindent=0pt
         \leftskip=0pt \rightskip=0pt plus15pt
         \parfillskip=0pt
         \lineskip=1pt\baselineskip=0pt
         #2}}}

 \def\FirstQuadrant{\hbox to 100pt{\vrule\vbox to 100pt{%
        \hbox to 100pt{\hfil}\vfil\hrule}\hss}}

  \SetLabels
    \R(.5*.2) $\zeta\,\cdot$\\
    (.9*-.10) $\xi$\\
    \R(-.03*.9) $\eta$\\
    \T(.5*.9) \Caption{70pt}{%
          \it The norm of
          $g(\xi+i\eta)$ is indicated on
          contours of this invisible surface.}\\
  \endSetLabels

  \AffixLabels{\FirstQuadrant}

  \end

  Note that the coordinates to use for labels are indicated on the
edges of the grid (when visible) corresponding to the conventional
x- and y- axes of the Cartesian plane. By default the grid is
1-by-1. However, by the command \Edges{100}, you can change this
to 100-by-100 and many users find this alternative most
convenient. Place the command \Edges{...} in your style file (or
header) since its effect is is global. Other possible edge values
are 10 and 1000.

  If you use the command \Edges{...} at all, do so with care.  For
if you accidentally delete an \Edges{...} command your labels will
abruptly be badly misplaced and may logically but mysteriously
generate "dimension too big" errors under TeX and "off page" errors
under your driver.  

  You can dictate the edgescale for an individual figure by giving
the scale in brackets immediately after \SetLabels.  Thus, to
import into an article using say \Edge{100} a figure labelled using
another edgescale, say the original 1-by-1 default, you can use
\SetLabels[1]...\endSetLabels.

GETTING IT DOWN PAT

     Complicated labeling deserves the same respect as
complicated mathematics.  Do not expect it to come out perfect the
first time!  What is needed in either case is a mechanism to
repeatedly typeset troublesome pieces.

     One mechanism is always available.  One does complicated
labelling in a separate "test" file involving just the figure being
labelled;  a texpert will know how to \dump TeX's current state as
a temporary format that restarts rapidly at each retry.  Usually,
one then pastes the completed labelled figure back into the main
TeX file, but, of course, one can also \input it as an auxiliary
file.

     If you do not have a TeXpert at handy, here is a first
approximation to an efficient setup. By deletions reduce a copy
of your article to just a few lines before and after the figure.
Now label the figure, and finally, copy and paste the labelled
figure to the original article. Then copy the next figure to label
into this testbed and repeat. The TeXpert can improve the  speed
at which TeX starts up, by compiling a format specifically for
your article; just one caution: best NOT include in the format
ephemeral details of setup like \Set<mydriver>ArtSpecials (from
boxedeps.tex because this reads  figure dimensions which you may
change during your work session.

     An improved mechanism to repeatedly typeset troublesome
pieces is now available on the Macintosh; it is called LinoTeX;
see the same ftp sources.  It could be set up on many types
of computer.

     Before using labelfig.tex to attach labels to a graphics
object inserted using boxedeps.tex or BoxedArt.tex, make it a
firm rule to carefully adjust the bounding box using the trimming
commands of these packages, and also at least tentatively scale
and position the object. Beware of changing the grid inadvertently
after the labels have been positioned.  For example, correcting
the bounding box of a PostScript graphics object can foul up the
labels by changing the coordinate grid to which the labels are
attached. This is particularly true for the trimming  commands of
boxedeps.tex and BoxedArt.tex. However, as noted already, change
of scale is much less disruptive, and modest adjustments should be
well tolerated.

     Sometimes the labels protrude so far from the bounding box
of a figure that the figure has to be repositioned.  Best do this
by ad hoc spacing, say using \hglue and \vglue; altering the
bounding box would create a vicious circle.

     Remember that you are responsible for preventing labels
from overlapping. You are responsible for all label typography
including size and style. A label is really just about anything
that can be put in a TeX box. Note that spaces at the beginning
and end of labels will normally be suppressed; if you really want
them you must protect them with TeX braces.

     This package temporarily sets the \mathsurround parameter
of TeX to zero  while the labels are being affixed. This is done
because nonzero \mathsurround space would influence the position
of left and right aligned labels; then, when a texpert or printer
modifies mathsurround, diagram labeling might be disastrously
altered. There is a small price to pay involving labels that are
formatted as caption boxes including mathematics: you  may want or
need to specify an explicit mathsurround space within the caption
box; it will not influence anything outside.

     Those hostile to the use of * as separator between
the X and Y coordinates of label insertion points, are free to
impose another using \SetXYSeparator{<the new separator>}.  
Americans may prefer "," to "*" since they never use a 
comma as a decimal point; on the other hand, * may be more visible.

APPENDIX (I)  MERGING labelfig.tex LABELS INTO AN EPSF GRAPHICS OBJECT.

     As promised in the introduction, here is a recipe useful for
publishers. It works at least on Macintosh and at least for vectorized
graphics and Adobe type1 fonts.  (There is surely a similar recipe for
PCs under MSWindows.)

 (a)  Use boxedeps.tex utility to integrate the figure given by the eps
file, "x.eps" say, with a visible frame around it.  See
\ShowDisplacementBoxes command in boxedeps.tex.  To get precise results
automatically it is important to use the \Trim... commands of
boxedeps.tex making the "DisplacementBox" neatly fit the figure.

 (b)  Use the TeX printer driver and LaserWriter (versions >= 8.1.1) to
export to an EPSF the DVI page containing the integrated, labelled
figure. You now have an EPS file  "xx.eps"  that contains too much, and at
the wrong scale, and at wrong position.

 (c)  Convert the EPSF to an Adode Illustrator format EPSF using
the shareware utility called epsConvert by Sam Weiss
1993-- (currently $25).

 (d)  In Illustrator (or a compatible program), group the labels and the
"DisplacementBox"; copy them to the clipboard and paste them into "x.ps".
This step requires that all the label fonts be "visible to the Macintosh.

 (e)  Translate and scale the pasted group consisting of the labels plus
the "DisplacementBox" so as to make the "DisplacementBox" the bounding
box of (labelless) figure represented by "x.eps".  At this point the
labels will be correctly placed on the figure "x.eps".

 (f)  Ungroup and delete the "DisplacementBox".  The result is the
desired single EPS file, "x+.eps" say, It contains the original figure
plus its labels.  

     Using grouping and ungrouping appropriately in "x+.eps", a
publisher's staff can very efficiently improve label positions etc.

APPENDIX II)  SOME EXOTIC APPLICATIONS

     The grid of labelfig.tex is analogous to a light-table in
classical page makeup with wax or latex glue.  In principle, you
can use it to compose any page from its indivisible parts.  This
even has some of the artisanal charm of classical paste-up
provided you have a fast screen preview to make the process
"interactive".

     In practice labelfig.tex is a tool for nonstandard jobs.
Here are a few going beyond the labelling already discussed.

(I)  GRAPHICS INTEGRATION.

     This is accomplished by treating the imported graphics
objects as labels.  The underlying graphics object is then
typically an empty  \vbox to <dimension>{\vfill} in a TeX
\midinsert...\endinsert construction.  A label line
might be of the form

   (.1*.1) \\

The exact form of the special command varies from driver to
driver.  However, in the case of encapsulated PostScript graphics
(EPSF norm), by relying on boxedeps.tex, one can have the
following standard syntax (independant of driver  (see
boxedeps.doc for details.
  
  (.1*.1) \BoxedEPSF{MyFigure scaled <scale in mils>}\\

This may be slow since it requires TeX to read the PostScript
file to read bounding box using many complex macros.  So you
may want to try

  (.1*.1) \EPSFSpecial{MyFigure}{<scale in mils>}\\

which is fast and driver independant, but it squashes the
bounding box, normally to its lower left corner.

     Similarly for graphics of the Macintosh PICT norm ---
using BoxedArt.tex (same sources) in place of boxedeps.tex.

     This approach to integration is to be recommended when
one is assembling a composite graphics object.

 (II)  COMMUTATIVE DIAGRAM ENHANCEMENT

     Commutative diagrams or arrays of mathematical objects
connected by arrows of various sorts are common in mathematics.
The mathematical objects require the use of TeX.  Recently TeX
acquired a good collection of arrows of all slopes --- that of
LamSTeX --- plus pwerful macros to build the diagrams.

     However, even the LamSTeX collection is often
inadequate; it lacks for example double shafted arrows, dotted
arrows and curved arrows. Fortunately it is possible to produce
such arrows on an individual basis using sophisticated graphics
programs such as Illustrator and AldusFreehand (both serving
the EPSF norm) or using Metafont (with its public domain norm).
Since the creation of each new arrow is a work of love, you
probably want to limit the number of arrows by using LamSTeX
for most arrows. The 40K commutative diagram module of LamSTeX
has been adapted to work with AmSTeX and a copy may be posted
with LabelFig and related files. Unfortunately no one has yet
offered a version that works with Plain TeX or LaTeX.

       Suffice it here to say that when the exotic arrow has
been somehow imported into TeX, labelfig.tex treats it as a
label that one affixes to the commutative diagram.  Two other
steps will be treated in separate notes, namely the matter of
extracting the dimension specifications for the arrow and the
construction of the arrow --- for these steps are far from
unique and often depend intimately on your computer environment. 
Notes for the Macintosh-Textures-Illustrator combination are
found in the file ExoticArrows.doc.

 (III) NESTING 

Ingenuity pays off in exploiting labelfig.tex. One can
mix graphics and typography quite freely.  labelfig.tex is good
for freeform or overlapping arrangements, while boxedeps.tex (or
BoxedArt.tex) is best for regimented non-overlapping
arrangements --- and the two can be combined.

     The default behavior of labelfig.tex is not ideal 
for nesting objects, because to prevent trouble for beginners
the register for labels is globally cleared when \AffixLabels
concludes.  But there are switches available

      \LabelsGlobal      \LabelsLocal

which change this.  To understand this, extend the above test 
by something like:

 \LabelsLocal
 \SetLabels
    (.5*.5) AAA\\
 \endSetLabels

 {%%% Watch for influence of braces!!
 \SetLabels
    (.5*.5) ZZZ\\
 \endSetLabels
   \AffixLabels{\FirstQuadrant}
 }

   \AffixLabels{\FirstQuadrant}

     There are however potential pitfalls.  Neither
labelfig.tex nor boxedeps.tex has been tested under extreme
conditions. Problems may occur if their procedures are
indiscriminately nested. For boxedeps.tex (not labelfig.tex)
there is a precise cause for worry, namely many of its
variables are "global", which means that TeX braces will not
provide the protection one might expect.

COMMAND SUMMARY FOR labelfig.tex

  Here [...] means optional (one or zero)
       [...]* means any number of such constructs

  \SetLabels
    [[<P>](<X><Sep><Y>) <label> \\]*
  \endSetLabels
  \ShowGrid  % this makes the grid visible (once)
  \AffixLabels{<the figure>}

   --- <P> is tack position, one of eleven or empty
              order irrelevant

                   \L\T      \T      \R\T

                   \L\E      \E      \R\E

                     \L               \R

                   \L\B      \B      \R\B

   --- (<X><Sep><Y>) insertion point;
  <Sep> is separator, = * by default;
  \SetXYSeparator{<Sep>} changes it.
   <X> and <Y> are real numbers

  --- <label> a label to attach 

  --- <the figure> the figure to label 

  \GlobalLabels (default)     
  \LocalLabels  setting for nested constructs.

 \Grids makes ALL grids appear; \HideGrid then makes just next disappear.
 \noGrids returns to default.  The commands are always global.

 \GridLineWidth{<dimension>} adjusts width of grid lines. Default is very
small, to give "hairline" effect. If your grid lines are missing try
setting \GridLineWidth{1pt}.

 \Edges#1 globally changes the edge size of all grids to the numerical 
value #1, which must be 1, 10, 100, or 1000.  The default is 1.

VERSION HISTORY.
 --- Jan 1993: \Edges#1 and [??] option after \SetLabels
 --- July 1992: \Grids, \noGrids, \HideGrid;
       Gridlines become hairlines; \GridLineWidth{<dimension>}.
 --- Oct 1991, Jan 1992: \SetXYSeparator{<Sep>},  \LabelsGlobal,
       \LabelsLocal.
 --- July 1991: first release

Address for bugs and other feedback:

        Raymond S\'eroul
        IREM and Lab. de Typographie Informatise
        Univ. Rene Descartes
        Strasbourg

    Tel 33-88-41-63-45
    Email:  A18645@FRCCSC21.BITNET

        Laurent Siebenmann
        Mathematique, Bat. 425,
        Univ de Paris-Sud,
        91405-Orsay,
        France

    Tel 33-1-6941-7949; 
    Email: lcs@topo.math.u-psud.fr

\newif\ifimportant
\importanttrue %\importanttrue  or \importantfalse

\newcommand {\Ebb}{{\mathbb{E}}}
\newtheorem{proposition}{Proposition}
\newtheorem{theorem}{Theorem}
\newtheorem{corollary}{Corollary}
\newtheorem{lemma}{Lemma}

\newtheorem{remark}{Remark}

\usepackage[dvips]{graphics}
\usepackage[dvips]{graphicx}
\usepackage{amsmath,amssymb,amsfonts,latexsym,verbatim,color,epsfig,psfrag}%epsf,
\usepackage{times,multirow,multicol, array}
\usepackage{algorithm, algorithmic}
\usepackage{caption,subcaption}
\usepackage{cite}
\usepackage{url}
\usepackage{dblfloatfix}
\usepackage{booktabs}
\usepackage{setspace}
\usepackage{soul} % for strikethrough
\usepackage{bbm}
\usepackage{epstopdf}
\usepackage{lipsum,amsmath}
\usepackage{bigints}
\usepackage{tikz}

\newcommand{\colvec}[2][.8]{%
  \scalebox{#1}{%
    \renewcommand{\arraystretch}{.8}%
    $\begin{bmatrix}#2\end{bmatrix}$%
  }
}

\usepackage{stackengine}
\stackMath
\newcommand\tsup[2][2]{%
 \def\useanchorwidth{T}%
  \ifnum#1>1%
    \stackon[-1.3ex]{\tsup[\numexpr#1-1\relax]{#2}}{\mathchar"307E}%
  \else%
    \stackon[-1ex]{#2}{\mathchar"307E}%
  \fi%
}

\usepackage{accents}
\newlength{\dhatheight}

\usepackage[letterpaper,hmargin=1in,top=0.85in, bottom=0.97in]{geometry}
\definecolor{gray}{rgb}{0.5,0.5,0.5}

\definecolor{islamicgreen}{rgb}{0.0, 0.56, 0.0}

\begin{document}

\bstctlcite{IEEEexample:BSTcontrol}

% paper title can use linebreaks  within to get better formatting as desired
\title{\LARGE Training Signal Design for Sparse Channel Estimation in Intelligent Reflecting Surface-Assisted Millimeter-Wave Communication}

\author{Song Noh,~\IEEEmembership{Member,~IEEE},
Heejung Yu,~\IEEEmembership{Senior Member,~IEEE},\newline
and Youngchul Sung$^\dagger$,~\IEEEmembership{
Senior Member,~IEEE} 
\thanks{
S. Noh is with the Department of Information and Telecommunication Engineering, Incheon National University, Incheon 22012, South Korea (e-mail: songnoh@inu.ac.kr). \newline
\indent H. Yu is with the Department of Electronics and Information Engineering, Korea University, Sejong 30019, South Korea (e-mail: heejungyu@korea.ac.kr). \newline
\indent Y. Sung {\em (corresponding author)} is with the School of Electrical Engineering, Korea Advanced Institute of Science and Technology, Daejeon 305-701, South Korea (e-mail: ysung@ee.kaist.ac.kr).
}}

\maketitle

\vspace{-4em}
\begin{abstract}
In this paper, the problem of training signal design for intelligent reflecting surface (IRS)-assisted millimeter-wave (mmWave) communication under a sparse channel model is considered. The problem is approached based on  the Cram$\acute{\text{e}}$r-Rao lower bound (CRB) on  the mean-square error (MSE) of channel estimation. By exploiting the sparse structure of mmWave channels, the CRB for the channel parameter composed of path gains and path angles is derived in closed form under Bayesian and hybrid parameter assumptions.  Based on the derivation and analysis, an IRS reflection pattern design method is proposed by minimizing the CRB as a function of design variables under constant modulus constraint on reflection coefficients.   
Numerical results validate the effectiveness of the proposed design method  for sparse mmWave channel estimation.  
\end{abstract}

\vspace{-2em}
%%%%%%%%%%%%%%%%%%%%%%%%%%%%%%%%%%%%%%%%%%%%%%%%%%%%%%%%%%%%%%%%%%%%%%%%
\section{Introduction} \label{sec:introduction}
%%%%%%%%%%%%%%%%%%%%%%%%%%%%%%%%%%%%%%%%%%%%%%%%%%%%%%%%%%%%%%%%%%%%%%%%

IRSs  have gained much attention as one of the potential technologies for 6G   under various names  such as reconfigurable intelligent surfaces (RISs), reflector-arrays or intelligent walls. The objective of   communication system design up to now was  to design optimal signal waveforms and encoding/decoding schemes for given wireless channels, but IRSs have changed the paradigm of this conventional communication system design. IRSs aim to realize intelligent wireless channels by controlling the radio propagation model rather than optimizing  transmission-and-reception schemes under given channels \cite{Liaskos&Nie&Tsioliaridou&etal:ComMag18, Basar&Renzo&etal:Access19}. 
An IRS consists of an array of passive scattering elements and the phase and/or other characteristics of the  signal reflected by each element is controlled. The controllable radio propagation model is beneficial for wireless communication to improve signal quality and coverage [5], [6], and various aspects of the IRS technology have been investigated.
The propagation and pathloss model with IRS was  studied \cite{Ozdogan&Bjornson&Larsson:WCL20} and the design of transmit beamformer and  IRS phase shifters  was examined with various objectives and constraints. For example, the  problem of sum rate maximization under a transmit power constraint was studied in \cite{Guo&Liang&Chen&Larsson:TWC20} and 
the maximization of minimum signal-to-interference-plus-noise ratio (SINR) subject to a given power constraint was considered in \cite{Nadeem&Kammoun&Chaaban&etal:TWC20}.  
The transmit power and phase shift of IRS were optimized to improve energy efficiency in downlink multiuser MISO channels with IRS in \cite{Huang&Zappone&etal:TWC18}. 
The secrecy rate performance associated with IRS was investigated in  \cite{Dong&Wang:WCL20, Cui&Zhang&Zhang:WCL19, Yu&Xu&Sun&Ng&Schober:JSAC20, Qiao&Alouini:WCL20}. 
Simultaneous wireless information and power transfer (SWIPT) was also considered in the context of IRS in \cite{Wu&Zhang:WCL20}. 
In particular, IRSs can be very useful in mmWave channels since the propagation is directive and the propagation loss is large in this band.    IRSs can provide diversity paths to improve the link quality to fully harness the large bandwidth available in this band 
 \cite{Wu&Zhang:20COMMAG, Ying&Demirhan&Alkhateeb:20arXiv}. In this regard, the authors in \cite{Wang&Fang&Yuan&Chen&Duan&Li:20TVT}  designed a transmit precoder and the IRS reflection pattern for data transmission in  mmWave channels.  For such design of transmit beamformer and IRS phase shifters, channel state information (CSI) is required and most previous works assume that perfect CSI of the IRS and the direct link is available. However, in practice,  channels should be estimated. Channel estimation in IRS-assisted communication is not simple because 
an IRS is composed of passive reflecting elements which cannot send their own pilot signal and this fact makes the available channel estimation methods devised for relay channels not directly applicable. In this paper, we consider mmWave communication, which can potentially get much benefit from IRSs, and investigate 
training signal design for channel estimation in IRS-assisted mmWave communication.

\vspace{-1em}
%%%%%%%%%%%%%%%%%%%%%%%%%%%%%%%%%%%%%%%%
\subsection{Related Works and Contributions}

%* Individual channel estimation
We here focus on works on channel estimation in IRS-assisted communication relevant to our work. One line of approaches to channel estimation in IRS-assisted systems is to estimate individual channels, i.e., one from the transmitter to the IRS  and the other from the IRS to the receiver  \cite{Zhang&Qi&Li&Lu:SPAWC20, Hu&Dai:20arXiv}. Under the assumption that the channel between the base station and the IRS varies  slowly  as compared to the channel between the IRS and a user, the authors in \cite{Zhang&Qi&Li&Lu:SPAWC20, Hu&Dai:20arXiv} proposed two-step channel estimation frameworks.  These approaches enable the precoding schemes in \cite{Huang&Zappone&Debbah&Yuen:18ICASSP,Ye&Guo&Alouini:20TWC} to be feasible, which require the knowledge of individual channels. 
%*  cascaded channel estimation
Another approach is to estimate the cascaded channel composed of the transmitter-to-IRS and IRS-to-receiver channels by treating the cascaded channel as a single effective channel 
\cite{Mishra&Johansson:ICASSP19,He&Yuan:WCL20, Jensen&Carvalho:ICASSP20, Zheng&Zhang:WCL20} (this approach is adopted in this study).  In \cite{Mishra&Johansson:ICASSP19}, the authors proposed an one-by-one channel estimation protocol in which only a single IRS element is activated while other elements remain off in each step.  In \cite{He&Yuan:WCL20}, the authors  investigated a cascaded channel framework, exploiting  sparse matrix factorization and matrix completion techniques in a downlink massive multiple-input multiple-output (MIMO) system. Under the assumption of a rank-deficient channel, the pilot symbols and the IRS reflection pattern having on-off states are generated randomly by using Gaussian and Bernoulli distributions, respectively.  In \cite{Jensen&Carvalho:ICASSP20}, the authors designed an optimal channel estimation scheme based on minimization of the Cram$\acute{\text{e}}$r-Rao lower bound (CRB) under a dense channel model  and showed that an optimal IRS activation patterns was given by  the columns of the discrete Fourier transform (DFT) matrix.  In \cite{Zheng&Zhang:WCL20}, a transmission protocol to perform channel estimation and IRS optimization successively  was proposed for an IRS-based orthogonal frequency division multiplexing (OFDM) system.

% Drawback of the existing work
The majority of the existing works assume that the number of training symbols for channel estimation is large enough as compared to the number of (grouped) IRS reflecting elements \cite{Zhang&Qi&Li&Lu:SPAWC20, Hu&Dai:20arXiv, Mishra&Johansson:ICASSP19,He&Yuan:WCL20,Jensen&Carvalho:ICASSP20,Zheng&Zhang:WCL20}.  However, it can be difficult to satisfy this assumption in practical systems because an IRS typically adopts a large number of reflecting elements to achieve high passive beamforming gain.  The contributions of this paper are summarized as follows.

% Contributions
\begin{itemize}
    \item We incorporated the channel sparsity in mmWave channels into the signal model to reduce the number of parameters to be estimated. The parameter in this case is given by path gains and path angles, and the considered framework enables channel estimation with the reduced number of pilot symbols less than the number of the reflecting elements in the IRS.
    
    \item We derived the CRB under  two assumptions: One is that both  path gains and path angles are random parameters with known distributions and the other is that the path gains are random but the path angles are deterministic and unknown.
    
    \item In the Bayesian parameter case, we derived a condition to minimize the Bayesian CRB and showed that the Fisher information ``density" for angle estimation was  given by a quadratic form composed of the derivative of the array response and the IRS reflection pattern matrix square and this determined the quality of angle estimation across the angle domain.
    
    \item In the hybrid parameter case, treating the path gains as random nuisance parameters, we formulated the problem of optimal IRS reflection pattern matrix design by minimizing the CRB on angle estimation, and solved this optimization problem by using the projected gradient method (PGM) \cite{Bertsekas:76AC,Bubeck:17book}.

   \item With numerical evaluation, we demonstrated that the proposed design method exploiting channel sparsity yielded noticeable gain  in sparse mmWave channels when the number of training symbols was less than the number of the reflecting elements in the IRS.

\end{itemize}

\vspace{-0.5em}
%%%%%%%%%%%%%%%%%%%%%%%%%%%%%%%%%%%%%%%%
\subsection{Notations and Organization}

Vectors and matrices are written in boldface with matrices in capitals. All vectors are column
vectors. For a matrix $\Abf$,  $\mathbf{A}^T$, $\mathbf{A}^H$, and $\mathbf{A}^\dagger$ indicate the transpose, Hermitian transpose, and Moore-Penrose inverse of $\mathbf{A}$,  respectively.  $\Abf \succeq 0 $ means that the matrix $\Abf$ is a Hermitian positive semidefinite matrix. 
 $[\Abf]_{i,j}$ denotes the  $(i,j)$-th element of $\Abf$.  $[\Abf]_{i_1:i_2,j_1:j_2}$ denotes the submatrix of $\Abf$  given by the intersection of rows $i_1,\ldots,i_2$ and columns $j_1,\ldots,j_2$.
  For a vector $\abf$, $[\abf]_i$ denotes the $i$-th element of $\abf$, and $\|\abf\|$ represents vector $\ell_2$-norm.  $\mathbf{0}_{n}$ and $\mathbf{1}_n$  are the zero vector of length $n$ and the vector of length $n$ composed of all one elements, respectively.
$\ebf_n(i)$ denotes the $i$-th column vector of the identity matrix of size $n\times n$.  $\Ibf_{n}$ is the identity matrix of size $n\times n$.
$\text{diag}(\Abf)$ denotes a vector of the elements on the diagonal of $\Abf$ and
$\text{diag}(\abf)$ denotes a diagonal matrix whose diagonal is $\abf$.  
${\mathbb E}\{\cdot\}$ denotes statistical expectation.  $\abf\sim\mathcal{CN}(\mubf, \Sigmabf)$ means that random vector $\abf$ is complex circularly-symmetric Gaussian distributed with mean vector $\mubf$ and covariance matrix $\Sigmabf$.  $\abf \sim \text{Unif}[a,b]$ means that the elements of $\abf$ are randomly and uniformly distributed over the interval $[a,b]$. 
${\mathbb{C}}$ denotes  the set of complex numbers.  The symbol $\odot$ and $\otimes$ denote the Hadamard product and the Kronecker product, respectively. 
$\text{Re}(\cdot)$ $\text{Im}(\cdot)$ denote the real part and the imaginary part, respectively.  $\text{tr}(\cdot)$ denotes the trace operator.  $\text{vec}(\cdot)$ denotes the vectorization to stack the columns of the input matrix.  $\iota :=\sqrt{-1}$.

This paper is organized as follows. The system model is described in Section \ref{sec:system_model}.
In Section \ref{sec:bayesian_crb_based_design}, the  IRS reflection pattern design is considered under the Bayesian CRB. In Section \ref{sec:hybrid_crb_based_design},  the  IRS reflection pattern design is considered under the hybrid CRB. Numerical results are provided in Section \ref{sec:numerical_result}, followed by conclusions in Section \ref{sec:conclusion}.

%%%%%%%%%%%%%%%%%%%%%%%%%%%%%%%%%%%%%%%%%%%%%%%%%%%%%%%%%%%%%%%%%%%%%%%%
\section{System Model} \label{sec:system_model}
%%%%%%%%%%%%%%%%%%%%%%%%%%%%%%%%%%%%%%%%%%%%%%%%%%%%%%%%%%%%%%%%%%%%%%%%

\begin{figure}[t]
\SetLabels
\L(0.155*-0.12) \footnotesize (a) System model\\
\L(0.685*-0.12) \footnotesize (b) Channel model\\
\endSetLabels
\leavevmode
\begin{psfrags}
\centerline{
\psfrag{(tx)}[c]{\small \hspace{-0.7em} TX} %
\psfrag{(rx)}[c]{\small\hspace{-0.7em} RX} %
\psfrag{(w1)}[c]{\small \hspace{-0.7em} $w_{1k}$} %
\psfrag{(w2)}[c]{\small \hspace{-0.7em} $w_{2k}$} %
\psfrag{(wN)}[c]{\small \hspace{-0.2em}$ w_{Nk}$} %
\psfrag{(alpha0)}[c]{\small Direct path $\alpha_0$} %
\psfrag{(ht1)}[c]{\small \hspace{-0.7em} $h_{1}^{(t)}$} %
\psfrag{(ht2)}[c]{\small \hspace{-0.7em} $h_{2}^{(t)}$} %
\psfrag{(htN)}[c]{\small \hspace{-0.7em} $h_{N}^{(t)}$} %
\psfrag{(hr1)}[l]{\small \hspace{-0.7em} $h_{1}^{(r)}$} %
\psfrag{(hr2)}[l]{\small \hspace{-0.7em} $h_{2}^{(r)}$} %
\psfrag{(hrN)}[l]{\small \hspace{-0.7em} $h_{N}^{(r)}$} %
\psfrag{(at1)}[c]{\small \hspace{-1.7em} $\alpha_1^{(t)} \ubf_N(\psi_1^{(t)})$} %
\psfrag{(at2)}[c]{\small \hspace{1.6em} $\alpha_{L^{(t)}}^{(t)} \ubf_N(\psi_{L^{(t)}}^{(t)})$} %
\psfrag{(ar1)}[c]{\small \hspace{0.7em} $\alpha_1^{(r)} \ubf_N(\psi_1^{(r)})$} %
\psfrag{(arN)}[c]{\small \hspace{-0.7em} $\alpha_{L^{(r)}}^{(r)} \ubf_N(\psi_{L^{(r)}}^{(r)})$} %
\psfrag{(irs)}[c]{\small IRS (ULA with $N$ reflecting elements)} %
\psfrag{(hybrid)}[l]{\normalsize $\fbf_k^{(i)}$} %
%\ShowGrid
\strut\AffixLabels{
\includegraphics[scale=0.8]{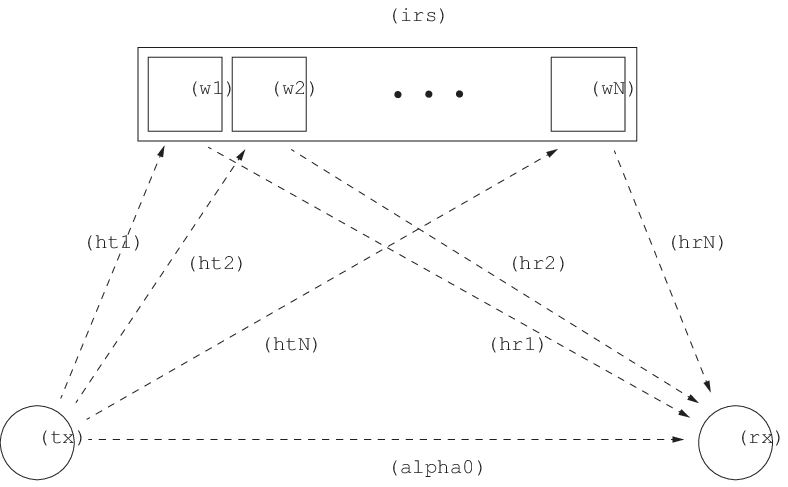}\hspace{3em}
\includegraphics[scale=0.8]{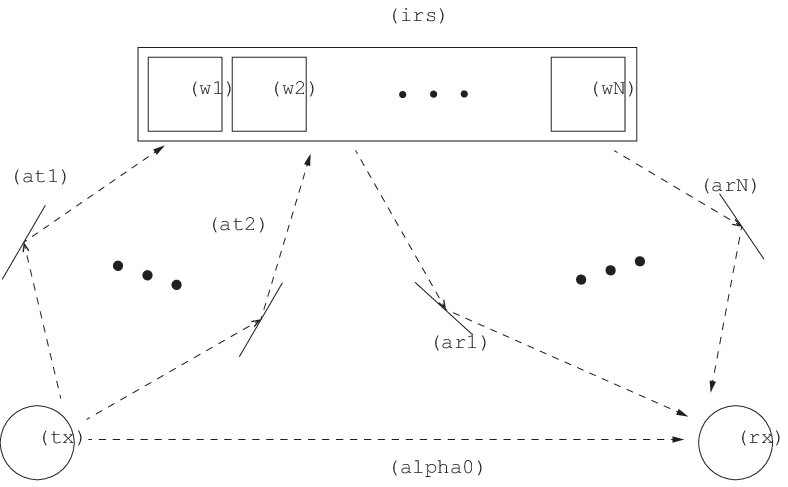}
}}
\vspace{0.3em}
\caption{The considered IRS-assisted point-to-point communication link.}
\label{fig:system_model}
\end{psfrags}
\vspace{-1.0em}
\end{figure}

We consider an IRS system composed of a transmitter, an IRS, and a receiver, as shown in Fig. \ref{fig:system_model}.
We assume that the transmitter and the receiver have a single antenna and the IRS has a uniform linear array (ULA) of  $N$ passive reflecting elements.\footnote{\
Extension to the case of multiple receive antennas is possible by treating $\hbf^{(r)}$ as the beamformed effective channel.}  We assume that there exists a direct channel path from the transmitter to the receiver, where the direct-path channel gain is denoted by $\alpha_0 \in {\mathbb{C}}$, and there exist channel links between the transmitter and the IRS $\hbf^{(t)}=[h_1^{(t)}, \cdots, h_N^{(t)} ]^T\in\mathbb{C}^{N}$ and between the IRS and the receiver $\hbf^{(r)}= 
[ h_1^{(r)},\cdots, h_N^{(r)} ]^T\in\mathbb{C}^{N}$. 
We assume that the considered IRS system operates in the mmWave band and hence adopt a geometry-based sparse channel model relevant to mmWave communication \cite{Sayeed&Raghavan:07STSP,Seo&Sung&Lee&Kim:15arXiv,Wang&Fang&Yuan&Chen&Duan&Li:20TVT, Ying&Demirhan&Alkhateeb:20arXiv, Zhang&Xu&Xu&Ng&Sun:20CL}. Thus,  the channel vectors of $\hbf^{(t)}$ and $\hbf^{(r)}$  are modelled  as 
\begin{equation} \label{eq:sparseChAssum}
\hbf^{(t)}
= 
\sum_{i=1}^{L^{(t)}} \alpha_i^{(t)}\ubf_N(\psi_i^{(t)})
~~~\mbox{and}~~~    
\hbf^{(r)}
=
\sum_{j=1}^{L^{(r)}} \alpha_j^{(r)}\ubf_N(\psi_j^{(r)}),
\end{equation}
where $\alpha_i^{(t)}$ and $\psi_i^{(t)}$ are the gain and angle-of-departure (AoD) of the $i$-th path of the channel from the transmitter to the IRS, $\alpha_j^{(r)}$ and $\psi_j^{(r)}$ are the gain and angle-of-arrival (AoA) of the $j$-th path of the channel from the IRS to the receiver, $L^{(t)}$ and $L^{(r)}$ are the numbers of multi-paths of $\hbf^{(t)}$ and $\hbf^{(r)}$, respectively, and   $\ubf_N(\psi)$ is  the array response vector of an $N$-element ULA, given by  
$\ubf_N(\psi) = 
\left[1, e^{\iota \pi \psi},e^{\iota \pi 2\psi}, \cdots, e^{\iota \pi (N-1)\psi}\right]^T$.
Here,   $\psi$ is the normalized path angle  given by $\psi = \frac{2d\sin(\phi)}{\lambda}$,  $d$ is the spacing between adjacent reflecting elements at the IRS, $\lambda$ is the carrier wavelength, and $\phi \in [-\pi/2,\pi/2]$ is the unnormalized physical path angle. We assume critical spatial sampling at the IRS, i.e., $d= \frac{\lambda}{2}$, and hence $\psi \in [-1,1]$.

% FIGURE

We consider training-based channel estimation with a training period composed of $K$ symbols in the beginning of each channel coherence interval.
At the $k$-th symbol time during the training period (i.e., $1\le k \le K$), the transmitter sends a training symbol $s_k$ with $\mathbb{E}\{|s_k|^2 \} = \rho$, and the $N$ passive reflecting elements at the IRS reflect the incoming signal from the transmitter with a complex gain vector $\wbf_k = [w_{1k}, \ldots, w_{Nk}]^T$ with  $|w_{nk}| = \sqrt{\beta}$ for some $\beta\in(0, 1]$.   Then, the received signal $y_k$ at the receiver at the $k$-th symbol time is given by \cite{Ying&Demirhan&Alkhateeb:20arXiv, Wang&Fang&Yuan&Chen&Duan&Li:20TVT, Zhang&Xu&Xu&Ng&Sun:20CL}
\begin{equation}  \label{eq:yimodel}
y_k = 
\left(\alpha_0 + \sum_{n=1}^N w_{nk}^*h^{(t)}_n h^{(r)}_{n} \right) s_k + n_k
=
\left(\alpha_0 +  \wbf_k^H \left(\hbf^{(t)}\odot \hbf^{(r)}\right)\right) \sqrt{\rho} + n_k, 
\end{equation}
where 
$n_k \sim \Cc\Nc(0,\sigma_n^2)$ is circularly-symmetric complex Gaussian noise.  The IRS-assisted cascaded channel vector $\hbf$ from the transmitter to the receiver  is defined as  
\begin{align}
\hbf
&:=
\hbf^{(t)}  \odot \hbf^{(r)} 
=
\sum_{i=1}^{L^{(t)}} \sum_{j=1}^{L^{(r)}}
\alpha_i^{(t)}\alpha_j^{(r)}\ubf_N\left(\psi_i^{(t)}+\psi_j^{(r)}\right) 
=
\sum_{\ell =1}^L \alpha_\ell \ubf_N(\psi_\ell)=\Ubf_{\boldsymbol \psi}[\alpha_1,\cdots,\alpha_L]^T, \label{eq:def_cascaded_channel_vec}
\end{align}
where 
$L=L^{(t)}L^{(r)}$ denotes the number of paths in the cascaded channel $\hbf$, 
$\alpha_\ell~(=\alpha_i^{(t)}\alpha_j^{(r)})$ and $\psi_\ell~(=\psi_i^{(t)}+\psi_j^{(r)})$ denote the effective gain and angle of the $\ell$-th path of the cascaded channel $\hbf$, respectively, and $\Ubf_{\boldsymbol \psi} = [\ubf_N(\psi_1), \cdots, \ubf_N(\psi_L)]\in\mathbb{C}^{N\times L}$.  Then, based on \eqref{eq:yimodel}, the received signal during the entire training period can be written  in vector form as 
\begin{align}
\ybf 
&=
\Sbf
\underbrace{
\left[\begin{array}{cc}
\mathbf{1}_K & \Wbf^H
\end{array}\right]}_{=:\widetilde{\Wbf}^H}
\left[\begin{array}{c}
\alpha_0 \\
\hbf
\end{array}\right]
+ \nbf %\\
=
\Sbf \widetilde{\Wbf}^H 
\left[\begin{array}{cc}
1 & \mathbf{0}_L^T\\
\mathbf{0}_N  & \Ubf_{\boldsymbol\psi}
\end{array}\right] \boldsymbol \alpha + \nbf,
%=: \mbf + \nbf 
\label{eq:finalsignalmodel}
\end{align}
where $\ybf=[y_1,y_2,\ldots,y_K]^T$, $\nbf=[n_1,n_2,\ldots,n_K]^T$, $\Sbf=\mbox{diag}(s_1,\cdots,s_K)$, $\Wbf = [\wbf_1, \cdots, \wbf_K] \in\mathbb{C}^{N \times K}$,  and $\alphabf = [\alpha_0, \alpha_1, \ldots, \alpha_L]^T$.  We will refer to $\Wbf$ as the {\em reflection pattern matrix} used at the IRS during the training period.

The goal of channel estimation is to obtain  the channel parameters $\alphabf$ and $\psibf=[\psi_1,\cdots,\psi_L]^T$ based on the received signal $\ybf$.  Once the channel $\hbf$  is estimated during the training period, some control information can be sent from the receiver to the IRS to match the reflection coefficients $\{w_{ij}\}$ to $\hbf$ in order to maximize the data rate or other desired performance measure based on the received signal model \eqref{eq:yimodel} during the data transmission period.

\vspace{0.5em}
\begin{remark}
Due to the existence of the direct path, the signal model \eqref{eq:finalsignalmodel} incorporating the channel sparsity in the mmWave band has a special structure. It is not a simple linear model in terms of the unknown parameters $\{\alpha_0,\alpha_1,\cdots,\alpha_L,\psi_1,\cdots,\psi_L \}$ or a signal model associated with simple AoA estimation. It is a mixed nonlinear model in terms of the channel path-gain parameters  $\alpha_0, \alpha_1,\cdots, \alpha_L$ and the path-angle parameters $\psi_1,\cdots,\psi_L$.
\end{remark}
\vspace{0.5em}

\begin{remark}
Note that the IRS is typically equipped  a large  number $N$ of passive reflecting elements to maximize its reflection power. In this case,  the number of parameters to be estimated is potentially large, i.e.,  $\hbf \in\mathbb{C}^{N}$ with large $N$. However, by exploiting the sparse scattering nature of mmWave channels (i.e., the value of $L$ is relatively small), we can reduce the required time for training because of the reduced size of channel parameters $\alphabf\in\mathbb{C}^{L+1}$ and $\psibf \in \mathbb{R}^{L}$. 
\end{remark}
\vspace{0.5em}

For the rest of this paper, we  consider  the problem of optimal training signal design under the signal model \eqref{eq:finalsignalmodel}.  Under the  signal model \eqref{eq:finalsignalmodel}, we have two design variables $\Sbf$ and $\Wbf$, where $\Sbf=\mbox{diag}(s_1,\cdots,s_K)$ is  the  training symbol sequence at the transmitter and $\Wbf$ is the IRS reflection pattern matrix during the training period.  In the context of IRS, we set $s_k = \sqrt{\rho}$, $1\le k \le K$, i.e., $\Sbf = \sqrt{\rho}\Ibf_K$ during the training period for analytical tractability and  focus on the design of the IRS reflection pattern matrix $\Wbf$ in the remainder of this paper. Then, the goal  
 is to design the IRS reflection pattern matrix $\Wbf$ so that channel estimation based on the designed $\Wbf$ yields best performance under a reasonable criterion.  
   Among several existing criteria for training signal design \cite{Hassibi&Hochwald:03IT,Zhang&Qi&Li&Lu:SPAWC20, Hu&Dai:20arXiv, Mishra&Johansson:ICASSP19,He&Yuan:WCL20,Jensen&Carvalho:ICASSP20,Zheng&Zhang:WCL20}, we adopt the CRB on the MSE of unbiased channel estimation \cite{Carvalho&Slock:97SPAWC,Jensen&Carvalho:ICASSP20}, and  optimize the IRS reflection pattern matrix $\Wbf$ by minimizing the CRB as a function of $\Wbf$. The CRB for given parameter $\thetabf$ is expressed as \cite{Kay:book}
 \begin{equation}  \label{eq:CRBintro}
 \Ebb_{\boldsymbol y | \boldsymbol \theta} \{  
 (\thetabf - \hat{\thetabf}(\ybf)) (\thetabf - \hat{\thetabf}(\ybf))^H \}  \succeq \Jbf(\thetabf)^{-1},
 \end{equation}
where $\thetabf$ is the unknown parameter vector, given by $\thetabf = [\alphabf^T, \psibf^T]^T$ in our case, $\hat{\thetabf}(\ybf)$ is any unbiased estimator of $\thetabf$, and $\Jbf(\thetabf)$ is the Fisher information matrix (FIM) given by the covariance matrix of the score function $\frac{\partial}{\partial \boldsymbol \theta} \ln p(\ybf|\thetabf)$, i.e., $\Jbf(\thetabf) =  \Ebb_{\boldsymbol y | \boldsymbol \theta} \left\{ \left(\frac{\partial}{\partial \boldsymbol \theta} \ln p(\ybf|\thetabf)\right) \left(\frac{\partial}{\partial \boldsymbol \theta} \ln p(\ybf|\thetabf)\right)^H     \right\}$.

%%%%%%%%%%%%%%%%%%%%%%%%%%%%%%%%%%%%%%%%%%%%%%%%%%%%%%%%%%%%%%%%%%%%%%%%
\section{Analysis from Bayesian CRB} \label{sec:bayesian_crb_based_design}
%%%%%%%%%%%%%%%%%%%%%%%%%%%%%%%%%%%%%%%%%%%%%%%%%%%%%%%%%%%%%%%%%%%%%%%%

  The derivation of FIM and corresponding CRB depends on the assumption on the parameter $\thetabf = [\alphabf^T, \psibf^T]^T$. In this section, to gain insights into the IRS reflection pattern matrix design with analytical tractability in the sparse channel case, we  consider the Bayesian approach that  assumes a prior distribution on $\thetabf$ and  considers the MSE performance averaged over the prior distribution on $\thetabf$.
The Bayesian CRB is obtained by taking expectation $\Ebb_{\boldsymbol \theta}$ over $p(\thetabf)$ on both sides of \eqref{eq:CRBintro} as \begin{equation}  \label{eq:BayesianCRBintroEQ}
\Ebb_{\boldsymbol \theta} \Ebb_{\boldsymbol y | \boldsymbol \theta} \{  
 (\thetabf - \hat{\thetabf}(\ybf)) (\thetabf - \hat{\thetabf}(\ybf))^H \} \stackrel{(a)}{\succeq} \Ebb_{\boldsymbol \theta} \{ \Jbf(\thetabf)^{-1} \}  \stackrel{(b)}{\succeq} \left(\Ebb_{\boldsymbol \theta}  \{\Jbf(\thetabf)\}  \right)^{-1},
\end{equation}
where Step (b) is valid due to Jensen's inequality on positive-definite matrix inverse. Hence, the inverse of the averaged FIM or Bayesian FIM $\Ebb_{\boldsymbol \theta}  \{\Jbf(\thetabf)\}$ provides a tractable lower bound on the channel estimation MSE averaged over  both $\ybf$ and $\thetabf$.

%%%%%%%%%%%%%%%%%%%%%%%%%%%%%%%%%%%%%%%%%%%%%%%%%%%%%%%%%%%%%%%%%%%%%%%%
\subsection{Derivation of Bayesian CRB}

In high-frequency channels such as mmWave channels, the gain and  angle of a path are typically uncorrelated \cite{Ayach&Rajagopal&AduSurra&Pi&Heath:14WC,Adhikaryetal:14JSAC,GarciaMorales&Femenias&Roy&Castor&Zorzi:20Access}. Hence, we assume  that $\alphabf$ and $\psibf$ are statistically independent with marginal distributions  $p(\alphabf)$ and $p(\psibf)$, respectively. 
From the signal model  \eqref{eq:finalsignalmodel} with additive Gaussian noise,  the joint probability density function (pdf) is written as 
\begin{align}
p(\ybf,\thetabf)=p(\ybf, \alphabf, \psibf)
&=
p(\ybf|\alphabf,\psibf) p(\alphabf)p(\psibf)
=
\frac{1}{\left(\pi \sigma_n^2\right)^K} \text{exp}\left(-\frac{1}{\sigma_n^2}\left\| \ybf - \mbf \right\|^2  \right) p(\alphabf)p(\psibf), \label{eq:like_y_alpha}
\end{align}
where $\mbf := \sqrt{\rho} \widetilde{\Wbf}^H   \left[\alpha_0, \hbf^T\right]^T$, as seen in \eqref{eq:finalsignalmodel}. 
Note that $\alphabf$ is complex-valued whereas $\psibf$ is real-valued.  Hence, it is convenient to consider the real-valued version $\tilde{\thetabf}$ of the parameter vector $\thetabf$,  defined as
%\begin{equation}  \label{eq:brevethetabf}
   $\tilde{\thetabf} := \left[\text{Re}\{\alphabf^T\}, \text{Im}\{\alphabf^T\}, \psibf^T\right]^T$.
%\end{equation}
By using the property of the mapping  $\xbf \mapsto \colvec{\text{Re}\{\mathbf{x}\}\\ \text{Im}\{\mathbf{x}\} }$  for a complex vector $\xbf$,  the Bayesian CRB for the real-valued version of the parameter is given by \cite{Carvalho&Slock:97SPAWC}
\begin{align} \label{eq:bayesianCRBineq}
\mathbb{E}_{\ybf, \boldsymbol\theta} 
\biggl\{ 
\left(\tilde{\thetabf} - g(\ybf)\right) \left(\tilde{\thetabf} - g(\ybf)\right)^H
\biggr\} 
\succeq 
\Jbf_{\tilde{\boldsymbol\theta} \tilde{\boldsymbol\theta}}^{-1},
\end{align}
for any unbiased channel estimator $g(\ybf)$ for $\tilde{\thetabf}$, where  the real-valued version Bayesian FIM $\Jbf_{\tilde{\boldsymbol\theta} \tilde{\boldsymbol\theta}}$ is given by \cite{Carvalho&Slock:97SPAWC}
\begin{align}
&\hspace{10em}
\Jbf_{\tilde{\boldsymbol\theta} \tilde{\boldsymbol\theta}}=
\tilde{\Mbf}
\left[\begin{array}{ccc}
\Jbf_{\boldsymbol\alpha \boldsymbol\alpha} 	& \Jbf_{\boldsymbol\alpha \boldsymbol\alpha^*} 		& \Jbf_{\boldsymbol\alpha \boldsymbol\psi} \\
\Jbf_{\boldsymbol\alpha^* \boldsymbol\alpha}	& \Jbf_{\boldsymbol\alpha^* \boldsymbol\alpha^*} 	& \Jbf_{\boldsymbol\alpha^* \boldsymbol\psi} \\
\Jbf_{\boldsymbol\psi \boldsymbol\alpha}		& \Jbf_{\boldsymbol\psi \boldsymbol\alpha^*} 			& \Jbf_{\boldsymbol\psi  \boldsymbol\psi} \\
\end{array}\right]
\tilde{\Mbf}^H, \label{eq:def_real_fim} \\
%\end{align}
%\begin{align}
&\hspace{4em}\tilde{\Mbf} =
\left[\begin{array}{cc}
\Mbf & \mathbf{0}_{2(L+1)} \mathbf{0}_L^T \\
\mathbf{0}_L \mathbf{0}_{2(L+1)}^T  & \Ibf_{L}
\end{array}\right],~~~~~~
\Mbf
=
\frac{1}{2} 
\left[\begin{array}{ccc}
\Ibf_{L+1} & \Ibf_{L+1} \\
-j\Ibf_{L+1} & j\Ibf_{L+1} 
\end{array}\right],  \label{eq:def_Mtilde_M}  \\
%\end{align}
%\begin{align}
\Jbf_{\boldsymbol\alpha \boldsymbol\alpha}
&=
-\mathbb{E}_{\ybf,\boldsymbol\theta} \left\{ \frac{\partial}{\partial \boldsymbol\alpha^*} 
\left( \frac{\partial \ln p(\ybf, \boldsymbol\alpha, \boldsymbol\psi)}{\partial \boldsymbol\alpha^*} \right)^H \right\},~~~~~
\Jbf_{\boldsymbol\alpha \boldsymbol\alpha^*}
=
-\mathbb{E}_{\ybf, \boldsymbol\theta} \left\{ \frac{\partial}{\partial {\boldsymbol\alpha^*}} 
\left( \frac{\partial \ln p(\ybf, \boldsymbol\alpha, \boldsymbol\psi)}{\partial {\boldsymbol\alpha}} \right)^H \right\}, \label{eq:Jalphaalpha}
\end{align}
and $\Jbf_{\boldsymbol\alpha  \boldsymbol\psi}$, $\Jbf_{\boldsymbol\alpha^*  \boldsymbol\psi}$,  $\Jbf_{\boldsymbol\psi  \boldsymbol\alpha}$,  $\Jbf_{\boldsymbol\psi  \boldsymbol\alpha^*}$ and  $\Jbf_{\boldsymbol\psi  \boldsymbol\psi}$ are similarly defined by considering that $\boldsymbol\psi$ is a real-valued vector.  
Here,   $\mathbb{E}_{\ybf, \boldsymbol\theta}$ denotes the expectation with respect to the joint pdf $p(\ybf,\boldsymbol \theta)=p(\ybf,\alphabf,\psibf)$,  $\Mbf$ in \eqref{eq:def_Mtilde_M} is the complex-to-real conversion matrix \cite{Carvalho&Slock:97SPAWC,Omar&Slock&Bazzi:11PIMRC}, and $\frac{\partial}{\partial {\boldsymbol\alpha}}$ and   $\frac{\partial}{\partial {\boldsymbol\alpha^*}}$ in  \eqref{eq:Jalphaalpha} are the Wirtinger complex derivative \cite{Brandwood:83IEEP}. Note that the Bayesian FIM does not depend on the parameters $\alphabf$ and $\psibf$, and the parameter subscript in \eqref{eq:def_real_fim} is for matrix partition purpose. 

To enable derivation of Bayesian CRB with still incorporating many meaningful distributions \cite{Wang&Fang&Yuan&Chen&Duan&Li:20SPL,Lin&Yu&Zhu&Schober:20Arxiv}, we assume that the direct path gain $\alpha_0$ has non-zero mean $\mu_0$, the reflected path gains $\alpha_\ell$, $\ell=1,\cdots,L$ have zero mean, and  all the path gains $\alpha_\ell$, $\ell=0,1,\cdots,L$ have the same  finite second-order central moment $\sigma^2$, and that 
the path angle is uniformly distributed over $[\Delta_1, \Delta_2]\subset [-1,1]$,  i.e., $\psi_\ell \stackrel{i.i.d.}{\sim}  \mbox{Unif}[\Delta_1,\Delta_2]$, $\ell=1,\cdots,L$.
Under this prior assumption, the Bayesian CRB is given by  the following theorem.

\vspace{0.5em}
\begin{theorem}\label{the:crb_bayesian_general_psi}
For the system model in \eqref{eq:finalsignalmodel} with $\mathbb{E}\{\alphabf\} = \mubf = [\mu_0, \mathbf{0}_{L}^T]^T$, $\mathbb{E}\{(\alphabf -\mubf)(\alphabf -\mubf)^H\} = \sigma^2\Ibf_{L+1}$, and $\psi_\ell \stackrel{i.i.d.}{\sim} \mbox{Unif}[\Delta_1,\Delta_2]$, $\ell=1,\cdots,L$,  the Bayesian CRB for any unbiased estimator of $\tilde{\thetabf}$ is given by $\text{tr}\left(\Jbf_{\tilde{\boldsymbol\theta} \tilde{\boldsymbol\theta}}^{-1}\right)$, where $\Jbf_{\tilde{\boldsymbol\theta} \tilde{\boldsymbol\theta}} = \Jbf_{\tilde{\boldsymbol\theta} \tilde{\boldsymbol\theta}, D} + \Jbf_{\tilde{\boldsymbol\theta} \tilde{\boldsymbol\theta}, P}$. The likelihood part  $\Jbf_{\tilde{\boldsymbol\theta} \tilde{\boldsymbol\theta}, D}$ and the prior part $\Jbf_{\tilde{\boldsymbol\theta} \tilde{\boldsymbol\theta}, P}$   are given by 
\begin{align}
\Jbf_{\tilde{\boldsymbol\theta} \tilde{\boldsymbol\theta}, D}
&=
\tilde{\Mbf}
\left[\begin{array}{cc|c}
\Jbf_{\boldsymbol\alpha \boldsymbol\alpha, D} 	& \mathbf{0}_{L+1} \mathbf{0}_{L+1}^T 					& \mathbf{0}_{L+1}\mathbf{0}_{L}^T \\
\mathbf{0}_{L+1} \mathbf{0}_{L+1}^T				& \Jbf_{\boldsymbol\alpha \boldsymbol\alpha, D}^* 	& \mathbf{0}_{L+1}\mathbf{0}_{L}^T \\ \hline
\mathbf{0}_{L}\mathbf{0}_{L+1}^T 					& \mathbf{0}_{L}\mathbf{0}_{L+1}^T  					& \Jbf_{\boldsymbol\psi  \boldsymbol\psi, D} \\
\end{array}\right]
\tilde{\Mbf}^H  \label{eq:thm_bayesian_J_D}\\
\Jbf_{\tilde{\boldsymbol\theta} \tilde{\boldsymbol\theta}, P}
&=
\tilde{\Mbf}
\left[\begin{array}{cc|c}
\Jbf_{\boldsymbol\alpha \boldsymbol\alpha, P} 		& \mathbf{0}_{L+1}\mathbf{0}_{L+1}^T 				 	& \mathbf{0}_{L+1}\mathbf{0}_{L}^T  \\
\mathbf{0}_{L+1}\mathbf{0}_{L+1}^T 			  	& \Jbf_{\boldsymbol\alpha^* \boldsymbol\alpha^*, P}	& \mathbf{0}_{L+1}\mathbf{0}_{L}^T  \\ \hline
\mathbf{0}_{L}\mathbf{0}_{L+1}^T  					& \mathbf{0}_{L}\mathbf{0}_{L+1}^T   					& \Jbf_{\boldsymbol \psi \boldsymbol \psi,P}
\end{array}\right]
\tilde{\Mbf}^H, \label{eq:thm_bayesian_J_P}
\end{align}
where
\begin{align}
%---------------------------------------------------------------------------
% J_{\alpha \alpha}
\Jbf_{\boldsymbol\alpha \boldsymbol\alpha, D}
&=
\frac{\rho}{\sigma_n^2} \hspace{-0.2em}
\left[\hspace{-0.2em}\begin{array}{cc}
K 						& (\boldsymbol\varphi_{\psi}^H \bar{\wbf})^* \mathbf{1}_L^T \\
(\boldsymbol\varphi_{\psi}^H \bar{\wbf}) \mathbf{1}_L   & 
\left(\displaystyle \sum_{m=0}^{N-1}\sum_{n=0}^{N-1} \varphi_{\psi}((n-m)\pi) [\Qbf]_{m+1, n+1} \right) \Ibf_L + 
\boldsymbol\varphi_{\psi}^H \Qbf \boldsymbol\varphi_{\psi}  (\mathbf{1}_L\mathbf{1}_L^T - \Ibf_L)
\end{array}\hspace{-0.2em}\right],  \label{eq:thm_crb_bayesian_Jaa_D} \\
%---------------------------------------------------------------------------
% J_{\psi \psi}
\Jbf_{\boldsymbol\psi \boldsymbol\psi, D}
&=
\frac{2\rho \sigma^2}{\sigma_n^2} 
\left( \pi^2  \sum_{m=1}^{N-1} \sum_{n=1}^{N-1} mn \varphi_\psi(\pi(n-m)) [\Qbf]_{m+1,n+1} \right)
\Ibf_{L}, \label{eq:thm_crb_bayesian_Jpp_D}  \\
%---------------------------------------------------------------------------
% J_{\alpha \alpha, P}
\Jbf_{\boldsymbol\alpha \boldsymbol\alpha, P}
&=
-\mathbb{E}\left\{
\frac{\partial}{\partial\boldsymbol\alpha^*} \left(\frac{\partial \ln p(\boldsymbol\theta)}{\partial \boldsymbol\alpha^*}\right)^H \right\}, ~
\Jbf_{\boldsymbol\alpha^* \boldsymbol\alpha^*, P}
=
-\mathbb{E}\left\{
\frac{\partial}{\partial\boldsymbol\alpha} \left(\frac{\partial \ln p(\boldsymbol\theta)}{\partial \boldsymbol\alpha}\right)^H \right\}, ~
%---------------------------------------------------------------------------
% J_{\psi \psi, P}
\Jbf_{\boldsymbol\psi \boldsymbol\psi, P}
=
\mathbf{0}_L \mathbf{0}_L^T, \label{eq:Theorem1JththP} \\
% w
\bar{\wbf} &= \left[\bar{w}_1, \ldots, \bar{w}_N \right]^T =  \sum_{k=1}^K \wbf_k,~~~~~ \Qbf=\Wbf\Wbf^H,~~~\mbox{and} \label{eq:theorem1_barWandQ} \\
%\end{align}
%\begin{align}
% varphi_psi
\boldsymbol\varphi_\psi 
&=
\left[{\small \begin{array}{c}
\vspace{-0.5em} 1  \\ \varphi_{\psi}(\pi) \\ \vspace{-0.5em} \vdots  \\  \varphi_{\psi}\left((N-1)\pi\right) 
\end{array}}\right] ~~~\text{with}~~~
\varphi_{\psi}(n\pi) 
=
\frac{2 e^{\iota n\pi \left(\frac{\Delta_2+\Delta_1}{2}\right)}}{ n\pi (\Delta_2 - \Delta_1)}
\sin\left(n\pi \left(\frac{\Delta_2 - \Delta_1}{2}\right)\right).
\label{eq:Theorem1_charac_psi_v2} 
\end{align}
\end{theorem}
{\em Proof:} See Appendix  \ref{subsec:crb_bayesian_general_psi}.
\vspace{0.5em}

\begin{comment}
Note that the likelihood part of the Bayesian FIM is characterized by the first and  second moments of $\alphabf$ and the characteristic function $\varphi_\psi(\cdot)$ of $\psibf$.
\end{comment}

%\vspace{-1em}
%%%%%%%%%%%%%%%%%%%%%%%%%%%%%%%%%%%%%%%%%%%%%%%%%%%%%%%%%%%%%%%%%%%%%%%%
\subsection{Analysis with the Least Information Prior Assumption} \label{subsec:BayesDesign}

Solving the optimization problem for   the reflection pattern matrix by minimizing the Bayesian CRB in Theorem \ref{the:crb_bayesian_general_psi}  
is not easy in general cases.  In this subsection, we consider the least information  prior case under the assumption of Theorem \ref{the:crb_bayesian_general_psi}
and investigate the reflection pattern matrix design in this case. To this end, we assume that the direct path gain is modelled by Rician fading (i.e., $\alpha_0 \sim \mathcal{CN}(\mu_0, \sigma^2)$), the indirect path gain  is modelled by Rayleigh fading (i.e., $\alpha_\ell \sim \Cc\Nc(0,\sigma^2)$ for $\ell=1,\ldots, L$), and the path angle is uniformly distributed  over the entire angle domain (i.e., $\psi_\ell {\sim}  \mbox{Unif}[-1,1]$, $\ell=1,\cdots,L$). These prior distributions of $\alpha_\ell$ and $\psi_\ell$ have maximum entropy under the finite second moment and   finite support assumptions, respectively.
Based on this prior assumption, the Bayesian FIM is given by the following corollary.

\vspace{0.5em}
\begin{corollary}\label{cor:crb_bayesian}
For the signal model  \eqref{eq:finalsignalmodel} with $\alphabf \sim\mathcal{CN}(\mubf, \sigma^2\Ibf_{L+1})$, $\mubf = [\mu_0, \mathbf{0}_{L}^T]^T$, and $\psi_\ell \stackrel{i.i.d.}{\sim} \mbox{Unif}[-1,1]$, $\ell=1,\cdots,L$, 
the diagonal submatrices of $\Jbf_{\tilde{\boldsymbol\theta} \tilde{\boldsymbol\theta}, D}$ in \eqref{eq:thm_bayesian_J_D} and $\Jbf_{\tilde{\boldsymbol\theta} \tilde{\boldsymbol\theta}, P}$ in \eqref{eq:thm_bayesian_J_P} are given by 
\begin{align}
%---------------------------------------------------------------------------
% J_{\alpha \alpha}
\Jbf_{\boldsymbol\alpha \boldsymbol\alpha, D}
&=
\frac{\rho}{\sigma_n^2} 
\left[\begin{array}{cc}
K 						& \bar{w}_1^* \mathbf{1}_L^T \\
\bar{w}_1 \mathbf{1}_L   &   \text{tr}(\Qbf) \Ibf_L + [\Qbf]_{1,1} (\mathbf{1}_L\mathbf{1}_L^T - \Ibf_L)
\end{array}\right], \label{eq:cor_crb_bayesian_Jaa_D} \\
%---------------------------------------------------------------------------
% J_{\psi \psi}
\Jbf_{\boldsymbol\psi \boldsymbol\psi, D}
&=
\frac{2\rho \sigma^2}{\sigma_n^2} 
\left(\pi^2  \sum_{m=1}^{N-1} m^2  [\Qbf]_{m+1,m+1}\right)
\Ibf_{L}, \label{eq:cor_crb_bayesian_Jpp_D}  \\
%---------------------------------------------------------------------------
% J_{\alpha \alpha, P}
\Jbf_{\boldsymbol\alpha \boldsymbol\alpha, P}
&=
\Jbf_{\boldsymbol\alpha^* \boldsymbol\alpha^*, P}
=
\frac{1}{\sigma^2}
\Ibf_{L+1}, ~~~\text{and}~~~\Jbf_{\boldsymbol \psi \boldsymbol \psi,P} = \mathbf{0}_{L}\mathbf{0}_{L}^T.  \label{eq:cor_JththP}
\end{align}
\end{corollary}
{\em Proof:} See Appendix  \ref{subsec:crb_bayesian}.
\vspace{0.5em}

Due to the block diagonal structure of $\Jbf_{\tilde{\boldsymbol\theta} \tilde{\boldsymbol\theta}}$,  the inverse of the Bayesian FIM is given by \begin{align}
% inverse
\Jbf_{\tilde{\boldsymbol\theta} \tilde{\boldsymbol\theta}}^{-1}
&=
\underbrace{\tilde{\Mbf}
\left[\begin{array}{cc}
2\Ibf_{2(L+1)} & \\
&\Ibf_{L}
\end{array}\right]}_{=\tilde{\Mbf}^{-H}}
\left[\begin{array}{cc|c}
\Jbf_{\boldsymbol\alpha \boldsymbol\alpha}^{-1}    & \mathbf{0}_{L+1}\mathbf{0}_{L+1}^T 			 		& \mathbf{0}_{L+1}\mathbf{0}_{L}^T \\
\mathbf{0}_{L+1}\mathbf{0}_{L+1}^T 				& \Jbf_{\boldsymbol\alpha \boldsymbol\alpha}^{*-1}  	& \mathbf{0}_{L+1}\mathbf{0}_{L}^T \\ \hline
\mathbf{0}_{L}\mathbf{0}_{L+1}^T	 				& \mathbf{0}_{L}\mathbf{0}_{L+1}^T  					& \Jbf_{\boldsymbol\psi  \boldsymbol\psi}^{-1} 
\end{array}\right]
\underbrace{\left[\begin{array}{cc}
2\Ibf_{2(L+1)} & \\
&\Ibf_{L}
\end{array}\right]
\tilde{\Mbf}^H}_{=\tilde{\Mbf}^{-1}}, \label{eq:hybrid_crb_blk_inverse_n1}
\end{align}
where $\Jbf_{\boldsymbol\alpha \boldsymbol\alpha}~=\Jbf_{\boldsymbol\alpha \boldsymbol\alpha, D}+\Jbf_{\boldsymbol\alpha \boldsymbol\alpha, P}$, $\Jbf_{\boldsymbol\psi  \boldsymbol\psi}~=\Jbf_{\boldsymbol\psi  \boldsymbol\psi, D}+\Jbf_{\boldsymbol\psi  \boldsymbol\psi, P}=\Jbf_{\boldsymbol\psi  \boldsymbol\psi, D}$, and we used the fact that 
 the inverse of $\Mbf$ in  \eqref{eq:def_Mtilde_M} is $\Mbf^{-1} = 2 \Mbf^H$.

Now, let us consider the IRS reflection pattern design that minimizes the Bayesian CRB given by the trace of \eqref{eq:hybrid_crb_blk_inverse_n1}  while maintaining  
the constant modulus constraint on the reflection pattern matrix $|[\Wbf]_{p,q}| = \sqrt{\beta}$ for all $p, q$.

%%%%%%%%%%%%%%%%%%%%%%%%%%%%
\begin{lemma}\label{lem:eigen_decom_bayesian_crb}
The inverse of the  matrix  $\Jbf_{\boldsymbol\alpha \boldsymbol\alpha} = \Jbf_{\boldsymbol\alpha \boldsymbol\alpha, D}+ \Jbf_{\boldsymbol\alpha \boldsymbol\alpha, P}$ of size $(L+1) \times (L+1)$ in \eqref{eq:hybrid_crb_blk_inverse_n1}  is eigen-decomposed as
\begin{align}
% \Jbf_{\alpha \alpha}^{-1}
\Jbf_{\boldsymbol\alpha \boldsymbol\alpha}^{-1}
&=
\Bbf \times \text{diag}\left(
K + \frac{\tau-\kappa}{2}, 
K + \frac{\tau+\kappa}{2}, 
\left(K + \tau - L[\Qbf]_{1,1} \right)\mathbf{1}_{L-1}^T
\right)^{-1}
\times \Bbf^H, \label{eq:lem_Jaa_decom_n0}
\end{align}
where  $\Bbf$ is a unitary matrix defined in \eqref{eq:lem_Jaa_decom_n2} in Appendix \ref{subsec:eigen_decom_bayesian_crb},  
\begin{align}
\tau 
&=
\frac{\sigma_n^2}{\rho\sigma^2} + \text{tr}(\Qbf)  + (L-1)[\Qbf]_{1,1} - K, ~~\text{and}~~
\kappa 
= 
\sqrt{\tau^2 + 4L|\bar{w}_1|^2}. \label{eq:def_tau_kappa}
\end{align}
Here,  $\Qbf=\Wbf\Wbf^H$ as defined in \eqref{eq:theorem1_barWandQ} and $\bar{w}_1 = \sum_{k=1}^K  [\Wbf]_{1,k}$.
\end{lemma}

{\em Proof:}  See Appendix \ref{subsec:eigen_decom_bayesian_crb}.

\vspace{0.5em}

\begin{comment}
\begin{remark} 
Note that all three distinct eigenvalues of $\Jbf_{\boldsymbol\alpha \boldsymbol\alpha}$ in \eqref{eq:lem_Jaa_decom_n0} are strictly positive. 
Since $\Qbf \succeq 0$ by its definition,  the eigenvalue $K+\tau-L[\Qbf]_{1,1} = \frac{\sigma_n^2}{\rho \sigma^2} + \text{tr}(\Qbf) - [\Qbf]_{1,1} > 0$. The eigenvalue $K+\frac{\tau-\kappa}{2}$ is  positive because of the inequality 
$
\left(2K+\tau\right)^2 
=
\tau^2 + 4K(K+\tau)
\stackrel{(a)}{>} 
\tau^2 + 4L|\bar{w}_1|^2
= 
\kappa^2
$ where the inequality (a) is valid by  the Cauchy-Schwartz inequality and the definition of $\tau$, i.e., 
\begin{align}
|\bar{w}_1|^2 
= 
\left|\sum_{k=1}^K  [\Wbf]_{1,k}\right|^2 
\le
K \sum_{k=1}^K \left|[\Wbf]_{1,k}\right|^2
=
K[\Qbf]_{1,1}
&<
K\frac{K+\tau}{L}. \label{eq:bar_w_1_sq_ineq}
\end{align}
(Here, $[\Qbf]_{1,1} < (\frac{\sigma_n^2}{\rho\sigma^2}+\sum_{i=2}^{N}[\Qbf]_{i,i}+L[\Qbf]_{1,1})/L=(K+\tau)/L$.)
Since  $\kappa \ge 0$ in \eqref{eq:def_tau_kappa} and $K+ \frac{\tau -\kappa}{2} >0$,  the remaining eigenvalue $K+\frac{\tau+\kappa}{2}$ is positive. 
\end{remark}
\end{comment}

\vspace{0.5em}
\begin{proposition} \label{pro:beam_design_bayesian_crb}
The IRS reflection pattern $\Wbf$ minimizing the Bayesian CRB from Corollary \ref{cor:crb_bayesian} satisfies $\sum_{k=1}^K  [\Wbf]_{1,k} = 0$. 
\end{proposition}
{\em Proof:} By  Lemma \ref{lem:eigen_decom_bayesian_crb}, the Bayesian CRB from Corollary \ref{cor:crb_bayesian}  
can be rewritten as
\begin{align}
\text{tr}\bigl(\Jbf_{\tilde{\boldsymbol\theta} \tilde{\boldsymbol\theta}}^{-1}\bigr)
&\stackrel{(a)}{=}
\text{tr}\left(
\left[\begin{array}{ccc}
2\Jbf_{\boldsymbol\alpha\boldsymbol\alpha}^{-1} & \mathbf{0}_{L+1}\mathbf{0}_{L+1}^T& \mathbf{0}_{L+1}\mathbf{0}_{L}^T\\
\mathbf{0}_{L+1}\mathbf{0}_{L+1}^T &2\Jbf_{\boldsymbol\alpha\boldsymbol\alpha}^{*-1} & \mathbf{0}_{L+1}\mathbf{0}_{L}^T\\
\mathbf{0}_{L}\mathbf{0}_{L+1}^T & \mathbf{0}_{L}\mathbf{0}_{L+1}^T &\Jbf_{\boldsymbol\psi\boldsymbol\psi}^{-1}\\
\end{array}\right]
\right) \label{eq:equationTheo21}\\
&\stackrel{(b)}{=}
4 \left(
\frac{1}{K+\frac{\tau-\kappa}{2}} + 
\frac{1}{K+\frac{\tau+\kappa}{2}} +
\frac{L-1}{K+\tau - L[\Qbf]_{1,1}}
\right)
+
\frac{\sigma_n^2}{\rho \sigma^2} \frac{L}{2\pi^2 \sum_{m=1}^{N-1} m^2[\Qbf]_{m+1, m+1}} \\
&\stackrel{(c)}{=}
4 \left(
\frac{1}{K+\frac{\tau-\kappa}{2}} + 
\frac{1}{K+\frac{\tau+\kappa}{2}} +
\frac{L-1}{K+\tau - LK\beta}
\right)
+
\frac{\sigma_n^2}{\rho \sigma^2} \frac{3L}{\pi^2 K\beta (N-1)N(2N-1)}, \label{eq:tr_JthetaR_inv_hybrid}
\end{align}
where Step $(a)$ is computation of 
\eqref{eq:hybrid_crb_blk_inverse_n1} by using $\text{tr}(\Abf\Bbf\Cbf)=\text{tr}(\Bbf\Cbf\Abf)$ and $\tilde{\Mbf}^H\Mbf = \text{diag}(\frac{1}{2}\Ibf_{2(L+1)},$ $\Ibf_L)$ due to  $\Mbf^{-1}=2\Mbf^H$, 
  for Step $(b)$
we use the eigenvalues of $\Jbf_{\boldsymbol\alpha\boldsymbol\alpha}$ in Lemma \ref{lem:eigen_decom_bayesian_crb} and the result of $\Jbf_{\boldsymbol\psi\boldsymbol\psi}$ in \eqref{eq:cor_crb_bayesian_Jpp_D}
from Corollary \ref{cor:crb_bayesian}, and for Step $(c)$ we use the constant modulus constraint  
on $\Wbf$, i.e., $[\Qbf]_{m,m} = \sum_{k=1}^K \left|[\Wbf]_{m,k}\right|^2 = K\beta,~\forall m$ and the series sum $\sum_{m=1}^{N-1}m^2=(N-1)N(2N-1)/6$.  Hence, 
the dependence of 
 $\text{tr}\bigl(\Jbf_{\tilde{\boldsymbol\theta} \tilde{\boldsymbol\theta}}^{-1}\bigr)$ in \eqref{eq:tr_JthetaR_inv_hybrid} on $\Wbf$ is only through $\tau$ and $\kappa$.
 As seen in \eqref{eq:def_tau_kappa}, furthermore, the dependence of $\tau$ on $\Qbf$  disappears due to the constant modulus constraint  
on $\Wbf$ because $\text{tr}(\Qbf)=NK\beta$ and $[\Qbf]_{1,1}=K\beta$ for all $\Wbf$ satisfying the constant modulus constraint. Therefore, the remaining dependence of the CRB on $\Wbf$ is only through $\kappa$, which is a function of 
$|\bar{w}_1|$ only, as seen in \eqref{eq:def_tau_kappa}.
The derivative of the CRB $\text{tr}\bigl(\Jbf_{\tilde{\boldsymbol\theta} \tilde{\boldsymbol\theta}}^{-1}\bigr)$ in \eqref{eq:tr_JthetaR_inv_hybrid} (as a function of $\kappa$) with respect to (w.r.t.) $\kappa$ is given by
\begin{equation}
\frac{d}{d\kappa}\text{tr}\bigl(\Jbf_{\tilde{\boldsymbol\theta} \tilde{\boldsymbol\theta}}^{-1}\bigr)= \frac{1}{2} \frac{\left(K+\tau/2 \right)\kappa}{\left(K+\tau/2-\kappa/2 \right)^2\left(K+\tau/2+\kappa/2 \right)^2}  \ge 0.
\end{equation}
Hence, the  CRB $\text{tr}\bigl(\Jbf_{\tilde{\boldsymbol\theta} \tilde{\boldsymbol\theta}}^{-1}\bigr)$ in \eqref{eq:tr_JthetaR_inv_hybrid} is a monotone increasing function of $\kappa$. Therefore, the minimum CRB $\bigl(\Jbf_{\tilde{\boldsymbol\theta} \tilde{\boldsymbol\theta}}^{-1}\bigr)$ is attained at the smallest value of $\kappa$ and in turn the smallest $\kappa$ is obtained with $|\bar{w}_1|= 0$ from \eqref{eq:def_tau_kappa}. \hfill$\blacksquare$

\vspace{0.5em}

Note that in this case, $\mbox{tr}(\Jbf_{\boldsymbol \psi \boldsymbol \psi}^{-1})$ is given by a constant due to the symmetric prior assumption  $\psi_\ell \stackrel{i.i.d.}{\sim} \mbox{Unif}[-1,1]$ as seen in \eqref{eq:equationTheo21} - \eqref{eq:tr_JthetaR_inv_hybrid}. Hence, the overall CRB reduction by the condition in Proposition \ref{pro:beam_design_bayesian_crb}  is by improving the estimation performance for $\alphabf$.  The received signal \eqref{eq:finalsignalmodel} during the training period (without noise term)  can be rewritten as 
\begin{align}
\ybf 
&=\sqrt{\rho} \left( \alpha_0 \mathbf{1}_K + \Wbf^H \hbf \right) =  \sqrt{\rho} \left(
\alpha_0 \mathbf{1}_K  +  
\left(\sum_{\ell = 1}^L \alpha_\ell \right)
[\Wbf]_{1,1:K}^H
+
[\Wbf]_{2:N,1:K}^H
[\Ubf_{\boldsymbol\psi}]_{2:N,1:L}
\left[\begin{array}{c}
\alpha_1 \\ \vdots \\ \alpha_L
\end{array}\right]
\right), \label{eq:finalsignalmodel_v2} 
\end{align}
since the first element of the ULA response vector $\ubf_N(\psi_\ell)$ is  one.
Due to the condition $\bar{w}_1=\sum_{k=1}^K [\Wbf]_{1,k}=0$ of Proposition \ref{pro:beam_design_bayesian_crb}, the inner product between the first column $\alpha_0 \mathbf{1}_K$ and the second column $(\sum_{\ell = 1}^L \alpha_\ell )
[\Wbf]_{1,1:K}^H$ in the right-hand side (RHS) of \eqref{eq:finalsignalmodel_v2} are orthogonal since $\left(\mathbf{1}_K^T [\Wbf^T]_{1:K,1}\right)^* = \left( \sum_{k=1}^K [\Wbf]_{1,k} \right)^*  = (\bar{w}_1)^*=0$. Hence, we have reduced interference from $\{\alpha_1, \ldots, \alpha_L\}$ to $\alpha_0$, which can facilitate the estimation of $\alphabf$.

%%%%%%%%%%%%%%%%%%%%%%%%%%%%%%%%%%%%%%%%%%%%%%%%%%%%%%%%%%%%%%%%%%%%%%%%
\subsubsection{Quantitative Analysis} \label{subsec:InterpBayesDesign}

% FIGURE
\begin{figure}[h]
\SetLabels
\L(0.263*-0.07) \footnotesize (a)  \\
\L(0.785*-0.07) \footnotesize (b)  \\
% \L(0.73*-0.15) \footnotesize (same legend as in (a)) \\
\endSetLabels
\leavevmode
%\ShowGrid
\strut\AffixLabels{
\includegraphics[scale=0.55]{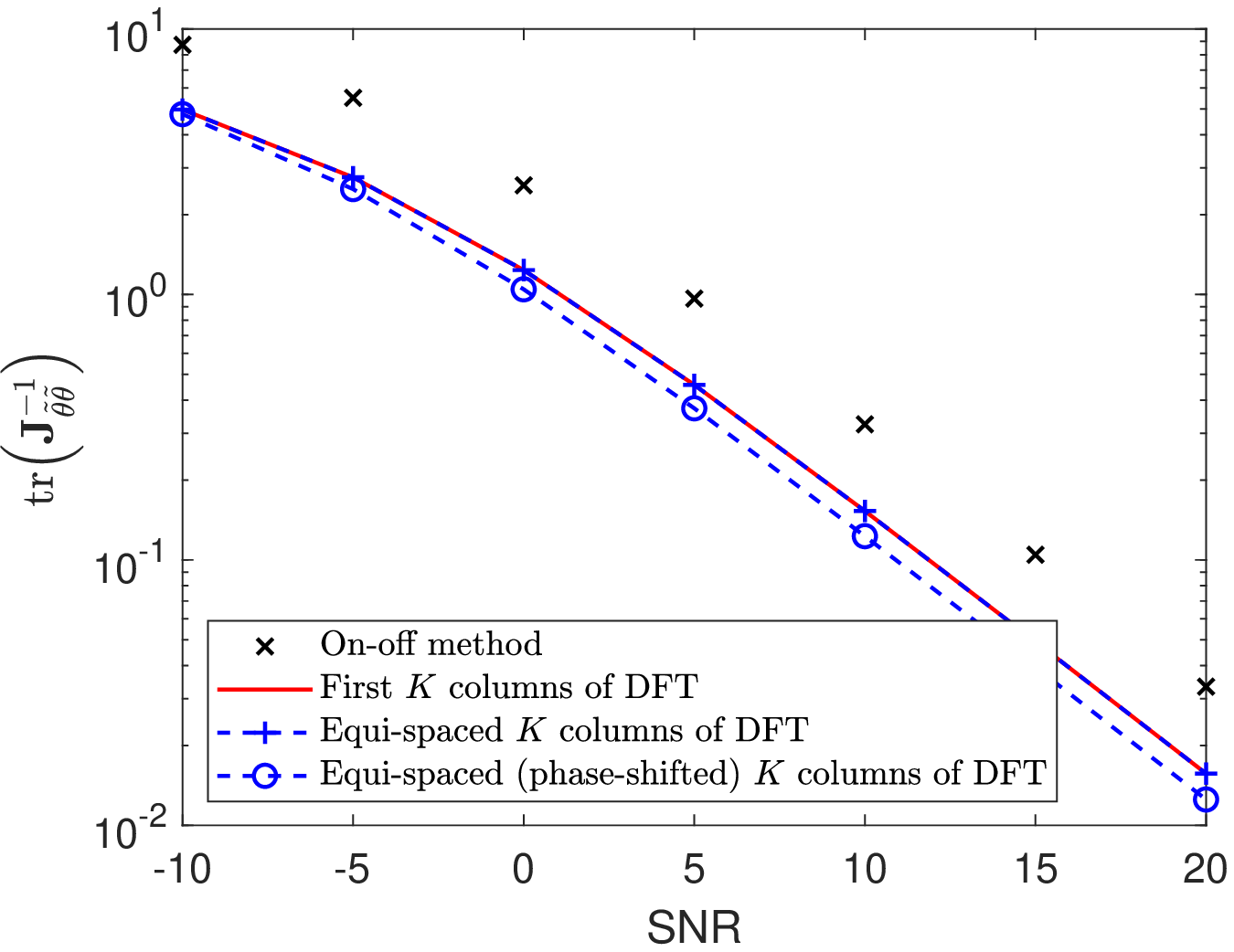}\hspace{2.05em}
\includegraphics[scale=0.55]{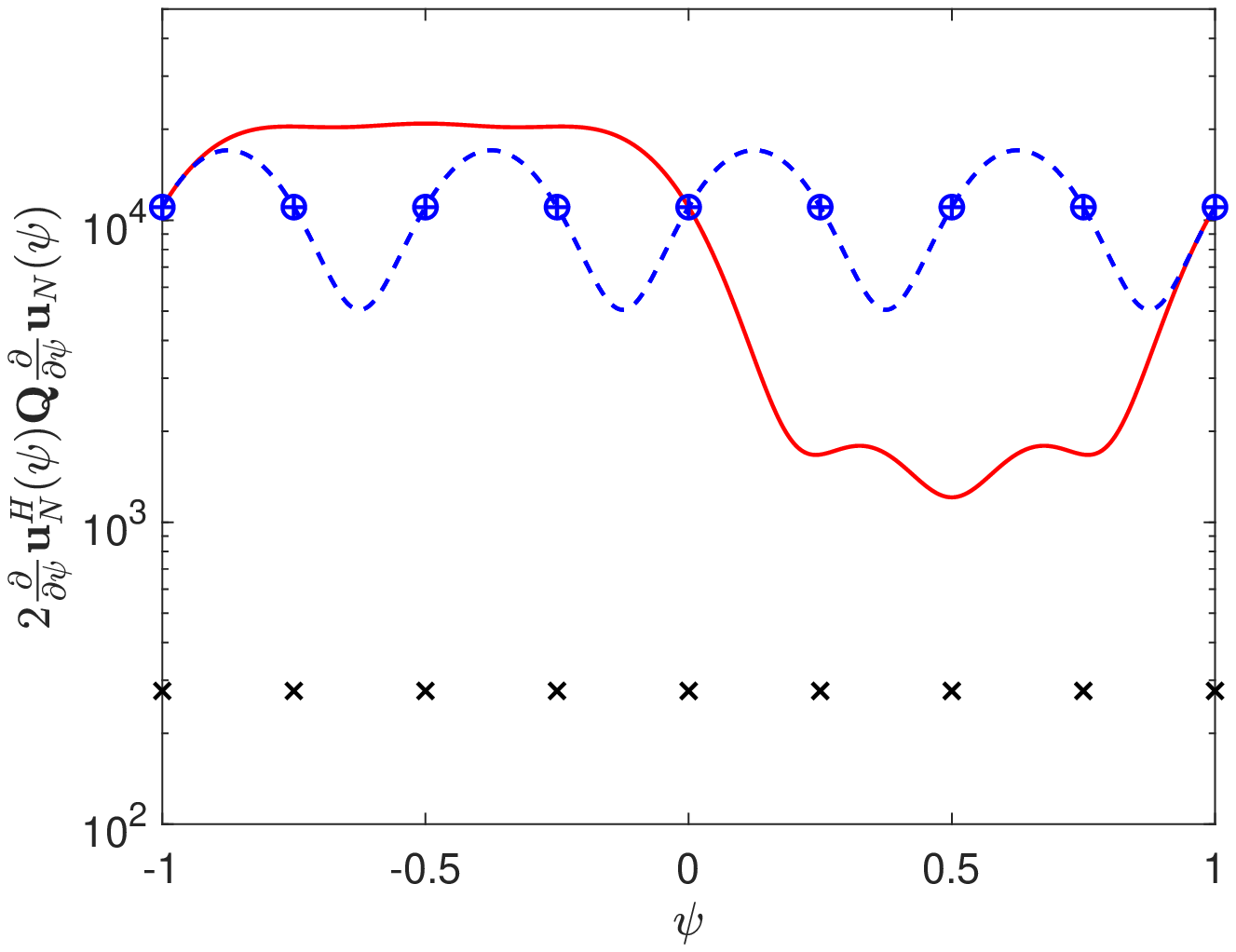}} 
\vspace{0.4em} 
\caption{Bayesian CRB performance: $N=8$, $K=4$, and  $L=2$: 
(a) Bayesian CRB for the overall parameter $\thetabf=[\alphabf^T, \psibf^T]^T$ and
(b) Density of $\Jbf_{\boldsymbol\psi\boldsymbol\psi,D}$ over the angle domain (the same legend as in (a)).}
\label{fig:comparison_bayesian_crb_over_snr} 
\end{figure}

% MINIPAGE
\begin{comment}
\begin{figure}[t]
\begin{minipage}[b]{0.475\linewidth}
\strut\AffixLabels{
\includegraphics[scale=0.57]{figures/crb_bayesian_theta_K4_N8_L2_v4.eps}} 
\vspace{-0.7em} 
\caption{Bayesian CRB for the overall parameter $\thetabf = [\alphabf, \psibf]$ for several IRS reflection patterns ($N=8$, $K=4$, and  $L=2$)}
\label{fig:comparison_bayesian_crb_over_snr} 
\end{minipage}
\quad \hspace{0.5em}
\begin{minipage}[b]{0.475\linewidth}
\strut\AffixLabels{
\includegraphics[scale=0.57]{figures/snapshot_fim_bayesian_psi_K4_N8_L2_v4.eps}} 
\vspace{-0.7em} 
\caption{Density of $\Jbf_{\boldsymbol\psi\boldsymbol\psi,D}$ over the angle domain ($N=8$, $K=4$, and  $L=2$)} 
\label{fig:bayesian_crb_snapshot_over_psi} 
\end{minipage}
\end{figure}
\end{comment}

The sum condition $\sum_{k=1}^K[\Wbf]_{1,k}=0$ in Proposition \ref{pro:beam_design_bayesian_crb} can be applied to previous designs including the DFT-based reflection patterns \cite{Jensen&Carvalho:ICASSP20, Zheng&Zhang:WCL20}. This can be done by adjusting  the first row elements of the reflection patterns so that the sum condition is satisfied. 
Fig. \ref{fig:comparison_bayesian_crb_over_snr}(a) shows the Bayesian CRB performance of several IRS reflection patterns for the estimation of the overall parameter $\alpha_0,\alpha_1,\cdots,\alpha_L,\psi_1,\cdots,\psi_L$  
%when $N=8$, $K=4$, and $L=2$.
when the number of IRS elements  $N=8$, the number of training symbols $K=4$, and the number of paths $L=2$. 
Here, the signal-to-noise ratio (SNR) is defined as $\mbox{SNR}=\rho/\sigma_n^2$.
The on-off method  switching the on-off states of the reflecting elements \cite{Mishra&Johansson:ICASSP19} shows degraded performance over the entire range of SNR. This is because  the on-off switching method does not exploit all  reflecting elements simultaneously for channel estimation.
The orthogonal reflection patterns including the first $K$ columns \cite{Jensen&Carvalho:ICASSP20, Zheng&Zhang:WCL20} and equi-spaced $K$ columns of the $N\times N$ DFT matrix yield improved performance as compared to the on-off method. 
We implemented the condition of Proposition \ref{pro:beam_design_bayesian_crb} on the equi-spaced $K$ columns on the DFT matrix by  progressively shifting the phases of the $K$ column vectors by $e^{j \frac{2\pi}{K}(k-1)}$, $k=1,2,\cdots,K$ so that $\sum_{k=1}^K [\Wbf]_{1,k}=0$ (note that the first row of a DFT matrix is an all-one vector). Indeed, the phase-shifted equi-spaced $K$ columns of the DFT matrix shows further performance improvement, as seen in Fig. \ref{fig:comparison_bayesian_crb_over_snr}(a). 
Fig. \ref{fig:comparison_bayesian_crb_over_snr}(b) shows the quantity $2\frac{\partial}{\partial \psi} \ubf_N^H(\psi) \Qbf \frac{\partial}{\partial \psi} \ubf_N(\psi)$ as a function 
of $\psi$ for the IRS reflection patterns considered in Fig.  \ref{fig:comparison_bayesian_crb_over_snr}(a). Note that 
$\Jbf_{\boldsymbol \psi \boldsymbol \psi, D}$  in \eqref{eq:cor_crb_bayesian_Jpp_D}  whose inverse constitutes the lower diagonal block of $\text{tr}\bigl(\Jbf_{\tilde{\boldsymbol\theta} \tilde{\boldsymbol\theta}}^{-1}\bigr)$ in \eqref{eq:equationTheo21} determining the angle estimation performance  
is given by $\Jbf_{\boldsymbol \psi \boldsymbol \psi, D}= \frac{\rho\sigma^2}{\sigma_n^2} \Ebb_\psi \{  2\frac{\partial}{\partial \psi} \ubf_N^H(\psi) \Qbf \frac{\partial}{\partial \psi} \ubf_N(\psi)  \} \Ibf_L$, as seen in the proof of Theorem 1 in Appendix. Hence, $2\frac{\partial}{\partial \psi} \ubf_N^H(\psi) \Qbf \frac{\partial}{\partial \psi} \ubf_N(\psi)$ shows the density of Fisher information over the angle domain before taking the  expectation over $\psi_\ell \sim \mbox{Unif}[-1,1]$.  High Fisher information density at a certain angle means high estimation performance (i.e., low estimation error) at that angle.   It is seen that the Fisher information density  is angle-dependent for the considered IRS reflection patterns.  
The first $4$ columns of the $8 \times 8$ DFT matrix has high Fisher information density for  $\psi\in[-1,0]$ and low Fisher information density for $\psi\in[0,1]$. This is because the first $4$ columns are equivalent to the orthogonal steering vectors to the look-angles in $[-1,0]$.
On the other hand, the equi-spaced $4$ columns of the $8 \times 8$ DFT matrix spreads out Fisher information over the entire angle range with some ripples. The phase-shifted equi-spaced DFT columns satisfying the condition in Proposition \ref{pro:beam_design_bayesian_crb} has the same value of the integrated Fisher information as the equi-spaced DFT columns. 
% TABLE
% Requires the booktabs if the memoir class is not being used
\begin{table}[h]%[htbp]
%\footnotesize
\small
\centering
%\topcaption{Table captions are better up top} % requires the topcapt package
%\begin{tabular}{@{} lccc @{}} % Column formatting, @{} suppresses leading/trailing space
\begin{tabular}{lccc} % Column formatting, @{} suppresses leading/trailing space
	\toprule
%      	\multicolumn{2}{c}{Item} \\
   	\cmidrule(r){2-4} % Partial rule. (r) trims the line a little bit on the right; (l) & (lr) also possible
      	Method    & Maximum (dB) & Minimum (dB) & Average (dB) \\
     \midrule
		On-off method												& 24.4	& 24.4& 24.4\\
           First $K$ columns of DFT									& 43.2	& 30.8& 40.4\\
      	Equi-spaced $K$ columns of DFT 						& 42.3	& 37.0& 40.4\\
      	Equi-spaced (phase-shifted) $K$ columns of DFT 	& 42.3	& 37.0& 40.4\\
	\bottomrule
\end{tabular}
%\vspace{-0.0em}
\caption{Comparison of several IRS reflection patterns in terms of density of $\Jbf_{\boldsymbol\psi\boldsymbol\psi,D}$.}\label{tab:irs_pattern_comparison}
\vspace{-1.5em}
\end{table}
Table \ref{tab:irs_pattern_comparison} summarizes the maximum, minimum, and average (i.e., integrated) values of $2\frac{\partial}{\partial \psi} \ubf_N^H(\psi) \Qbf \frac{\partial}{\partial \psi} \ubf_N(\psi)$ over the angle domain for the same setup in Fig. \ref{fig:comparison_bayesian_crb_over_snr}.
Note that all the reflection patterns other than the on-off method satisfy the constant modulus constraint  and they have the same integrated Fisher information $\Ebb_\psi \{  2\frac{\partial}{\partial \psi} \ubf_N^H(\psi) \Qbf \frac{\partial}{\partial \psi} \ubf_N(\psi)  \}$ with $\psi_\ell \sim \mbox{Unif}[-1,1]$. At a particular path angle, however,  the angle estimation performance can be different for each reflection pattern because the Fisher information density $2\frac{\partial}{\partial \psi} \ubf_N^H(\psi) \Qbf \frac{\partial}{\partial \psi} \ubf_N(\psi)$ varies over the angle domain and has distinct peak-to-valley height, as seen in Fig. \ref{fig:comparison_bayesian_crb_over_snr}(b).  This  suggests that we can exploit the Fisher information density itself rather than its integrated value when we focus on the angle estimation performance considering the desired performance distribution over the angle domain. This will be investigated in the next section.

\vspace{-0.2em}
%%%%%%%%%%%%%%%%%%%%%%%%%%%%%%%%%%%%%%%%%%%%%%%%%%%%%%%%%%%%%%%%%%%%%%%%
\section{Hybrid CRB-based Reflection Pattern Design} \label{sec:hybrid_crb_based_design}
%%%%%%%%%%%%%%%%%%%%%%%%%%%%%%%%%%%%%%%%%%%%%%%%%%%%%%%%%%%%%%%%%%%%%%%%

To take advantage of the Fisher information density before taking expectation, we need deterministic parameter assumption. However, the full deterministic assumption for both path gains and path angles makes the derivation of CRB  very difficult. We circumvent this difficulty with a hybrid parameter assumption. 
Since  in sparse mmWave channels, the estimation of path angles is crucial to collect the propagation power spread in the space with beamforming and yield high SNR \cite{Alkhateeb&Ayach&Leus&Heath:14STSP,Xiaoetal:17JSAC}, we now take the path angle parameter $\psibf$ as a deterministic  parameter (hence the expectation over $\psibf$ is not taken) whereas we treat the path gain parameter $\alphabf$ as a random nuisance parameter.

\begin{comment}
 Note that in 
\eqref{eq:BayesianCRBintroEQ}, $\Ebb_{\boldsymbol \theta}\{ \Jbf(\theta)^{-1} \}$ is a lower bound on the average MSE directly obtainable from the original deterministic CRB \eqref{eq:CRBintro}, and the Bayesian CRB $(\Ebb_{\boldsymbol \theta}\{ \Jbf(\theta) \})^{-1}$  is a lower bound on the lower bound $\Ebb_{\boldsymbol \theta}\{ \Jbf(\theta)^{-1} \}$. Jensen's inequality used at Step (b) of \eqref{eq:BayesianCRBintroEQ} can be loose when the underlying distribution is not concentrated, and thus the Bayesian CRB 
can be a loose lower bound especially for the angle parameter $\psibf$ with maximum spread $\psi_\ell \sim \mbox{Unif}[-1,1]$.  
\end{comment}

%%%%%%%%%%%%%%%%%%%%%%%%%%%%%%%%%%%%%%%%%%%%%%%%%%%%%%%%%%%%%%%%%%%%%%%%
\subsection{Derivation of Hybrid CRB}

The FIM for the  estimation of the random vector $\alphabf$ and the deterministic vector $\psibf$ can be obtained by averaging $\ybf$ and $\alphabf$ out from the covariance matrix of the score function  conditioned on $\psibf$.  Then, the derived CRB becomes a function of the  path-angle parameter $\psibf$ and thus yields a lower bound on the MSE of estimation for given true underlying parameter $\psibf$. 
From the system model  \eqref{eq:finalsignalmodel}, the conditional pdf of $\ybf$ and $\alphabf$ given $\psibf$ is given by  
\begin{align}
p(\ybf, \alphabf;\psibf)
&=
p(\ybf|\alphabf;\psibf) p(\alphabf)
=
\frac{1}{\left(\pi \sigma_n^2\right)^K} \text{exp}\left(-\frac{1}{\sigma_n^2}\left\| \ybf - \mbf \right\|^2  \right) p(\alphabf), \label{eq:like_y_alphaHybrid}
\end{align}
where   $\mbf = \sqrt{\rho} \widetilde{\Wbf}^H   \left[\alpha_0, \hbf^T\right]^T$ and $\hbf=\Ubf_{\boldsymbol\psi}[\alpha_1,\cdots,\alpha_L]^T$.  The CRB in this hybrid case is provided in the following theorem.

\vspace{0.5em}
\begin{theorem}\label{the:crb_hybrid}
Under  the system model  \eqref{eq:finalsignalmodel} with random path-gain parameter $\alphabf \sim\mathcal{CN}(\mubf, \sigma^2\Ibf_{L+1})$,  $\mubf = [\mu_0, \mathbf{0}_{L}^T]^T$, and deterministic path-angle parameter $\psibf$, 
the hybrid CRB for any unbiased estimator of the overall parameter $\thetabf =[\alphabf^T, \psibf^T]^T$ is given by $\text{tr}\left(\breve{\Jbf}_{\tilde{\boldsymbol\theta} \tilde{\boldsymbol\theta}}^{-1}\right)$, where $\breve{\Jbf}_{\tilde{\boldsymbol\theta} \tilde{\boldsymbol\theta}} = \breve{\Jbf}_{\tilde{\boldsymbol\theta} \tilde{\boldsymbol\theta}, D} + \breve{\Jbf}_{\tilde{\boldsymbol\theta} \tilde{\boldsymbol\theta}, P}$.  The two component matrices    $\breve{\Jbf}_{\tilde{\boldsymbol\theta} \tilde{\boldsymbol\theta}, D}$ and $\breve{\Jbf}_{\tilde{\boldsymbol\theta} \tilde{\boldsymbol\theta}, P}$ of $\breve{\Jbf}_{\tilde{\boldsymbol\theta} \tilde{\boldsymbol\theta}}$ are given by 
\begin{align}
\breve{\Jbf}_{\tilde{\boldsymbol\theta} \tilde{\boldsymbol\theta}, D}
&=
\tilde{\Mbf}
\left[\begin{array}{cc|c}
\breve{\Jbf}_{\boldsymbol\alpha \boldsymbol\alpha, D} 		& \mathbf{0}_{L+1} \mathbf{0}_{L+1}^T 									& \mathbf{0}_{L+1} \mathbf{0}_{L}^T \\
\mathbf{0}_{L+1} \mathbf{0}_{L+1}^T									& \breve{\Jbf}_{\boldsymbol\alpha \boldsymbol\alpha, D}^* 	& \mathbf{0}_{L+1} \mathbf{0}_{L}^T \\ \hline
\mathbf{0}_{L} \mathbf{0}_{L+1}^T										& \mathbf{0}_{L} \mathbf{0}_{L+1}^T 										& \breve{\Jbf}_{\boldsymbol\psi  \boldsymbol\psi, D} \\
\end{array}\right]
\tilde{\Mbf}^H  ~~~\text{and}~~~ 
\breve{\Jbf}_{\tilde{\boldsymbol\theta} \tilde{\boldsymbol\theta}, P}
=
\Jbf_{\tilde{\boldsymbol\theta} \tilde{\boldsymbol\theta}, P},
\label{eq:thm_hybrid_J_P}
\end{align}
where
\begin{align}
%---------------------------------------------------------------------------
% J_{\alpha \alpha}
\breve{\Jbf}_{\boldsymbol\alpha \boldsymbol\alpha, D}
&=
\frac{\rho}{\sigma_n^2} 
\left[\begin{array}{cc}
K & \bar{\wbf}^H \Ubf_{\boldsymbol\psi} \\
\Ubf_{\boldsymbol\psi}^H \bar{\wbf}   &  \Ubf_{\boldsymbol\psi}^H \Qbf \Ubf_{\boldsymbol\psi}
\end{array}\right];~~~~~ 
%---------------------------------------------------------------------------
% J_{\psi \psi}
[\breve{\Jbf}_{\boldsymbol\psi \boldsymbol\psi, D}]_{p,q}
&=
\left\{\begin{array}{ll}
2\frac{\rho\sigma^2}{\sigma_n^2}
\frac{\partial \ubf_N^H(\psi_p)}{\partial \psi_p} \Qbf \frac{\partial \ubf_N(\psi_q)}{\partial \psi_q}, 
& \text{if } p=q \\
0, 
& \text{if } p\neq q 
\end{array} \right.;  \label{eq:thm_crb_hybrid_D_Jpp}
\end{align}
$\Jbf_{\tilde{\boldsymbol\theta} \tilde{\boldsymbol\theta}, P}$, $\bar{\wbf}$ and $\Qbf$ are given in \eqref{eq:thm_bayesian_J_P}  and \eqref{eq:theorem1_barWandQ}; and $\ubf_N(\cdot)$ is the ULA response vector. %in \eqref{eq:def_u_vec}.
\end{theorem}
{\em Proof:} See Appendix \ref{subsec:append_crb_hybrid}.
\vspace{0.5em}

Note that in the hybrid case, the $p$-th diagonal element of $\breve{\Jbf}_{\boldsymbol\psi \boldsymbol\psi, D}$ is nothing but the Fisher information density at $\psi_p$, mentioned in the previous section. Since  $\Jbf_{\tilde{\boldsymbol\theta} \tilde{\boldsymbol\theta}, P}~(=\breve{\Jbf}_{\tilde{\boldsymbol\theta} \tilde{\boldsymbol\theta}, P})$ is block-diagonal, as seen in \eqref{eq:thm_bayesian_J_P}, with its constituent matrices $\Jbf_{{\boldsymbol\alpha} {\boldsymbol\alpha}, P}=\frac{1}{\sigma^2}\Ibf_{L+1}$ and $\Jbf_{{\boldsymbol\psi} {\boldsymbol\psi}, P}={\mathbf{0}}_L {\mathbf{0}}_L^T$,   
the hybrid FIM $\breve{\Jbf}_{\tilde{\boldsymbol\theta} \tilde{\boldsymbol\theta}}$  has a block-diagonal form and we have separate sub-FIMs for the path gains and the path angles, as seen in \eqref{eq:thm_hybrid_J_P}. Thus, the MSE lower bound for the estimation of the path-angle parameter $\psibf$ is determined by the inverse of the lower diagonal block $\breve{\Jbf}_{\boldsymbol\psi\boldsymbol\psi, D}$ due to $\breve{\Jbf}_{{\boldsymbol\psi} {\boldsymbol\psi}, P}={\mathbf{0}}_L {\mathbf{0}}_L^T$, whereas the MSE lower bound for the estimation of the path-gain parameter $\alphabf$ is determined by the upper diagonal element $\breve{\Jbf}_{\boldsymbol \alpha \boldsymbol \alpha  }= \breve{\Jbf}_{\boldsymbol \alpha \boldsymbol \alpha,D  }+\breve{\Jbf}_{\boldsymbol \alpha \boldsymbol \alpha,P }$. 
Both of the FIM submatrices are functions of the design variable $\Qbf~(=\Wbf\Wbf^H)$ and the true angle parameter $\psibf$. Focusing on the path-angle parameter while treating the path-gain parameter as a nuisance parameter, we can optimize the IRS reflection matrix $\Wbf$ in order to yield the best angle-estimation CRB based on   $\breve{\Jbf}_{\boldsymbol\psi\boldsymbol\psi, D}$. (Note that in \eqref{eq:thm_hybrid_J_P}, the lower diagonal block of $\tilde{\Mbf}$ multiplied to $\breve{\Jbf}_{\boldsymbol\psi \boldsymbol\psi, D}$ is the identity matrix, as seen in  \eqref{eq:def_Mtilde_M}.)

\vspace{-1em}
%%%%%%%%%%%%%%%%%%%%%%%%%%%%%%%%%%%%%%%%%%%%%%%%%%%%%%%%%%%%%%%%%%%%%%%%
\subsection{Hybrid CRB-based Reflection Pattern}

As seen in  Theorem \ref{the:crb_hybrid},  the FIM submatrix  $\breve{\Jbf}_{\boldsymbol\psi\boldsymbol\psi, D}$ determining the performance of path angle estimation is a function of the design variable $\Qbf~(=\Wbf\Wbf^H)$ and the true underlying path angle $\psibf$. A key point to note is that the CRB $\mbox{tr}\left(\breve{\Jbf}^{-1}_{\boldsymbol\psi\boldsymbol\psi, D}\right)$ determines the estimation performance when the true path angles are $\psibf=[\psi_1,\cdots,\psi_L]^T$.  Hence, our design approach with the hybrid CRB is that we first design a set of targeted look-angles and then optimize $\Wbf$ so that $\Wbf$ yields best estimation performance when the true path angles are at the targeted look-angles. Here, we can exploit some side information about the path angles if any, e.g., its support information $\psi_\ell \in [\Delta_1,\Delta_2],~\forall \ell$. With such information,  we design the set of targeted look-angles as   $\xi_{L_T, \ell}\in\left\{\Delta_1+\frac{\Delta_2-\Delta_1}{L_T}(\ell-1)~|~ \ell = 1,\ldots, L_T \right\}$, i.e., the  targeted look-angles are evenly spaced over the entire angle range $[\Delta_1,\Delta_2]$ with $L_T$ angles.  Then, the optimization problem is formulated as 
\begin{eqnarray}
\min_{\Wbf} 
&&
\text{tr}\left(\breve{\Jbf}^{-1}_{\boldsymbol\psi \boldsymbol\psi,D}\left| \boldsymbol \psi=\xibf\defeq[\xi_{L_T,1},\cdots,\xi_{L_T,L_T}]\right.\right) \label{eq:opt_prob_hybrid_opt1}\\
\text{subject to} 
&&
|[\Wbf]_{p,q}| = \sqrt{\beta}, ~~~\forall p,q. \label{eq:opt_prob_hybrid_problem}
\end{eqnarray}

\vspace{0.5em}
\begin{remark}
Using the Cauchy-Schwartz inequality $\left(\sum_{\ell=1}^L \sqrt{\lambda_\ell} \frac{1}{\sqrt{\lambda_\ell}} \right)^2 \le \sum_{\ell=1}^{L}\lambda_\ell \cdot \sum_{\ell=1}^L \frac{1}{\lambda_\ell}$, where $\lambda_\ell = 2\frac{\rho\sigma^2}{\sigma_n^2}
\frac{\partial \ubf_N^H(\psi_\ell)}{\partial \psi_\ell} \Qbf \frac{\partial \ubf_N(\psi_\ell)}{\partial \psi_\ell}$,  we have the following inequality regarding the trace of the inverse of $\breve{\Jbf}_{\boldsymbol \psi \boldsymbol \psi,D}$:
\begin{align}
\text{tr}\left(\breve{\Jbf}_{{\boldsymbol\psi} {\boldsymbol\psi},D}^{-1} \right)
\ge 
L^2\left[
\text{tr}\left( \breve{\Jbf}_{{\boldsymbol\psi} {\boldsymbol\psi},D} \right)\right]^{-1}~~~\mbox{for any true }\psibf. \label{eq:obj_ftn_hybrid_crb_v1}
\end{align}
The equality is achieved 
if and only if $\lambda_1 = \cdots = \lambda_L$. This condition can be achieved by 
$\Qbf = \Wbf\Wbf^H = K\beta \Ibf_{N}$ if $K\ge N$.  That is, with $\Qbf=K\beta\Ibf_N$, 
$\lambda_\ell = 2K\beta \frac{\rho\sigma^2}{\sigma_n^2}
\frac{\partial \ubf_N^H(\psi_\ell)}{\partial \psi_\ell} \frac{\partial \ubf_N(\psi_\ell)}{\partial \psi_\ell}$ $=2K\beta \frac{\rho\sigma^2}{\sigma_n^2}\sum_{m=1}^{N-1}m^2
$
does not depend on the index $\ell$. (Please see in \eqref{eq:def_acute_u_n} in Appendix \ref{subsec:crb_bayesian_general_psi} regarding $\frac{\partial \ubf_N(\psi_\ell)}{\partial \psi_\ell}$.) So, we can see the optimality of $\Qbf=K\beta \Ibf_N$ obtained in the previous work \cite{Jensen&Carvalho:ICASSP20, Zheng&Zhang:WCL20} in the case of $K \ge N$ in this sense too.
\end{remark}

\vspace{0.5em}

Since our main focus in this paper is sparse mmWave channels, the situation is  $L \le K < N$. We consider  the optimization problem  \eqref{eq:opt_prob_hybrid_opt1} -  \eqref{eq:opt_prob_hybrid_problem} in the case of $K < N$.  The objective function \eqref{eq:opt_prob_hybrid_opt1} can be rewritten as
\begin{align}
\text{tr}\left(\breve{\Jbf}_{{\boldsymbol\psi} {\boldsymbol\psi}, D}^{-1}\left| \psibf = \xibf \right.\right)
&=
\frac{\sigma_n^2}{2\rho \sigma^2} 
\sum_{\ell=1}^{L_T}  \frac{1}
{\text{tr}\left(\Wbf^H \frac{\partial \ubf_N(\xi_{L_T,\ell})}{\partial \xi_{L_T,\ell}} 
\frac{\partial \ubf_N^H(\xi_{L_T,\ell})}{\partial \xi_{L_T,\ell}} \Wbf\right)} \\
&\stackrel{(a)}{=}
\frac{\sigma_n^2}{2\rho \sigma^2} 
\sum_{\ell=1}^{L_T}  \frac{1}
{\text{vec}\left(\frac{\partial \ubf_N^H(\xi_{L_T,\ell})}{\partial \xi_{L_T,\ell}} \Wbf\right)^H  
\text{vec}\left(\frac{\partial \ubf_N^H(\xi_{L_T,\ell})}{\partial \xi_{L_T,\ell}} \Wbf\right) } \\
&\stackrel{(b)}{=}
\frac{\sigma_n^2}{2\rho \sigma^2} 
\sum_{\ell=1}^{L_T}  \frac{1}
{\overrightarrow{\wbf}^H 
\left(\Ibf_K  \otimes  \frac{\partial \ubf_N(\xi_{L_T,\ell})}{\partial \xi_{L_T,\ell}} \frac{\partial \ubf_N^H(\xi_{L_T,\ell})}{\partial \xi_{L_T,\ell}}\right)
\overrightarrow{\wbf}  } ~\defeq~ f(\overrightarrow{\wbf}), \label{eq:obj_ftn_hybrid_crb_v2}
\end{align}
where  $\overrightarrow{\wbf} = \text{vec}(\Wbf)$, Step $(a)$ is valid due to $\text{tr}(\Abf^H\Bbf) = \text{vec}(\Abf)^H\text{vec}(\Bbf)$, and Step $(b)$ is valid because  $\text{vec}(\Abf\Xbf\Bbf) = (\Bbf^T \otimes \Abf) \text{vec}(\Xbf)$ and $(\Abf_1 \otimes \Bbf_1)(\Abf_2 \otimes \Bbf_2) =  (\Abf_1\Abf_2 \otimes \Bbf_1\Bbf_2)$.
Thus, the objective function is the sum of the inverses of  Rayleigh quotients since  $\overrightarrow{\wbf}^H\overrightarrow{\wbf}=NK\beta$ under the constant modulus constraint $|[\Wbf]_{p,q}|=\sqrt{\beta},\forall p,q$. It is known that even the optimization of the sum of Rayleigh quotients is not a simple problem. In our case, we have the sum of  the inverses of Rayleigh quotients and the elementwise constant modulus constraint in addition.   To tackle this optimization problem, we adopt the PGM  \cite{Bertsekas:76AC,Bubeck:17book}. %\cite{goldsteinPGM}
 PGM is an iterative method applying gradient descent and projection onto the constraint set in an alternating manner  and is  suited to complicated constraints. The overall algorithm is summarized in Algorithm \ref{alg:irs_pattern_pgm}.
The objective function  \eqref{eq:obj_ftn_hybrid_crb_v2} is differentiable and its Wirtinger complex gradient is given by
\begin{align}
\nabla_{\overrightarrow{\wbf}^*}  f(\overrightarrow{\wbf}) 
&= 
- \sum_{\ell=1}^{L_T} \frac{\left(\Ibf_K  \otimes  \frac{\partial \ubf_N(\xi_{L_T,\ell})}{\partial \xi_{L_T,\ell}} \frac{\partial \ubf_N^H(\xi_{L_T,\ell})}{\partial \xi_{L_T,\ell}}\right)}{\left[
\overrightarrow{\wbf}^H 
\left(\Ibf_K  \otimes  \frac{\partial \ubf_N(\xi_{L_T,\ell})}{\partial \xi_{L_T,\ell}} \frac{\partial \ubf_N^H(\xi_{L_T,\ell})}{\partial \xi_{L_T,\ell}}\right)
\overrightarrow{\wbf}
\right]^2}
\overrightarrow{\wbf}.  \label{eq:Wirtinger_grad_obj_ftn_v1}
\end{align}
We start with an initial point $\overrightarrow{\wbf}^{(0)}$. At the $i$-th iteration, we update the current $\overrightarrow{\wbf}^{(i)}$ by using gradient descent based on the gradient vector  \eqref{eq:Wirtinger_grad_obj_ftn_v1} and then project the updated vector onto the constraint set 
$\Cc  := \left\{\overrightarrow{\wbf}=[w_1, \ldots, w_{KN}]^T \in\mathbb{C}^{KN}: |w_i| = \sqrt{\beta} \right\}$.
We repeat this iteration. In the step size determination, we apply the technique in \cite{Watt&Borhani&Katsaggelos:book}  to mitigate the slow crawling problem of gradient decent near a local minimum and help escape from local stationary points of the objective function. That is, we normalize the magnitude of each component of the gradient vector by multiplying the $i$-th component of the gradient vector by its magnitude inverse, as seen in Line 5 of Algorithm \ref{alg:irs_pattern_pgm}.

% ALGORITHM
\begin{algorithm} [t]                      	 % enter the algorithm environment
\caption{The Proposed PGM-based IRS Reflection Pattern Design}	 % give the algorithm a caption
\label{alg:irs_pattern_pgm}            		 % and a label for \ref{} commands later in the document
\begin{algorithmic}[1]                          % enter the algorithmic environment

\REQUIRE  Stopping criterion $\epsilon$, step size $\delta$, and targeted look-angles $\{\xi_{L_T,\ell}\}$
\STATE Compute $\frac{\partial \ubf_N(\xi_{L_T,\ell})}{\partial \xi_{L_T,\ell}}$ for $\ell=1,\cdots,L_T$.
\STATE Set $i=0$ and initialize $\overrightarrow{\wbf}^{(0)}$.
\REPEAT
\STATE Compute the gradient vector \eqref{eq:Wirtinger_grad_obj_ftn_v1} at $\overrightarrow{\wbf}^{(i)}$.

\STATE Update $\overrightarrow{\wbf}^{(i+1)} = \Pi_{\Cc} \left( \overrightarrow{\wbf}^{(i)} - \rbf_{i} \odot \nabla_{\overrightarrow{\wbf}^*}  f(\overrightarrow{\wbf}^{(i)}) \right)$, where $\Pi_{\Cc}$ is the projection onto $\Cc$; $\rbf_i = \left[ \frac{\delta}{\left|\left[\nabla_{\overrightarrow{\wbf}^*}  f(\overrightarrow{\wbf}^{(i)})\right]_1\right|}, \cdots, \frac{\delta}{\left|\left[\nabla_{\overrightarrow{\wbf}^*}  f(\overrightarrow{\wbf}^{(i)})\right]_{NK}\right|}\right]^T $; and $\odot$ denotes elementwise multiplication.
\STATE $i = i+1$ 
\UNTIL{$\left\|\overrightarrow{\wbf}^{(i)} - \overrightarrow{\wbf}^{(i-1)}\right\|^2/\left\|\overrightarrow{\wbf}^{(i-1)}\right\|^2 \le \epsilon$}
\end{algorithmic}
\end{algorithm}

%%%%%%%%%%%%%%%%%%%%%%%%%%%%%%%%%%%%%%%%%%%%%%%%%%%%%%%%%%%%%%%%%%%%%%%
\section{Numerical Results}\label{sec:numerical_result}
%%%%%%%%%%%%%%%%%%%%%%%%%%%%%%%%%%%%%%%%%%%%%%%%%%%%%%%%%%%%%%%%%%%%%%%

In this section, we provide numerical results to evaluate the proposed IRS reflection pattern design.  During the simulation, we set the magnitude of the reflection coefficient of each element at the IRS as  $\beta=1$.    We used the channel model \eqref{eq:def_cascaded_channel_vec}, where the number of the IRS-assisted channel paths was set  $L\in\{1,\ldots, 4\}$,  $\alpha_0\sim \Cc\Nc(\mu_0,\sigma^2)$ and $\alpha_\ell \stackrel{i.i.d.}{\sim} \Cc\Nc(0,\sigma^2)$, $\ell=1,\cdots,L$ with $\mu_0 =  \sigma^2 = 1$. %, and  $\psi_\ell \stackrel{i.i.d.}{\sim} \mbox{Unif}[-1,1]$, $\ell=1,\cdots,L$.  
The SNR is given by  $\mbox{SNR} =\rho / \sigma_n^2$.

% FIGURE
\begin{figure}[t]
%\begin{minipage}[b]{0.475\linewidth}
\SetLabels
\L(0.25*-0.05) \footnotesize (a) \\
\L(0.77*-0.05) \footnotesize (b) \\
\L(0.25*-1.18) \footnotesize (c) \\
\L(0.77*-1.18) \footnotesize (d) \\
\endSetLabels
\leavevmode
%\ShowGrid
\strut\AffixLabels{
\includegraphics[scale=0.55]{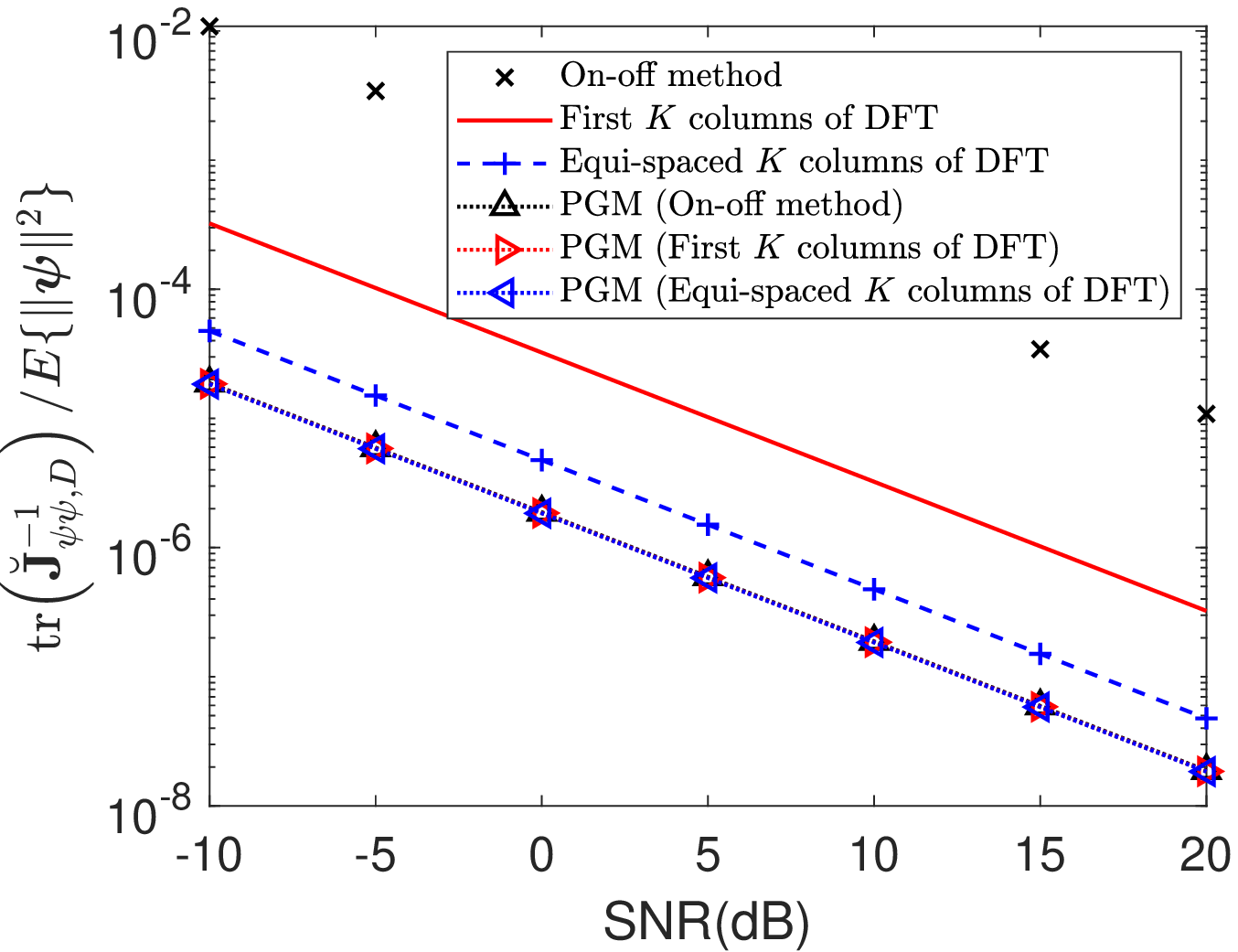}\hspace{1.5em} 
\includegraphics[scale=0.55]{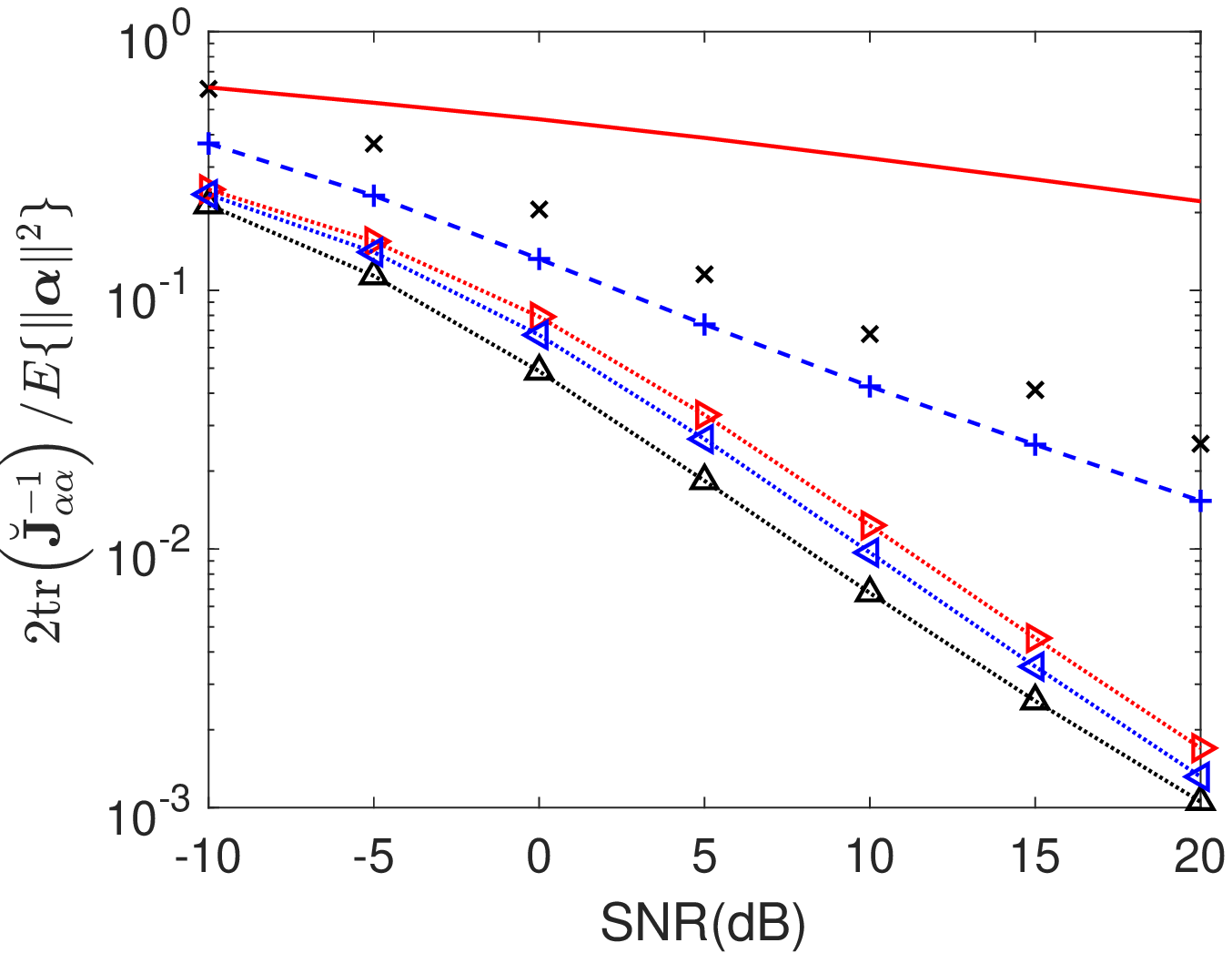} \\
}\\ \\

\vspace{-0.5em}
\strut\AffixLabels{
\hspace{0.7em}
\includegraphics[scale=0.55]{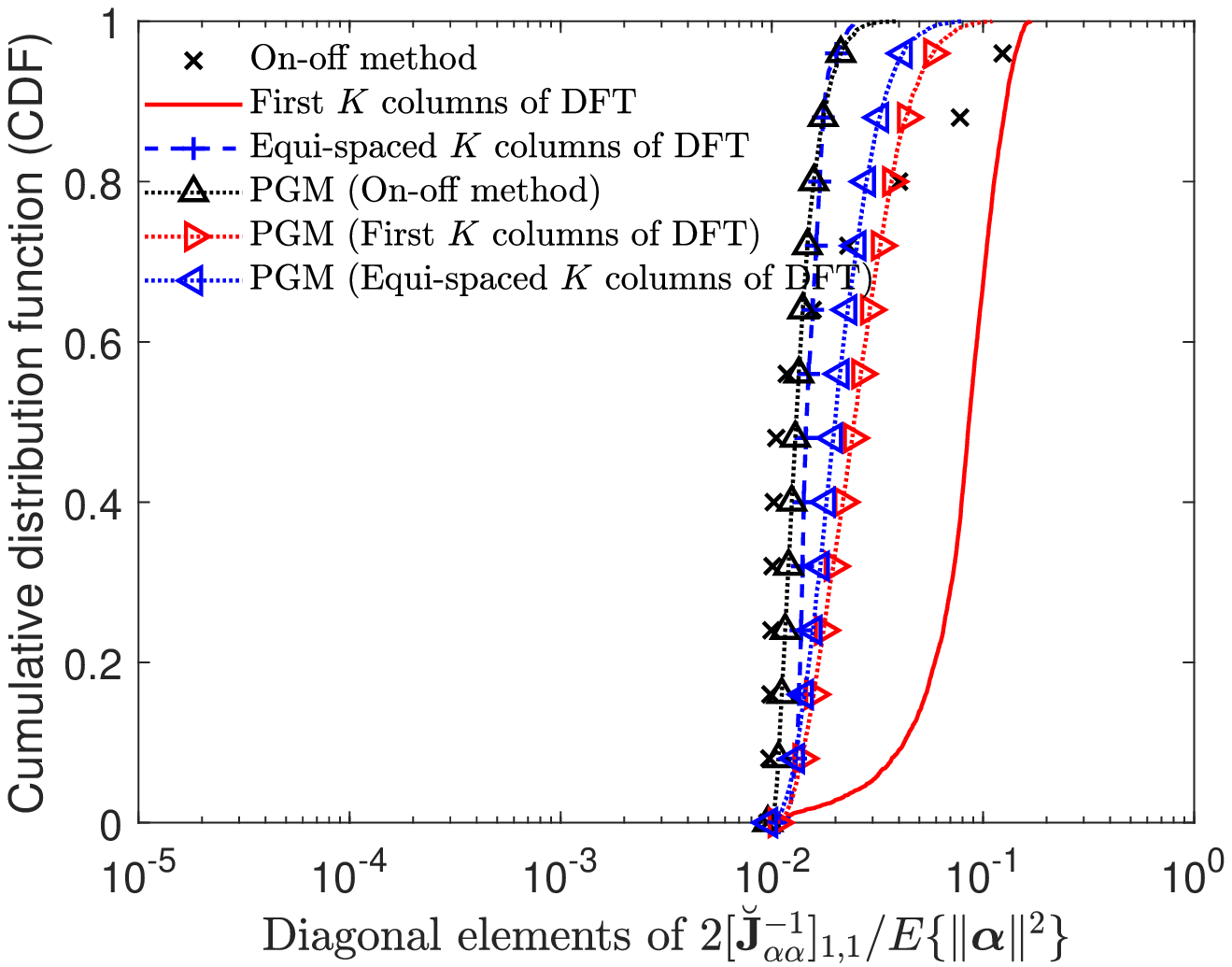}\hspace{2.1em} 
\includegraphics[scale=0.55]{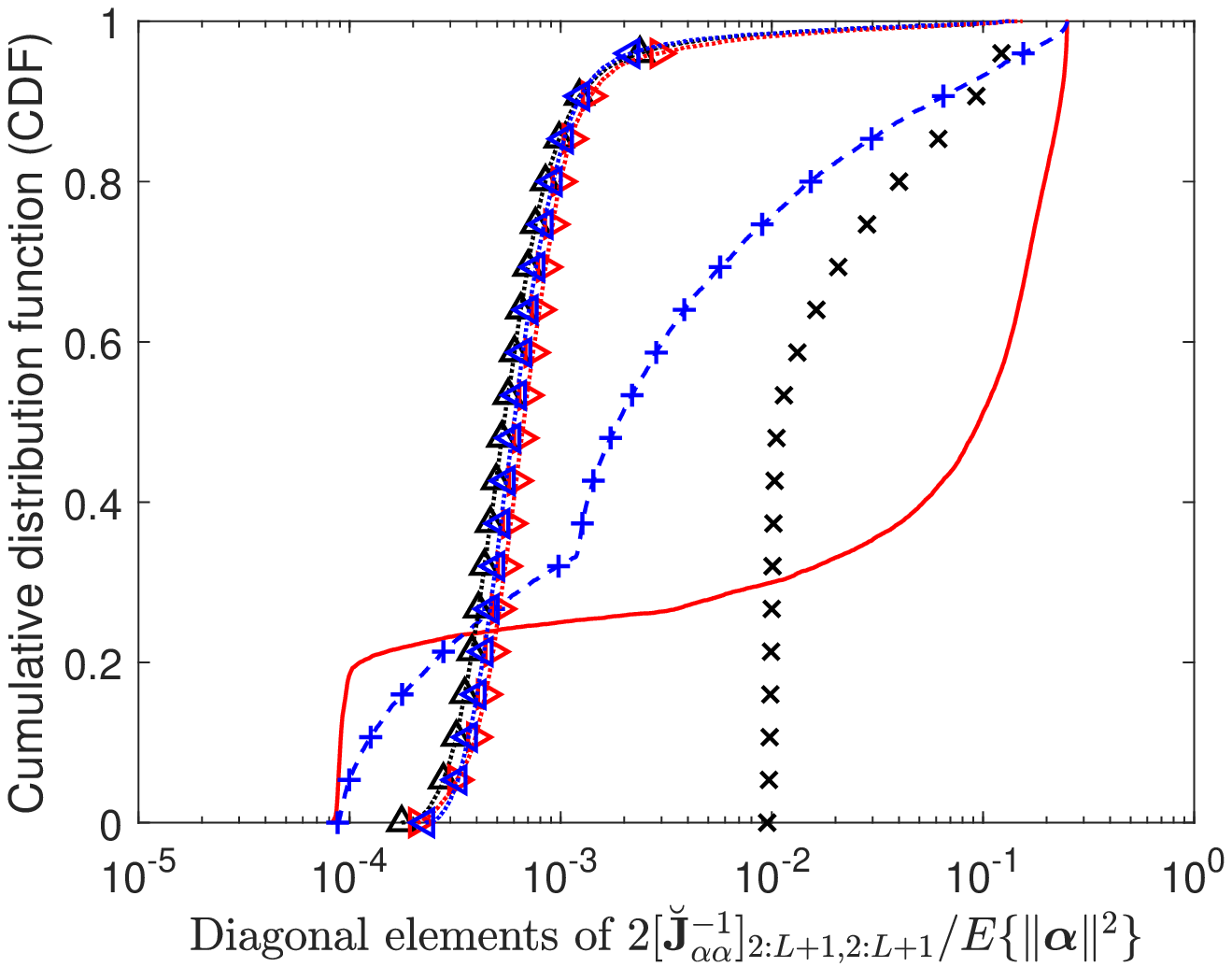}
}
\vspace{0.7em} 
\caption{Hybrid CRB comparision with several IRS reflection patterns versus SNR where $N=32$, $K=8$, and $L=3$ (PGM represents the proposed design with the initial pattern for Algorithm 1 which is shown in the parenthesis.): (a) $\mbox{tr}\bigl(\breve{\Jbf}_{\boldsymbol \psi \boldsymbol \psi,D}^{-1}\bigr)/\mathbb{E}\{\|\boldsymbol\psi\|^2\}$, 
(b) $\bigl(\mbox{tr}\bigl(\breve{\Jbf}_{\boldsymbol \alpha \boldsymbol \alpha}^{-1}\bigr) + \mbox{tr}\bigl(\breve{\Jbf}_{\boldsymbol \alpha^* \boldsymbol \alpha^*}^{-1}\bigr) \bigr)/\mathbb{E}\{\|\boldsymbol\alpha\|^2\}$, 
(c) $\bigl([\breve{\Jbf}_{\boldsymbol \alpha \boldsymbol \alpha}^{-1}]_{1,1} + [\breve{\Jbf}_{\boldsymbol \alpha^* \boldsymbol \alpha^*}^{-1}]_{1,1} \bigr)/\mathbb{E}\{\|\boldsymbol\alpha\|^2\}$ at $\text{SNR}=5$dB, and 
(d) $\text{diag}\bigl([\breve{\Jbf}_{\boldsymbol \alpha \boldsymbol \alpha}^{-1}]_{2:L+1,2:L+1} + [\breve{\Jbf}_{\boldsymbol \alpha^* \boldsymbol \alpha^*}^{-1}]_{2:L+1,2:L+1} \bigr)/\mathbb{E}\{\|\boldsymbol\alpha\|^2\}$ at $\text{SNR}=5$dB.  (The subfigures (a)-(d) have the same legend.)
}
\label{fig:comparison_hybrid_crb_over_snr} 
\hspace{-1em}
\end{figure}

% FIGURE
\begin{figure}[t]
\SetLabels
\L(0.17*-0.07) \footnotesize (a) Hybrid CRB for $\psibf$ \\
\L(0.61*-0.07) \footnotesize (b) Hybrid CRB for $\alphabf$ (same legend as in (a))\\
\endSetLabels
\leavevmode
%\ShowGrid
\strut\AffixLabels{
\includegraphics[scale=0.55]{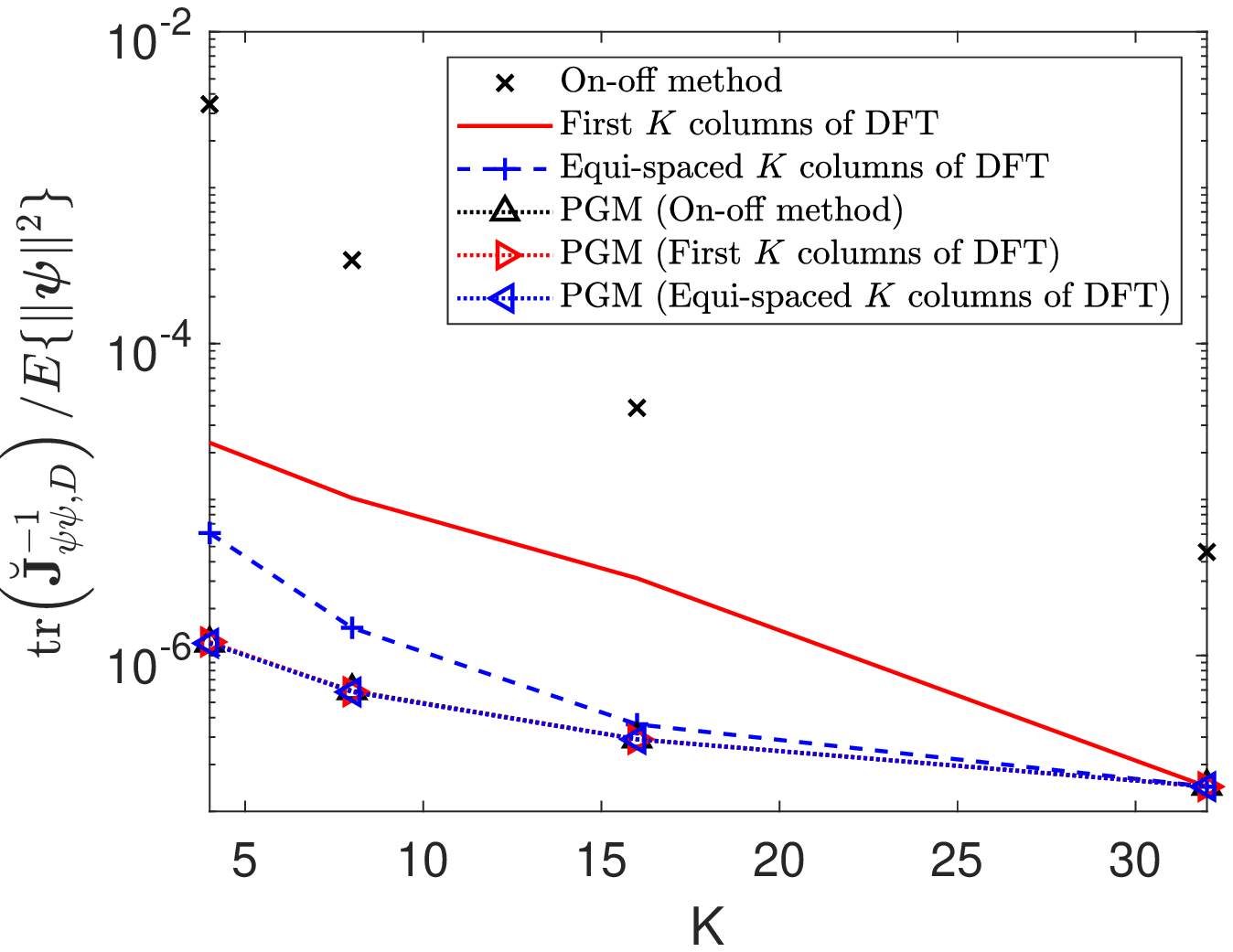} \hspace{1.25em}
\includegraphics[scale=0.55]{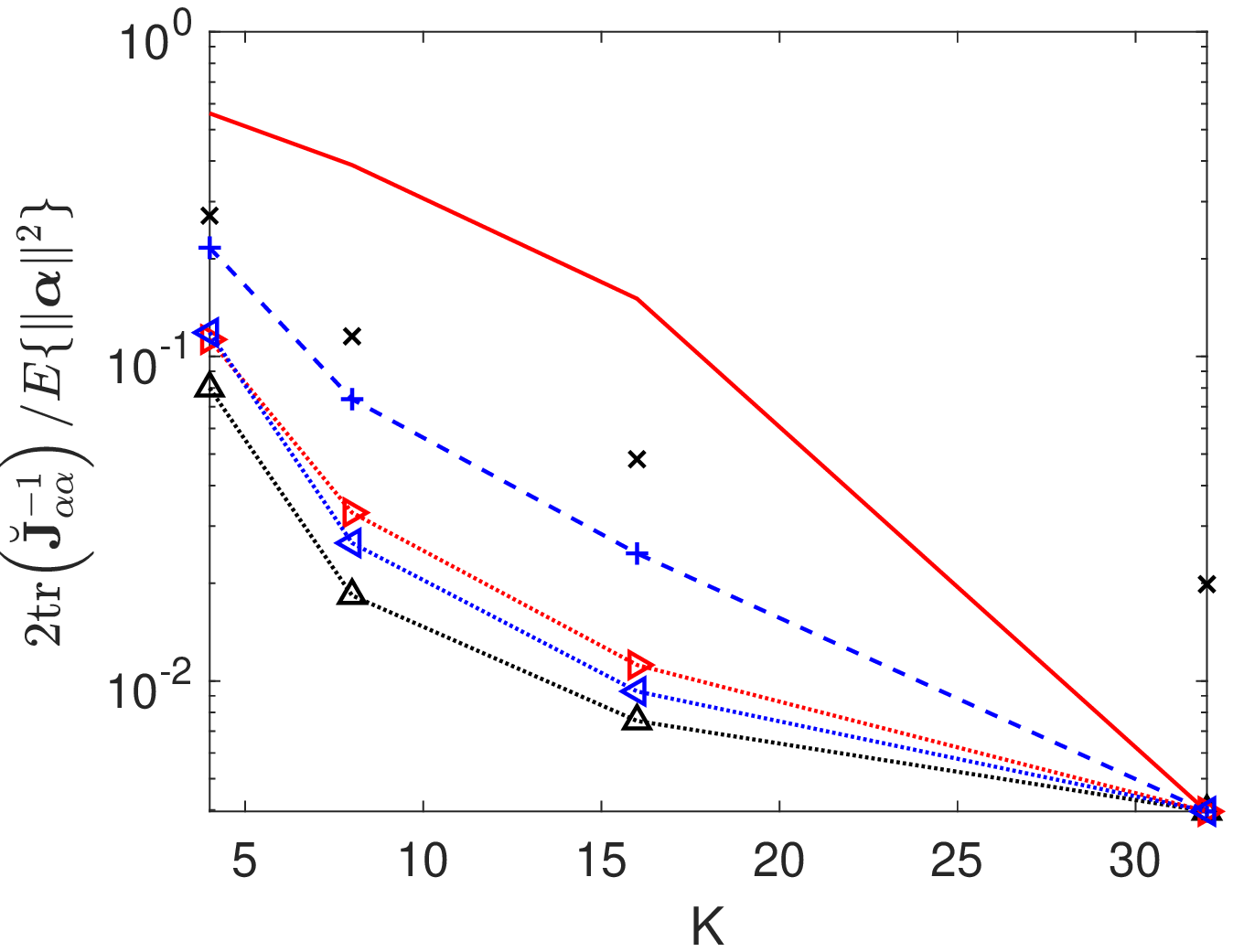}} 
\vspace{0.7em} 
\caption{Hybrid CRB comparision with several IRS reflection patterns versus $K$ where $N=32$, $L=3$, and $\text{SNR}=5$dB.}
\label{fig:comparison_hybrid_crb_over_k} 
\vspace{-0.8em}
\end{figure}
% FIGURE
\begin{figure}[t]
\SetLabels
\L(0.17*-0.07) \footnotesize (a) Hybrid CRB for $\psibf$ \\
\L(0.61*-0.07) \footnotesize (b) Hybrid CRB for $\alphabf$ (same legend as in (a))\\
\endSetLabels
\leavevmode
%\ShowGrid
\strut\AffixLabels{
\includegraphics[scale=0.55]{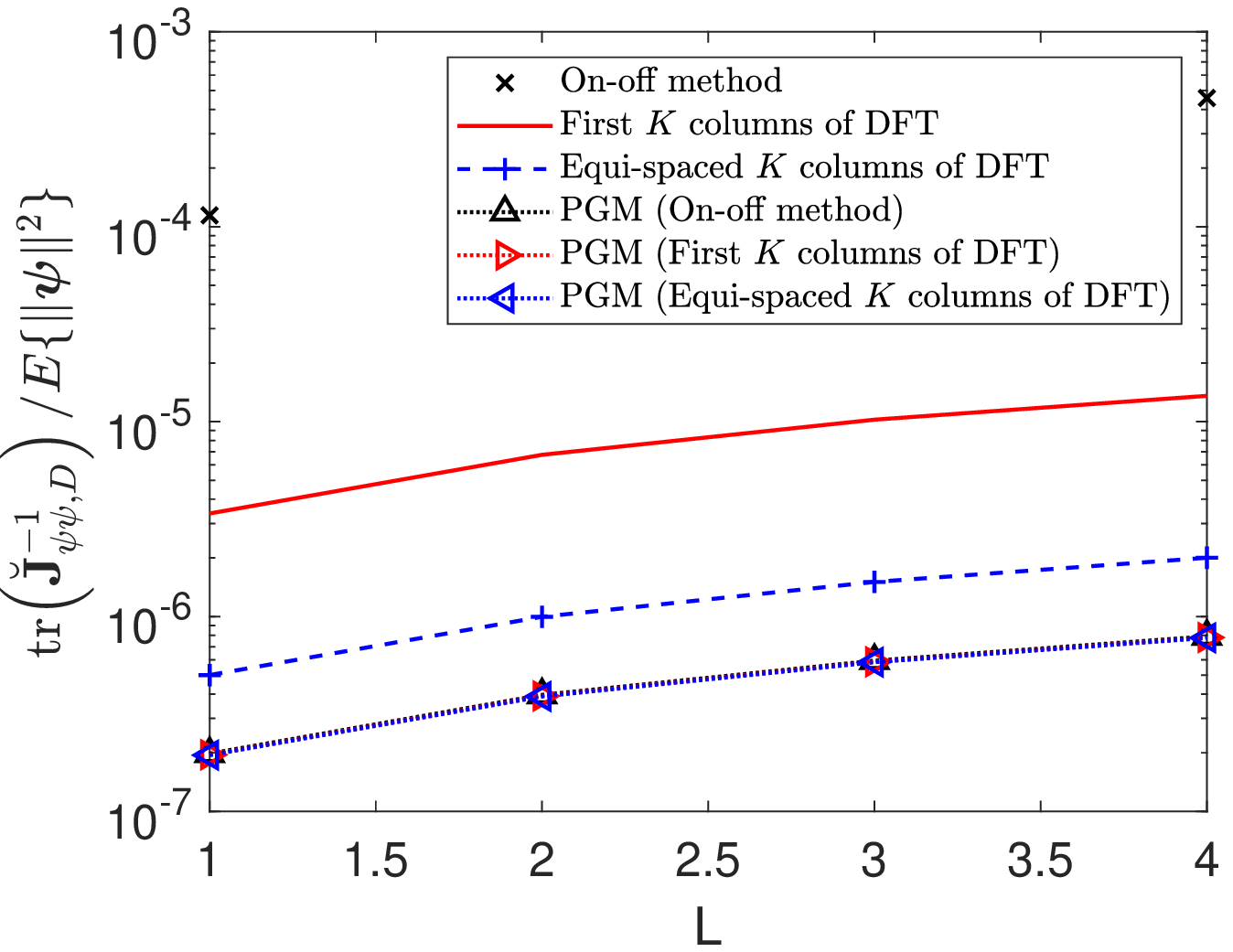} \hspace{1.25em}
\includegraphics[scale=0.55]{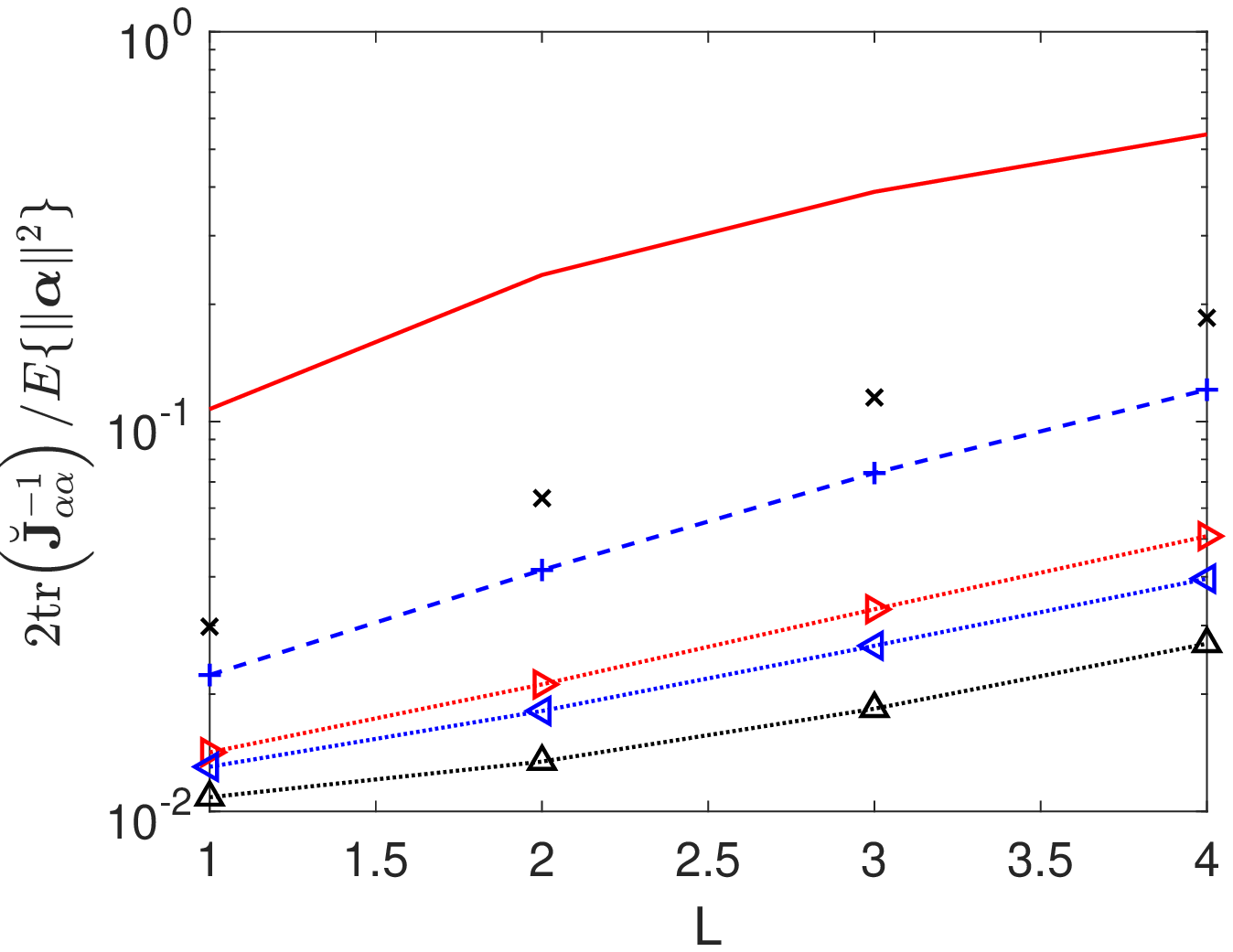}} 
\vspace{0.5em} 
\caption{Hybrid CRB comparision with several IRS reflection patterns versus $L$ where $N=32$, $K=8$, and $\text{SNR}=5$dB.}
\label{fig:comparison_hybrid_crb_over_l} 
\hspace{-1em}
\end{figure}

% Hybird CRB-based beam design
We considered the case of $N=32$ and $K=8$. We designed the IRS pattern matrix by running Algorithm \ref{alg:irs_pattern_pgm} with $\epsilon = 10^{-10}$,   $\delta=1$, and $L_T=N$ targeted look-angles evenly spaced over $[-1,1]$.  Then, we generated $\psi_\ell \stackrel{i.i.d.}{\sim}\text{Unif}[-1,1]$, $\ell=1,\cdots,L$ and computed $\mbox{tr}(\breve{\Jbf}^{-1}_{\boldsymbol \psi \boldsymbol \psi,D})$ and $\mbox{tr}(\breve{\Jbf}^{-1}_{\boldsymbol \alpha \boldsymbol \alpha})$ with $\breve{\Jbf}_{\boldsymbol \alpha \boldsymbol \alpha}=\breve{\Jbf}_{\boldsymbol \alpha \boldsymbol \alpha,D}+\breve{\Jbf}_{\boldsymbol \alpha \boldsymbol \alpha,P}$   by using the designed $\Wbf$ and the generated true angles $\psi_1,\cdots,\psi_L$. We repeated this procedure for $5000$ random channel realizations  and then took average over the $5000$ Monte Carlo runs. Hence, this angle estimation performance average  corresponds to $\Ebb_{\boldsymbol \psi}\{ \breve{\Jbf}^{-1}_{\boldsymbol \psi \boldsymbol \psi,D} \}$, i.e., the bound (a) not the bound (b) in \eqref{eq:BayesianCRBintroEQ}.   For other IRS pattern matrix designs, we applied the same Monte Carlo evaluation.  The result is shown in Fig. \ref{fig:comparison_hybrid_crb_over_snr}. It is seen in Fig. \ref{fig:comparison_hybrid_crb_over_snr}(a) that the equi-spaced $K$ DFT columns shows better performance than the on-off method and the first $K$ DFT columns since $\psi_\ell$ is distributed over the angle range $[-1,1]$. It is also seen that the proposed design yields noticeable gain over  the equi-spaced $K$ DFT columns. Note that the first $K$ DFT columns and the equi-spaced $K$ DFT columns show different performance here, whereas the two methods shows the same performance in Fig. \ref{fig:comparison_bayesian_crb_over_snr}(a).  This is because Fig. \ref{fig:comparison_hybrid_crb_over_snr}(a) shows the tighter bound  (a) in \eqref{eq:BayesianCRBintroEQ}, whereas Fig. \ref{fig:comparison_bayesian_crb_over_snr}(a) shows the looser bound (b) in \eqref{eq:BayesianCRBintroEQ}.
Fig. \ref{fig:comparison_hybrid_crb_over_snr}(b) shows the corresponding path-gain CRB $\mbox{tr}(\breve{\Jbf}^{-1}_{\boldsymbol \alpha \boldsymbol \alpha})$ averaged over the Monte Carlo runs. It is seen that  the first $K$ DFT columns shows severe performance degradation in path gain estimation. This is because this method has a heavy tail in the error distribution due to large gain estimation error occurring when the realized path angles are not within the coverage of the first $K$ DFT columns, as seen in Figs. \ref{fig:comparison_hybrid_crb_over_snr}(c) and (d).

Then, we evaluated the hybrid CRB performance of the IRS reflection patterns with different values of $K$.  
Fig. \ref{fig:comparison_hybrid_crb_over_k} shows that the designs using the orthogonal DFT matrix achieve the best performance when $K=N$ as expected, but there is noticeable gain by the proposed design method as $K$ decreases from $N$. Finally, we examined the performance of the proposed method with respect to the number of channel paths $L$ and the result is shown  in Fig. \ref{fig:comparison_hybrid_crb_over_l}. 
It is seen that the CRB performance degrades as $L$ increases due to the increased number of parameters to estimate. It is also seen that the design by  Algorithm \ref{alg:irs_pattern_pgm} yields better performance than other methods.

%%%%%%%%%%%%%%%%%%%%%%%%%%%%%%%%%%%%%%%%%%%%%%%%%%%%%%%%%%%%%%%%%%%%%%%
\section{Conclusion}\label{sec:conclusion}
%%%%%%%%%%%%%%%%%%%%%%%%%%%%%%%%%%%%%%%%%%%%%%%%%%%%%%%%%%%%%%%%%%%%%%%

We have considered the problem of training signal design for IRS-assisted mmWave communication under a sparse channel model. We have approached the problem based on CRB for channel estimation. With the full Bayesian approach, we have shown that the main factor affecting the angle estimation performance is the Fisher information density given by a quadratic form composed of the derivative of the ULA response and the IRS reflection pattern matrix square $\Qbf$. Based on this fact, we have approached the IRS pattern matrix design problem with a hybrid CRB under the assumption of random path gains and unknown deterministic path angles. We have proposed a PGM-based algorithm to solve optimal IRS pattern matrix design focusing on the path angle estimation critical in mmWave communication.  We have validated the effectiveness of the proposed design method with numerical evaluation.  
We can consider the following future works. First, in this paper we fixed $s_k= \sqrt{\rho}$ to simplify the problem. The joint design of the training symbol sequence and the IRS reflection pattern matrix remains as a future work. Next, development of estimation algorithms achieving the CRB under the signal model \eqref{eq:finalsignalmodel}  should be studied as future work.
%remains as a future work. %\cite{Tellambura:99COM}

%%%%%%%%%%%%%%%%%%%%%%%%%%%%%%%%%%%%%%%%%%%%%%%%%%%%%%%%%%%%%%%%%%%%%%%
\appendix 
%%%%%%%%%%%%%%%%%%%%%%%%%%%%%%%%%%%%%%%%%%%%%%%%%%%%%%%%%%%%%%%%%%%%%%%

%%%%%%%%%%%%%%%%%%%%%%%%%%%%%%%%%%%%%%%%%%%%%%%%%%%%%%%%%%%%%%%%%%%%%%%
\subsection{Proof of Theorem \ref{the:crb_bayesian_general_psi}}\label{subsec:crb_bayesian_general_psi}
%%%%%%%%%%%%%%%%%%%%%%%%%%%%%%%%%%%%%%%%%%%%%%%%%%%%%%%%%%%%%%%%%%%%%%%

{\bf Step 1:} To prove Theorem \ref{the:crb_bayesian_general_psi}, we first provide the following lemma. Note that in the Bayesian case we need the expectation of the second order derivatives of the logarithm of $p(\ybf,\thetabf)$ with $\thetabf=[\alphabf^T,\psibf^T]^T$, as seen in \eqref{eq:bayesianCRBineq} - \eqref{eq:Jalphaalpha}. The logarithm of $p(\ybf,\thetabf)$ is decomposed as  $\ln p(\ybf,\thetabf)= \ln p(\thetabf) + \ln p(\ybf|\thetabf)$ as seen in \eqref{eq:like_y_alpha}, and Lemma \ref{lem:sec_der_prob_data} provides the second order derivatives of $\ln p(\ybf|\thetabf)$.

\vspace{1em}

\begin{lemma} \label{lem:sec_der_prob_data}
The second-order derivatives of the log-likelihood function $\ln p(\ybf | \boldsymbol\theta)$ under  the observation model \eqref{eq:finalsignalmodel}  are given by 
\begin{align}
%---------------------------------------------------------------------------
% \alpha* \alpha*
&\frac{\partial}{\partial\boldsymbol\alpha^*} \left(\frac{\partial \ln p(\ybf | \boldsymbol\theta)}{\partial \boldsymbol\alpha^*}\right)^H
= -
\frac{\rho}{\sigma_n^2} 
\left[\begin{array}{cc}
K & \bar{\wbf}^H \Ubf_{\boldsymbol\psi} \\
\Ubf_{\boldsymbol\psi}^H \bar{\wbf}   &  \Ubf_{\boldsymbol\psi}^H \Qbf \Ubf_{\boldsymbol\psi}
\end{array}\right]  
%---------------------------------------------------------------------------
% \alpha \alpha
=
\left(\frac{\partial}{\partial\boldsymbol\alpha} \left(\frac{\partial \ln p(\ybf | \boldsymbol\theta)}{\partial \boldsymbol\alpha}\right)^H\right)^*  \label{eq:lem_sec_der_n1} \\
%\end{align}
%\begin{align}
%---------------------------------------------------------------------------
% \psi \psi
&\frac{\partial}{\partial\psi_p} \left(\frac{\partial \ln p(\ybf | \boldsymbol\theta)}{\partial \psi_q}\right)^H \nonumber\\ %=
&=%~~~
-\frac{1}{\sigma_n^2} \left( 
-(\ybf - \mbf)^H \frac{\partial^2 \mbf}{\partial \psi_p \partial \psi_q}
+\frac{\partial \mbf^H}{\partial \psi_p} \frac{\partial \mbf}{\partial \psi_q}
+\frac{\partial \mbf^H}{\partial \psi_q} \frac{\partial \mbf}{\partial \psi_p} %\right. \nonumber\\
%&\left. ~~~
-\frac{\partial^2 \mbf^H}{\partial \psi_p \partial \psi_q} (\ybf - \mbf) 
\right)   \label{eq:lem_sec_der_n2} \\
%\end{align}
%\begin{align}
%---------------------------------------------------------------------------
% \alpha* \psi
& \frac{\partial}{\partial \boldsymbol\alpha^*} \left(\frac{\partial \ln p(\ybf | \boldsymbol\theta)}{\partial \psi_\ell}\right)^H \nonumber \\ %=
&=%~~~
-\frac{1}{\sigma_n^2} \left( 
- \sqrt{\rho} \widetilde{\Ubf}_{\psi_{\ell}}^H \widetilde{\Wbf}
\left( \ybf -  
\sqrt{\rho} \widetilde{\Wbf}^H 
\left[\begin{array}{cc}
1 & \\
   & \Ubf_{\boldsymbol\psi}
\end{array}\right] 
\boldsymbol\alpha
\right) %\right. \nonumber \\
%&\left.~~~
+ 
\rho
\left[\begin{array}{cc}
K & \bar{\wbf}^H \\
\Ubf_{\boldsymbol\psi}^H \bar{\wbf}   &  \Ubf_{\boldsymbol\psi}^H \Qbf 
\end{array}\right]  
\widetilde{\Ubf}_{\psi_{\ell}}  \alphabf
\right)  \label{eq:lem_sec_der_n3} \\
%---------------------------------------------------------------------------
% \alpha \psi
&= \left(\frac{\partial}{\partial \boldsymbol\alpha} \left(\frac{\partial \ln p(\ybf | \boldsymbol\theta)}{\partial \psi_\ell}\right)^H\right)^* 
%---------------------------------------------------------------------------
% \psi \alpha*
= \left(\frac{\partial}{\partial \psi_\ell} \left(\frac{\partial \ln p(\ybf | \boldsymbol\theta)}{\partial \boldsymbol\alpha^*}\right)^H\right)^H % \\
%---------------------------------------------------------------------------
% \psi \alpha
= \left(\frac{\partial}{\partial \psi_\ell} \left(\frac{\partial \ln p(\ybf | \boldsymbol\theta)}{\partial \boldsymbol\alpha}\right)^H\right)^T  \label{eq:lem_sec_der_n4} \\
%---------------------------------------------------------------------------
% \alpha* \alpha
&\frac{\partial}{\partial\boldsymbol\alpha^*} \left(\frac{\partial \ln p(\ybf | \boldsymbol\theta)}{\partial \boldsymbol\alpha}\right)^H
=
\frac{\partial}{\partial\boldsymbol\alpha} \left(\frac{\partial \ln p(\ybf | \boldsymbol\theta)}{\partial \boldsymbol\alpha^*}\right)^H
=
\mathbf{0}_{L+1}\mathbf{0}_{L+1}^T,  \label{eq:lem_sec_der_n5}
\end{align}
where  
\begin{align}
\widetilde{\Ubf}_{\psi_\ell}
&\stackrel{\triangle}{=}
\left[\begin{array}{cc}
\mathbf{0}_{N+1},  &   [0,\cdots,0,\underbrace{1}_{\tiny \ell^{th}},0,\cdots,0] \otimes \left[ 0,~ \frac{\partial \ubf_N^T(\psi_\ell)}{\partial \psi_\ell}  \right]^T
\end{array}\right]  \in \mathbb{C}^{(N+1)\times(L+1)} \\
\frac{\partial \ubf_N(\psi)}{\partial \psi}
&\stackrel{\triangle}{=}
\iota\pi
\left[ 0, e^{\iota\pi \psi}, 2e^{\iota 2\pi \psi }, \ldots, (N-1)e^{\iota (N-1) \pi \psi } \right]^T,  \label{eq:def_acute_u_n}
\end{align}
and the variables of $\bar{\wbf}$ and $\Qbf$ are defined in \eqref{eq:theorem1_barWandQ}.
\end{lemma}

{\em Proof:} By differentiating $\ln p(\ybf | \boldsymbol\theta)$ in  \eqref{eq:like_y_alpha} w.r.t. $\alphabf$ and $\psibf$, we have
\begin{align}
%---------------------------------------------------------------------------
% alpha*
\frac{\partial \ln p(\ybf | \boldsymbol\theta)}{\partial \boldsymbol\alpha^*}
&=
-\frac{1}{\sigma_n^2} \left( - \sqrt{\rho}
\left[\begin{array}{cc}
1 & \\ 
   & \Ubf^H_{\boldsymbol\psi}
\end{array}\right]
\widetilde{\Wbf} \ybf
+ \rho
\left[\begin{array}{cc}
1 & \\
   & \Ubf^H_{\boldsymbol\psi}
\end{array}\right]
\widetilde{\Wbf}\widetilde{\Wbf}^H
\left[\begin{array}{cc}
1 & \\
   & \Ubf_{\boldsymbol\psi}
\end{array}\right] \alphabf
\right)  \nonumber \\ %\label{eq:first_order_prob_y_n1}\\
%---------------------------------------------------------------------------
% alpha
&= \left(\frac{\partial \ln p(\ybf | \boldsymbol\theta)}{\partial \boldsymbol\alpha}\right)^*  \label{eq:first_order_prob_y_n2} \\
%---------------------------------------------------------------------------
% psi
\frac{\partial \ln p(\ybf | \boldsymbol\theta)}{\partial \psi_\ell}
&=
-\frac{1}{\sigma_n^2} \left( 
-(\ybf - \mbf)^H \frac{\partial \mbf}{\partial \psi_\ell}
-\frac{\partial \mbf^H}{\partial \psi_\ell} (\ybf - \mbf) 
\right)  \label{eq:first_order_prob_y_n3}\\
&=
-\frac{1}{\sigma_n^2} \biggl( 
- \sqrt{\rho} \ybf^H \widetilde{\Wbf}^H \widetilde{\Ubf}_{\psi_\ell} \alphabf 
- \sqrt{\rho}\alphabf^H \widetilde{\Ubf}_{\psi_\ell}^H \widetilde{\Wbf} \ybf   \nonumber % \\
\end{align}
\begin{align}
&\hspace{7em}
+  \rho\alphabf^H
\left[\begin{array}{cc}
1 & \\
   & \Ubf^H_{\boldsymbol\psi}
\end{array}\right]
\widetilde{\Wbf}\widetilde{\Wbf}^H \widetilde{\Ubf}_{\psi_\ell} \alphabf
+ \rho \alphabf^H  \widetilde{\Ubf}_{\psi_\ell}^H \widetilde{\Wbf}\widetilde{\Wbf}^H
\left[\begin{array}{cc}
1 & \\
   & \Ubf_{\boldsymbol\psi}
\end{array}\right] \alphabf
 \biggr).  \label{eq:first_order_prob_y_n4}
\end{align}
By differentiating \eqref{eq:first_order_prob_y_n2} and \eqref{eq:first_order_prob_y_n3}  w.r.t. $\alphabf$ and $\psibf$ again, we have
\begin{align}
%---------------------------------------------------------------------------
% \alpha* \alpha*
\frac{\partial}{\partial\boldsymbol\alpha^*} \left(\frac{\partial \ln p(\ybf | \boldsymbol\theta)}{\partial \boldsymbol\alpha^*}\right)^H
&=
-\frac{\rho}{\sigma_n^2} 
\left[\begin{array}{cc}
1 & \\
   & \Ubf^H_{\boldsymbol\psi}
\end{array}\right] 
\left[\begin{array}{cc}
K & \bar{\wbf}^H \\
\bar{\wbf} & \Qbf
\end{array}\right] 
\left[\begin{array}{cc}
1 & \\
   & \Ubf_{\boldsymbol\psi}
\end{array}\right]  \\
%---------------------------------------------------------------------------
% \alpha \alpha
\frac{\partial}{\partial\boldsymbol\alpha} \left(\frac{\partial \ln p(\ybf | \boldsymbol\theta)}{\partial \boldsymbol\alpha}\right)^H
&=
-\frac{\rho}{\sigma_n^2} 
\left[\begin{array}{cc}
1 & \\
   & \Ubf^T_{\boldsymbol\psi}
\end{array}\right] 
\left[\begin{array}{cc}
K & \bar{\wbf}^T \\
\bar{\wbf}^* & \Qbf^*
\end{array}\right] 
\left[\begin{array}{cc}
1 & \\
   & \Ubf^*_{\boldsymbol\psi}
\end{array}\right]   \\
%---------------------------------------------------------------------------
%% \psi \psi
\frac{\partial}{\partial\boldsymbol\alpha^*} \left(\frac{\partial \ln p(\ybf | \boldsymbol\theta)}{\partial \boldsymbol\alpha}\right)^H
&=
\frac{\partial}{\partial\boldsymbol\alpha} \left(\frac{\partial \ln p(\ybf | \boldsymbol\theta)}{\partial \boldsymbol\alpha^*}\right)^T
=
\mathbf{0}_{L+1}\mathbf{0}_{L+1}^T,
\end{align}
and the expression for $\frac{\partial}{\partial\psi_p} \left(\frac{\partial \ln p(\ybf | \boldsymbol\theta)}{\partial \psi_q}\right)^H$ is given in \eqref{eq:lem_sec_der_n2}, which can further be detailed like \eqref{eq:first_order_prob_y_n4}. (The detailed expression is omitted due to space limitation.)
By using  \eqref{eq:first_order_prob_y_n2}  and \eqref{eq:first_order_prob_y_n4}, the mixed second-order partial derivative is similarly derived as 
\begin{align}
%---------------------------------------------------------------------------
% \alpha^* \psi
\frac{\partial}{\partial \boldsymbol\alpha^*} \left(\frac{\partial \ln p(\ybf | \boldsymbol\theta)}{\partial \psi_\ell}\right)^H
&=
-\frac{1}{\sigma_n^2} \left[ 
- \sqrt{\rho} \widetilde{\Ubf}_{\psi_{\ell}}^H \widetilde{\Wbf}
\left( \ybf -  
\sqrt{\rho} \widetilde{\Wbf}^H 
\left[\begin{array}{cc}
1 & \\
   & \Ubf_{\boldsymbol\psi}
\end{array}\right] 
\boldsymbol\alpha
\right) \right.  \nonumber \\  
&\left.~~~~~~~~~~~
+ 
\rho
\left[\begin{array}{cc}
1 & \\
   & \Ubf_{\boldsymbol\psi}^H
\end{array}\right]
\widetilde{\Wbf}\widetilde{\Wbf}^H
\widetilde{\Ubf}_{\psi_{\ell}}  \alphabf
\right],  \label{eq:second_order_prob_alpha_psi}
\end{align}
and the expressions for $\frac{\partial}{\partial \boldsymbol\alpha} \left(\frac{\partial \ln p(\ybf | \boldsymbol\theta)}{\partial \psi_\ell}\right)^H$, $\frac{\partial}{\partial \psi_\ell} \left(\frac{\partial \ln p(\ybf | \boldsymbol\theta)}{\partial \boldsymbol\alpha^*}\right)^H$, and $\frac{\partial}{\partial \psi_\ell} \left(\frac{\partial \ln p(\ybf | \boldsymbol\theta)}{\partial \boldsymbol\alpha}\right)^H$ are given  in \eqref{eq:lem_sec_der_n4}.  \hfill$\blacksquare$

\vspace{0.5em}

{\bf Step 2:}  The FIM $\Jbf_{\tilde{\boldsymbol\theta} \tilde{\boldsymbol\theta}}$ is obtained by applying $-\mathbb{E}_{\ybf,\boldsymbol\theta}\{\cdot\}$ to the second-order derivatives of  $\ln p(\ybf,\thetabf)= \ln p(\thetabf) + \ln p(\ybf|\thetabf)$.
First, consider the log-likelihood part $\Jbf_{\tilde{\boldsymbol\theta} \tilde{\boldsymbol\theta},D}$ for which the second-order derivatives are given in Lemma \ref{lem:sec_der_prob_data}. The corresponding submatrices are given as %follows:
\begin{align}
%---------------------------------------------------------------------------
% J_{\alpha \alpha}
\Jbf_{\boldsymbol\alpha \boldsymbol\alpha, D}
&= -
\mathbb{E}_{\ybf, \boldsymbol\theta}\left\{
\frac{\partial}{\partial\boldsymbol\alpha^*} \left(\frac{\partial \ln p(\ybf | \boldsymbol\theta)}{\partial \boldsymbol\alpha^*}\right)^H 
\right\} 
=
\frac{\rho}{\sigma_n^2} 
\left[\begin{array}{cc}
K & \nu_1^* \mathbf{1}_L^T \\
\nu_1 \mathbf{1}_L   &   \zeta_1 \Ibf_L + \eta_1 (\mathbf{1}_L\mathbf{1}_L^T - \Ibf_L)%\Ibf_L^c 
\end{array}\right] \label{eq:crb_bayesian_Jaa_D_proof} \\
%---------------------------------------------------------------------------
% J_{\alpha \psi}
\Jbf_{\boldsymbol\alpha \boldsymbol\psi, D}
&= -
\mathbb{E}_{\ybf, \boldsymbol\theta}\left\{
\frac{\partial}{\partial \boldsymbol\alpha^*} \left(\frac{\partial \ln p(\ybf | \boldsymbol\theta)}{\partial \boldsymbol\psi}\right)^H 
\right\} % \nonumber\\
= 
\frac{\rho}{\sigma_n^2}
\mathbb{E}_{\boldsymbol\theta}\left\{  \hspace{-0.2em}
\left[\hspace{-0.1em}\begin{array}{cc}
K & \bar{\wbf}^H  \\
\Ubf_{\boldsymbol \psi}^H\bar{\wbf}   &  \Ubf_{\boldsymbol \psi}^H\Qbf 
\end{array}\hspace{-0.1em}\right]  \hspace{-0.4em}
\left[\hspace{-0.1em}\begin{array}{c}
\mathbf{0}_L^T \\
\acute{\Ubf}_{\boldsymbol\psi}
\end{array}\hspace{-0.1em}\right] \hspace{-0.1em}
\text{diag}(\alpha_1, \ldots, \alpha_L) \hspace{-0.1em}
\right\} \nonumber%\\
\end{align}
\begin{align}
&=
\frac{\rho}{\sigma_n^2}  
\left[\begin{array}{c}
\nu_2^*\mathbf{1}_L^T\\
\zeta_2 \Ibf_L + \eta_2 (\mathbf{1}_L\mathbf{1}_L^T - \Ibf_L)
\end{array}\right] 
\text{diag}(\mathbb{E}\{\alpha_1\}, \ldots,  \mathbb{E}\{\alpha_L\})
\label{eq:crb_bayesian_Jap_D_proof_n2}\\ 
%---------------------------------------------------------------------------
% J_{\psi \psi}
\Jbf_{\boldsymbol\psi \boldsymbol\psi, D}
&= -
\mathbb{E}_{\ybf, \boldsymbol\theta}\left\{ 
\frac{\partial}{\partial\boldsymbol\psi} 
\left(\frac{\partial \ln p(\ybf | \boldsymbol\theta)}{\partial \boldsymbol\psi}\right)^H
\right\}  
= \frac{1}{\sigma_n^2}
\mathbb{E}_{\ybf, \boldsymbol\theta}\left\{  \frac{\partial \mbf}{\partial \psibf}\frac{\partial \mbf^H}{\partial \psibf} +\left(\frac{\partial \mbf}{\partial \psibf}\frac{\partial \mbf^H}{\partial \psibf}\right)^*  \right\} \nonumber\\
&=
\frac{2\rho}{\sigma_n^2}  \text{Re}\left\{  
\zeta_3   
\text{diag}(\mathbb{E}\{|\alpha_1|^2\}, \ldots,  \mathbb{E}\{|\alpha_L|^2\})  \nonumber \right. \\ 
&\left. \hspace{1em}+
\eta_3  
\text{diag}(\mathbb{E}\{\alpha_1\}, \ldots,  \mathbb{E}\{\alpha_L\})^H
(\mathbf{1}_L\mathbf{1}_L^T - \Ibf_L)
\text{diag}(\mathbb{E}\{\alpha_1\}, \ldots,  \mathbb{E}\{\alpha_L\})
\right\},  \label{eq:crb_bayesian_Jpp_D_proof}
\end{align}
where
\begin{align}
\nu_1
&=
\boldsymbol\varphi_{\psi}^H \bar{\wbf}, \hspace{2em}
\zeta_1
=
\sum_{m=0}^{N-1}\sum_{n=0}^{N-1} \varphi_{\psi}((n-m)\pi) [\Qbf]_{m+1, n+1}, \hspace{5.1em}
\eta_1
=
\boldsymbol\varphi_{\psi}^H \Qbf \boldsymbol\varphi_{\psi},  \label{eq:nu_zeta_eta_n1}\\
\nu_2
&=
\acute{\boldsymbol\varphi}_\psi^H \bar{\wbf}, \hspace{2em}
\zeta_2
= 
\iota\pi \sum_{m=0}^{N-1} \sum_{n=1}^{N-1} n \varphi_\psi((n-m)\pi)  [\Qbf]_{m+1,n+1}, \hspace{3.4em}
\eta_2 
= 
\boldsymbol\varphi_{\psi}^H \Qbf \acute{\boldsymbol\varphi}_{\psi},  \label{eq:nu_zeta_eta_n2}\\
&\hspace{6em}
\zeta_3
= 
 \pi^2  \sum_{m=1}^{N-1} \sum_{n=1}^{N-1} mn \varphi_\psi(\pi(n-m)) [\Qbf]_{m+1,n+1}, \hspace{2.5em}
\eta_3
=
 \acute{\boldsymbol\varphi}_{\psi}^H \Qbf \acute{\boldsymbol\varphi}_{\psi},  \label{eq:nu_zeta_eta_n3}\\
\acute{\boldsymbol\varphi}_\psi 
&\stackrel{\triangle}{=} \iota\pi \left[\begin{array}{c}
0 \\ \varphi_\psi(\pi) \\ \vdots \\ (N-1)\varphi_\psi\left((N-1)\pi\right) 
\end{array}\right]
~~\text{with the characteristic function}~\varphi_{\psi}(t) = \mathbb{E}\left\{e^{\iota t\psi}\right\}  \nonumber\\
\acute{\Ubf}_{\boldsymbol\psi}
&\stackrel{\triangle}{=}
\left[\begin{array}{ccc}
\frac{\partial \ubf_N(\psi_1)}{\partial \psi_1}  & \cdots & \frac{\partial \ubf_N(\psi_L)}{\partial \psi_L} 
\end{array}\right] \in\mathbb{C}^{N \times L}.  \label{eq:def_acute_U_psi}
\end{align}
%%%%%%%%%%%%%%%
Under the the parameter distribution assumptions of $\psi_\ell \sim \text{Unif}[\Delta_1,\Delta_2]$, 
$\mathbb{E}\{\alpha_\ell\}=0$, and $\mathbb{E}\{|\alpha_\ell |^2\}=\sigma^2$ for $\ell=1, \ldots, L$, we have 
\begin{align}
\varphi_\psi\left(n\pi\right)
&=
\left\{\begin{array}{ll}
1,  & \text{if } n=0 \\
\frac{2 e^{\iota n\pi \left(\frac{\Delta_2+\Delta_1}{2}\right)}}{ n\pi (\Delta_2 - \Delta_1)}
\sin\left(n\pi \left(\frac{\Delta_2 - \Delta_1}{2}\right)\right), & \text{if } n\ne 0
\end{array}\right.,  \label{eq:charac_general_psi}\\
\Jbf_{\boldsymbol\alpha \boldsymbol\psi, D}
&= 
\mathbf{0}_{L+1} \mathbf{0}_{L}^T, ~~~\text{and}~~~
\Jbf_{\boldsymbol\psi \boldsymbol\psi, D}
=
\frac{2\rho\sigma^2}{\sigma_n^2}  \zeta_3   \Ibf_L.
\end{align}
%%%%%%%%%%%%%%%
By Lemma \ref{lem:sec_der_prob_data}, $\Jbf_{\boldsymbol \alpha^*\boldsymbol \alpha^*}=(\Jbf_{\boldsymbol \alpha\boldsymbol \alpha})^*$, and  $\Jbf_{\boldsymbol\alpha^* \boldsymbol\psi, D}$,  $\Jbf_{\boldsymbol\psi \boldsymbol\alpha, D}$, and $\Jbf_{\boldsymbol\psi \boldsymbol\alpha^*, D}$ are given by the complex conjugate, Hermitian conjugate, and transpose of $\Jbf_{\boldsymbol\alpha \boldsymbol\psi, D}$, respectively, as seen in \eqref{eq:lem_sec_der_n4}. 
The remaining off-diagonal matrices $\Jbf_{\boldsymbol\alpha \boldsymbol\alpha^*, D}$ and $\Jbf_{\boldsymbol\alpha^* \boldsymbol\alpha, D}$ become zero matrices
by \eqref{eq:lem_sec_der_n5}. 
Hence, we have the FIM  $\Jbf_{\tilde{\boldsymbol\theta} \tilde{\boldsymbol\theta}, D}$, as shown in \eqref{eq:thm_bayesian_J_D} with
\eqref{eq:thm_crb_bayesian_Jaa_D} and \eqref{eq:thm_crb_bayesian_Jpp_D}.

Now consider the prior part $\Jbf_{\boldsymbol\theta \boldsymbol\theta, P}$.
From $\ln p(\thetabf)=\ln p(\alphabf)+\ln p(\psibf)$  with 
$\psi_\ell \sim \mbox{Unif}[\Delta_1,\Delta_2]$, the elements of the FIM $\Jbf_{\tilde{\boldsymbol\theta} \tilde{\boldsymbol\theta}, P}$ are derived similarly to those of  $\Jbf_{\tilde{\boldsymbol\theta} \tilde{\boldsymbol\theta}, D}$ as follows:
\begin{align}
%---------------------------------------------------------------------------
% \alpha* \alpha*
\Jbf_{\boldsymbol\alpha \boldsymbol\alpha, P}
&=
-\mathbb{E}\left\{
\frac{\partial}{\partial\boldsymbol\alpha^*} \left(\frac{\partial \ln p(\boldsymbol\theta)}{\partial \boldsymbol\alpha^*}\right)^H \right\}, \hspace{4em} 
\Jbf_{\boldsymbol\alpha^* \boldsymbol\alpha^*, P}
=
-\mathbb{E}\left\{
\frac{\partial}{\partial\boldsymbol\alpha} \left(\frac{\partial \ln p(\boldsymbol\theta)}{\partial \boldsymbol\alpha}\right)^H \right\}, \\ 
\Jbf_{\boldsymbol\psi \boldsymbol\psi, P}
&=
-\mathbb{E}\left\{
\frac{\partial}{\partial\boldsymbol\psi} \left(\frac{\partial \ln p(\boldsymbol\theta)}{\partial \boldsymbol\psi}\right)^H \right\} = \mathbf{0}_{L}\mathbf{0}_{L}^T,
\end{align}
and the remaining off-diagonal matrices of  $\Jbf_{\tilde{\boldsymbol\theta} \tilde{\boldsymbol\theta}, P}$  are all zero matrices. This completes the proof.   \hfill$\blacksquare$

\vspace{0.2em}

%%%%%%%%%%%%%%%%%%%%%%%%%%%%%%%%%%%%%%%%%%%%%%%%%%%%%%%%%%%%%%%%%%%%%%%
\subsection{Proof of Corollary \ref{cor:crb_bayesian}}\label{subsec:crb_bayesian}
%%%%%%%%%%%%%%%%%%%%%%%%%%%%%%%%%%%%%%%%%%%%%%%%%%%%%%%%%%%%%%%%%%%%%%%

By exploiting the distribution of the path angle $\psi_\ell \sim \text{Unif}[-1,1]$ and the definitions of $\boldsymbol\varphi_{\psi}$ in  \eqref{eq:Theorem1_charac_psi_v2} and $\acute{\boldsymbol\varphi}_\psi$ in \eqref{eq:def_acute_U_psi}, we have  $\boldsymbol\varphi_{\psi} = [1,0,\cdots,0]^T$ and $\acute{\boldsymbol\varphi}_\psi = \mathbf{0}_N$, because 
\begin{equation}
    \varphi_\psi\left(n\pi\right) = \frac{\sin(\pi n)}{\pi n}  =\left\{
    \begin{array}{ll}
    1, & \text{if } n =0\\
    0, & \text{if } n \ne 0
    \end{array}
    \right..
\end{equation} 
In this case, the constituent parameters of $\Jbf_{\boldsymbol\alpha \boldsymbol\alpha, D}$ and $\Jbf_{\boldsymbol\psi \boldsymbol\psi, D}$ in \eqref{eq:nu_zeta_eta_n1} and \eqref{eq:nu_zeta_eta_n3}  are given by $\nu_1=\bar{w}_1$, $\zeta_1=\text{tr}(\Qbf)$, $\eta_1=[\Qbf]_{1,1}$, $\zeta_3=\pi^2\sum_{m=1}^{N-1}[\Qbf]_{m+1,m+1}$, and $\eta_3=0$. By substituting these results  into (\ref{eq:crb_bayesian_Jaa_D_proof}) - (\ref{eq:crb_bayesian_Jpp_D_proof}), we have the FIMs $\Jbf_{\boldsymbol\alpha \boldsymbol\alpha, D}$ and $\Jbf_{\boldsymbol\psi \boldsymbol\psi, D}$, as shown in \eqref{eq:cor_crb_bayesian_Jaa_D} and \eqref{eq:cor_crb_bayesian_Jpp_D}. 

Based on the assumption of $\alpha_\ell\sim\mathcal{CN}(0, \sigma^2)$ for $\ell=1,\ldots, L$,  the diagonal submatrices  of the FIM $\Jbf_{\tilde{\boldsymbol\theta} \tilde{\boldsymbol\theta}, P}$ are derived similarly to those of  $\Jbf_{\tilde{\boldsymbol\theta} \tilde{\boldsymbol\theta}, D}$ as follows:
\begin{align}
%---------------------------------------------------------------------------
% \alpha* \alpha*
\Jbf_{\boldsymbol\alpha \boldsymbol\alpha, P}
&=
-\mathbb{E}\left\{
\frac{\partial}{\partial\boldsymbol\alpha^*} \left(\frac{\partial \ln p(\boldsymbol\theta)}{\partial \boldsymbol\alpha^*}\right)^H \right\} = \frac{1}{\sigma^2}\Ibf_{L+1} 
=\Jbf_{\boldsymbol\alpha^* \boldsymbol\alpha^*, P} ~~~\text{and}~~~
\Jbf_{\boldsymbol\psi \boldsymbol\psi, P}
= \mathbf{0}_L\mathbf{0}_L^T.
\end{align}
Hence, we have the claim.  \hfill$\blacksquare$

%%%%%%%%%%%%%%%%%%%%%%%%%%%%%%%%%%%%%%%%%%%%%%%%%%%%%%%%%%%%%%%%%%%%%%%
\subsection{Proof of Lemma \ref{lem:eigen_decom_bayesian_crb}}\label{subsec:eigen_decom_bayesian_crb}
%%%%%%%%%%%%%%%%%%%%%%%%%%%%%%%%%%%%%%%%%%%%%%%%%%%%%%%%%%%%%%%%%%%%%%%

The FIM $\Jbf_{\boldsymbol\alpha \boldsymbol\alpha} = \Jbf_{\boldsymbol\alpha \boldsymbol\alpha, D}+ \Jbf_{\boldsymbol\alpha \boldsymbol\alpha, P}$ with  (\ref{eq:thm_crb_bayesian_Jaa_D}) and (\ref{eq:Theorem1JththP})
can be rewritten as 
\begin{align}
\Jbf_{\boldsymbol\alpha \boldsymbol\alpha}
&=
\frac{\rho}{\sigma_n^2} \left(K\Ibf_{L+1} + 
\left[\begin{array}{cc}
0 & \bar{w}_1^* \mathbf{1}_L^T \\
\bar{w}_1 \mathbf{1}_L & \frac{\tau}{L} \mathbf{1}_L\mathbf{1}_L^T
\end{array}\right]
+
\left[\begin{array}{cc}
0 				& \mathbf{0}_L^T \\
\mathbf{0}_L & \left(\frac{\tau}{L} - [\Qbf]_{1,1} \right) \left( (L-1) \Ibf_L   -   \Ibf_L^c \right)
\end{array}\right]
\right), \label{eq:lem_Jaa_decom_n1}
\end{align}
where $\Ibf_L^c  \defeq \mathbf{1}_L\mathbf{1}_L^T - \Ibf_L$ and $\tau$ is defined in \eqref{eq:def_tau_kappa}. 
Then, by using {\em (F.1)} and {\em (F.2)} in Lemma \ref{lem:appendLem3} below, the second and the third terms in the RHS of  \eqref{eq:lem_Jaa_decom_n1} are jointly eigen-decomposed as 
\begin{align}
&\Ebf 
\left[\begin{array}{cc}
\frac{\tau - \kappa}{2}& 0 \\
0 &  \frac{\tau + \kappa }{2}
\end{array}\right]
\Ebf^H 
+
\left(\frac{\tau}{L} - [\Qbf]_{1,1}\right) 
\left[\begin{array}{c|c}
0 & \mathbf{0}_{L-1}^T \\
\fbf_1 & \Fbf_2
\end{array}\right]
\left[\begin{array}{c|c}
0 & \mathbf{0}_{L-1}^T \\ \hline
\mathbf{0}_{L-1}  &L \Ibf_{L-1}
\end{array}\right]
\left[\begin{array}{c|c}
0 & \mathbf{0}_{L-1}^T \\
\fbf_1 & \Fbf_2
\end{array}\right]^H \nonumber \\
&\stackrel{(a)}{=}
\Ebf 
\left[\begin{array}{cc}
\frac{\tau - \kappa}{2}& 0 \\
0 &  \frac{\tau + \kappa }{2}
\end{array}\right]
\Ebf^H 
+
\left[\begin{array}{c|c}
0 & \mathbf{0}_{L-1}^T \\
\fbf_1 & \Fbf_2
\end{array}\right]
\left[\begin{array}{c|c}
0 & \mathbf{0}_{L-1}^T \\ \hline
\mathbf{0}_{L-1} &\left(\tau - L [\Qbf]_{1,1} \right)  \Ibf_{L-1}
\end{array}\right]
\left[\begin{array}{c|c}
0 & \mathbf{0}_{L-1}^T \\
\fbf_1 & \Fbf_2
\end{array}\right]^H \nonumber \\
&=
\Ebf 
\left[\begin{array}{cc}
\frac{\tau - \kappa}{2}& 0 \\
0 &  \frac{\tau + \kappa }{2}
\end{array}\right]
\Ebf^H 
+
\left[\begin{array}{c}
 \mathbf{0}_{L-1}^T \\
 \Fbf_2
\end{array}\right]
\left[\begin{array}{c}
\left(\tau - L [\Qbf]_{1,1} \right)  \Ibf_{L-1}
\end{array}\right]
\left[\begin{array}{c}
 \mathbf{0}_{L-1}^T \\
 \Fbf_2
\end{array}\right]^H \nonumber \\
&=
\underbrace{\left[\begin{array}{c|c}
\multirow{2}{*}{\Ebf} & \mathbf{0}_{L-1}^T \\
& \Fbf_2
\end{array}\right]}_{=:\Bbf}
\left[\begin{array}{ccc}
\frac{\tau - \kappa}{2}& &\\
&  \frac{\tau + \kappa }{2}& \\
&& \left(\tau -L [\Qbf]_{1,1} \right)  \Ibf_{L-1}
\end{array}\right]
\left[\begin{array}{c|c}
\multirow{2}{*}{\Ebf} & \mathbf{0}_{L-1}^T \\
& \Fbf_2
\end{array}\right]^H, \label{eq:lem_Jaa_decom_n2}
\end{align}
where $\kappa = \sqrt{\tau^2 + 4L|\bar{w}_1|^2}\ge 0$ and $\Ebf$, $\fbf_1$ and $\Fbf_2$ are given in Lemma \ref{lem:appendLem3}. The equality $(a)$ holds by $(L-1)\Ibf_L-\Ibf_L^c = \Fbf_2\Fbf_2^H$.  Then, by 
\eqref{eq:appendF2} and
  \eqref{eq:appendLem3_64}, the matrix $\Bbf$ defined in  \eqref{eq:lem_Jaa_decom_n2} is unitary  due to the construction of $\Ebf$ and $\Fbf_2$ in Lemma \ref{lem:appendLem3}  and  
\begin{align}
\Ebf^H 
\left[\begin{array}{c}
\mathbf{0}_{L-1}^T \\
\Fbf_2 
\end{array}\right]
=
\left[\begin{array}{c}
e_{21}^*\mathbf{1}_L^T\Fbf_2 \\ 
e_{22}^*\mathbf{1}_L^T\Fbf_2 
\end{array}\right]
=
\left[\begin{array}{c}
e_{21}^*\sqrt{L}\fbf_1^T\Fbf_2 \\ 
e_{22}^*\sqrt{L}\fbf_1^T\Fbf_2 
\end{array}\right]
= 
\mathbf{0},
\end{align}
where  $e_{ij}$ are given in Lemma \ref{lem:appendLem3}. Hence, 
\eqref{eq:lem_Jaa_decom_n2} is the eigen-decomposition of the sum of the second and third terms of the RHS of \eqref{eq:lem_Jaa_decom_n1}. Then, by considering the first term $K\Ibf_{L+1}$ in the RHS of \eqref{eq:lem_Jaa_decom_n1}, the inverse of $\Jbf_{\boldsymbol\alpha \boldsymbol\alpha}$  is eigen-decomposed as shown in \eqref{eq:lem_Jaa_decom_n0}. 
\hfill$\blacksquare$

\vspace{0.5em}

\begin{lemma} \label{lem:appendLem3} For any $x\in\mathbb{C}$ and non-negative $\tau, L \in\mathbb{R}$, the following eigen-decompositions hold:
\begin{align}
&\emph{(F.1)}~~
\left[\begin{array}{cc}
0 & x^* \mathbf{1}_L^T \\
x\mathbf{1} & \frac{\tau}{L}\mathbf{1}_L \mathbf{1}_L^T
\end{array}\right]
=
\Ebf 
\left[\begin{array}{cc}
\frac{\tau - \kappa}{2}& 0\\
0&  \frac{\tau + \kappa }{2}
\end{array}\right]
\Ebf^H  \label{eq:F1}\\
%\end{align}
%\begin{align}
&\emph{(F.2)}~~
\Ibf_L^c   = \mathbf{1}_L\mathbf{1}_L^T - \Ibf_L
=
\underbrace{\left[\begin{array}{c|c}
\fbf_1 & \Fbf_2
\end{array}\right]}_{=:\Fbf}
\left[\begin{array}{c|c}
L-1 & \mathbf{0}_{L-1}^T \\ \hline
\mathbf{0}_{L-1} &-\Ibf_{L-1}
\end{array}\right]
\left[\begin{array}{c|c}
\fbf_1 & \Fbf_2
\end{array}\right]^H, \label{eq:appendF2}
\end{align}
where $\kappa = \sqrt{\tau^2+4L|x|^2}$, $\fbf_1 = \frac{1}{\sqrt{L}}\mathbf{1}_L$,  $\Fbf_2$  orthogonal to $\fbf_1$ is determined such that the matrix $\Fbf$ in \eqref{eq:appendF2} is unitary, and the $L\times 2$ matrix $\Ebf$ is given by 
\begin{align}
\Ebf
&
=
\left\{\begin{array}{ll}
\left[\begin{array}{cc}
-\frac{\sqrt{2L}}{\kappa} \frac{x^*}{\sqrt{1- \frac{\tau}{\kappa} }} 	& \frac{\sqrt{2L}}{\kappa} \frac{x^*}{\sqrt{1+ \frac{\tau}{\kappa} }} \\
\frac{1}{\sqrt{2L}} \sqrt{1-\frac{\tau}{\kappa}} \mathbf{1}_L  		& \frac{1}{\sqrt{2L}} \sqrt{1+\frac{\tau}{\kappa}} \mathbf{1}_L 
\end{array}\right], & \text{if } |x|^2 \neq 0 \\
\left[\begin{array}{cc}
1 					& 0  \\
\mathbf{0}_L  	& \frac{1}{\sqrt{L}} \mathbf{1}_L 
\end{array}\right], & \text{if } |x|^2 = 0 
\end{array}\right. \stackrel{\triangle}{=}
\left[\begin{array}{cc}
e_{11} & e_{12} \\
e_{21}\mathbf{1}_L & e_{22} \mathbf{1}_L
\end{array}\right].  \label{eq:appendLem3_64}
\end{align}
\end{lemma}

{\em Proof:} Proof is by direction computation. \hfill{$\blacksquare$}

%%%%%%%%%%%%%%%%%%%%%%%%%%%%%%%%%%%%%%%%%%%%%%%%%%%%%%%%%%%%%%%%%%%%%%%
\subsection{Proof of Theorem \ref{the:crb_hybrid}}\label{subsec:append_crb_hybrid}

Since the second-order derivative $\frac{\partial}{\partial\boldsymbol\alpha^*} \left(\frac{\partial \ln p(\ybf | \boldsymbol\theta)}{\partial \boldsymbol\alpha^*}\right)^H$ in \eqref{eq:lem_sec_der_n1} of Lemma \ref{lem:sec_der_prob_data} is not a function of $\alphabf$ and $\psibf$ is given,  the FIM for $\alphabf$ is given by 
\begin{align} 
%---------------------------------------------------------------------------
% J_{\alpha \alpha}
\breve{\Jbf}_{{\boldsymbol\alpha} {\boldsymbol\alpha}, D}
&= 
\frac{\rho}{\sigma_n^2} 
\left[\begin{array}{cc}
K & \bar{\wbf}^H \Ubf_{\boldsymbol\psi} \\
\Ubf_{\boldsymbol\psi}^H \bar{\wbf}   &  \Ubf_{\boldsymbol\psi}^H \Qbf \Ubf_{\boldsymbol\psi}
\end{array}\right]
= 
\breve{\Jbf}_{{\boldsymbol\alpha} {\boldsymbol\alpha}, D}^*.
\end{align} 
Applying   $-\mathbb{E}_{\ybf,\boldsymbol\alpha|\boldsymbol \psi}\{\cdot\}$ to \eqref{eq:second_order_prob_alpha_psi} and \eqref{eq:lem_sec_der_n2}, respectively, we obtain $\breve{\Jbf}_{\boldsymbol\alpha \boldsymbol\psi, D}$ and $\breve{\Jbf}_{\boldsymbol\psi \boldsymbol\psi, D}$ as follows:
\begin{align}
%---------------------------------------------------------------------------
% J_{\alpha \psi}
\Jbf_{\boldsymbol\alpha \boldsymbol\psi, D}
&= -
\mathbb{E}_{\ybf, \boldsymbol\alpha|\boldsymbol \psi}\left\{
\frac{\partial}{\partial \boldsymbol\alpha^*} \left(\frac{\partial \ln p(\ybf | \boldsymbol\alpha; \boldsymbol\psi)}{\partial \boldsymbol\psi}\right)^H 
\right\}
=\mathbf{0}_{L+1}\mathbf{0}_{L}^T
\label{eq:crb_hybrid_Jap_D_proof_n1}\\
%---------------------------------------------------------------------------
% J_{\psi \psi}
[\Jbf_{\boldsymbol\psi \boldsymbol\psi, D}]_{p,q}
&= -
\mathbb{E}_{\boldsymbol\alpha|\boldsymbol \psi}\left\{ \mathbb{E}_{\ybf | \boldsymbol\alpha,\boldsymbol \psi} \left\{
\frac{\partial}{\partial\psi_p} \left(\frac{\partial \ln p(\ybf | \boldsymbol\alpha; \boldsymbol\psi)}{\partial \psi_q}\right)^H
\right\}\right\} 
= 
\frac{2}{\sigma_n^2}
\mathbb{E}_{\boldsymbol\alpha|\boldsymbol \psi}\left\{ \text{Re} \left\{
\frac{\partial \mbf^H}{\partial \psi_p} \frac{\partial \mbf}{\partial \psi_q}
\right\}\right\} \nonumber \\
&=
\frac{\rho}{\sigma_n^2} 2
\mathbb{E}_{\boldsymbol\alpha|\boldsymbol \psi}\left\{ \text{Re} \left\{
\alpha_p^* \alpha_q 
\frac{\partial \ubf_N^H(\psi_p)}{\partial \psi_p}   \Qbf   \frac{\partial \ubf_N(\psi_q)}{\partial \psi_q} 
\right\}\right\} \label{eq:crb_hybrid_Jpp_D_proof_n1} \\
&=
\left\{\begin{array}{ll}
\frac{\rho  2\pi^2}{\sigma_n^2}   \mathbb{E}\left\{|\alpha_p|^2\right\}
\sum_{m=1}^{N-1} \sum_{n=1}^{N-1} mn  e^{\iota\pi \psi_p(n-m)}  [\Qbf]_{m+1,n+1}, 
& \text{if } p=q \\
\frac{\rho 2\pi^2  }{\sigma_n^2} 
\text{Re}\left\{\mathbb{E}\left\{\alpha_p^*\right\} \mathbb{E}\left\{\alpha_q\right\}    
\sum_{m=1}^{N-1} \sum_{n=1}^{N-1} mn  e^{\iota\pi (\psi_q n  - \psi_p m)}   [\Qbf]_{m+1,n+1}\right\}, 
& \text{if } p\neq q 
\end{array} \right.,  \nonumber 
\end{align}
where 
$\frac{\partial \ubf_N^H(\psi)}{\partial \psi}$  is defined in \eqref{eq:def_acute_u_n}. 
Substituting the assumption of independent Rayleigh fading $\alpha_\ell \sim \Cc\Nc(0,\sigma^2)$, $\ell=1,\cdots,L$ into  \eqref{eq:crb_hybrid_Jpp_D_proof_n1} yields  $[\Jbf_{\boldsymbol\psi \boldsymbol\psi, D}]_{p,q}=0$ for $p\neq q$. 
The remaining $\Jbf_{\tilde{\boldsymbol\theta} \tilde{\boldsymbol\theta}, D}$ and $\Jbf_{\tilde{\boldsymbol\theta} \tilde{\boldsymbol\theta}, P}$ can be derived in a similar way to that in the proof of Theorem  \ref{the:crb_bayesian_general_psi}.   \hfill$\blacksquare$

%%%%%%%%%%%%%%%%%%%%%%%%%%%%%%%%%%%%%%%%%%%%%%%%%%%%%%%%%%%%%%%%%%%%%%%%
%% References
%%%%%%%%%%%%%%%%%%%%%%%%%%%%%%%%%%%%%%%%%%%%%%%%%%%%%%%%%%%%%%%%%%%%%%%%
%\begin{spacing}{1.54}
\bibliographystyle{IEEEtran}
\bibliography{IEEEabrv,referenceBibs}

\begin{thebibliography}{10}
\providecommand{\url}[1]{#1}
\csname url@samestyle\endcsname
\providecommand{\newblock}{\relax}
\providecommand{\bibinfo}[2]{#2}
\providecommand{\BIBentrySTDinterwordspacing}{\spaceskip=0pt\relax}
\providecommand{\BIBentryALTinterwordstretchfactor}{4}
\providecommand{\BIBentryALTinterwordspacing}{\spaceskip=\fontdimen2\font plus
\BIBentryALTinterwordstretchfactor\fontdimen3\font minus
  \fontdimen4\font\relax}
\providecommand{\BIBforeignlanguage}[2]{{%
\expandafter\ifx\csname l@#1\endcsname\relax
\typeout{** WARNING: IEEEtran.bst: No hyphenation pattern has been}%
\typeout{** loaded for the language `#1'. Using the pattern for}%
\typeout{** the default language instead.}%
\else
\language=\csname l@#1\endcsname
\fi
#2}}
\providecommand{\BIBdecl}{\relax}
\BIBdecl

\bibitem{Liaskos&Nie&Tsioliaridou&etal:ComMag18}
C.~{Liaskos}, S.~{Nie} \emph{et~al.}, ``A new wireless communication paradigm
  through software-controlled metasurfaces,'' \emph{IEEE Commun. Mag.},
  vol.~56, no.~9, pp. 162 -- 169, Sep. 2018.

\bibitem{Basar&Renzo&etal:Access19}
E.~{Basar}, M.~{Di Renzo} \emph{et~al.}, ``{Wireless communications through
  reconfigurable intelligent surfaces},'' \emph{IEEE Access}, vol.~7, pp.
  116\,753 -- 116\,773, Aug. 2019.

\bibitem{Ozdogan&Bjornson&Larsson:WCL20}
{\"{O}}.~{\"{O}zdogan}, E.~{Bj\"{o}rnson}, and E.~G. {Larsson}, ``{Intelligent
  reflecting surfaces: Physics, propagation, and pathloss modeling},''
  \emph{IEEE Wireless Commun. Lett.}, vol.~9, no.~5, pp. 581 -- 585, May 2020.

\bibitem{Guo&Liang&Chen&Larsson:TWC20}
H.~{Guo}, Y.~{Liang} \emph{et~al.}, ``{Weighted sum-rate maximization for
  reconfigurable intelligent surface aided wireless networks},'' \emph{IEEE
  Trans. Wireless Commun.}, vol.~19, no.~5, pp. 3064 -- 3076, May 2020.

\bibitem{Nadeem&Kammoun&Chaaban&etal:TWC20}
Q.~{Nadeem}, A.~{Kammoun} \emph{et~al.}, ``Asymptotic {Max-Min SINR} analysis
  of reconfigurable intelligent surface assisted {MISO} systems,'' \emph{IEEE
  Trans. Wireless Commun. (Early Access)}, pp. 1 -- 1, 2020.

\bibitem{Huang&Zappone&etal:TWC18}
C.~{Huang}, A.~{Zappone} \emph{et~al.}, ``Reconfigurable intelligent surfaces
  for energy efficiency in wireless communication,'' \emph{IEEE Trans. Wireless
  Commun.}, vol.~18, no.~8, pp. 4157 -- 4170, Aug. 2019.

\bibitem{Dong&Wang:WCL20}
L.~{Dong} and H.~{Wang}, ``{Secure MIMO transmission via intelligent reflecting
  surface},'' \emph{IEEE Wireless Commun. Lett.}, vol.~9, no.~6, pp. 787 --
  790, Jun. 2020.

\bibitem{Cui&Zhang&Zhang:WCL19}
M.~{Cui}, G.~{Zhang}, and R.~{Zhang}, ``{Secure wireless communication via
  intelligent reflecting surface},'' \emph{IEEE Wireless Commun. Lett.},
  vol.~8, no.~5, pp. 1410 -- 1414, May 2019.

\bibitem{Yu&Xu&Sun&Ng&Schober:JSAC20}
X.~{Yu}, D.~{Xu} \emph{et~al.}, ``Robust and secure wireless communications via
  intelligent reflecting surfaces,'' \emph{IEEE J. Sel. Areas Commun. (Early
  Access)}, pp. 1 -- 1, 2020.

\bibitem{Qiao&Alouini:WCL20}
J.~{Qiao} and M.~{Alouini}, ``Secure transmission for intelligent reflecting
  surface-assisted mmwave and terahertz systems,'' \emph{IEEE Wireless Commun.
  Lett. (Early Access)}, pp. 1 -- 1, 2020.

\bibitem{Wu&Zhang:WCL20}
Q.~{Wu} and R.~{Zhang}, ``{Weighted sum power maximization for intelligent
  reflecting surface aided SWIPT},'' \emph{IEEE Wireless Commun. Lett.},
  vol.~9, no.~5, pp. 586 -- 590, May 2020.

\bibitem{Wu&Zhang:20COMMAG}
Q.~Wu and R.~Zhang, ``{Towards smart and reconfigurable environment:
  Intelligent reflecting surface aided wireless network},'' \emph{IEEE Commun.
  Mag.}, vol.~58, no.~1, pp. 106 -- 112, Jan. 2020.

\bibitem{Ying&Demirhan&Alkhateeb:20arXiv}
X.~Ying, U.~Demirhan, and A.~Alkhateeb, ``{Relay aided intelligent
  reconfigurable surfaces: Achieving the potential without so many antennas},''
  \emph{arXiv:2006.06644}, Jun. 2020.

\bibitem{Wang&Fang&Yuan&Chen&Duan&Li:20TVT}
P.~Wang, J.~Fang \emph{et~al.}, ``{Intelligent reflecting surface-assisted
  millimeter wave communications: Joint active and passive precoding design},''
  \emph{IEEE Trans. Veh. Technol. (Early Access)}, pp. 1 -- 1, Oct. 2020.

\bibitem{Zhang&Qi&Li&Lu:SPAWC20}
J.~{Zhang}, C.~{Qi} \emph{et~al.}, ``{Channel estimation for reconfigurable
  intelligent surface aided massive MIMO system},'' in \emph{Proc. IEEE Int.
  Workshop Signal Process. Advances for Wireless Commun. (SPAWC)}, Atlanta, GA,
  May 2020.

\bibitem{Hu&Dai:20arXiv}
C.~Hu and L.~Dai, ``{Two-timescale channel estimation for reconfigurable
  intelligent surface aided wireless communications},''
  \emph{arXiv:1912.07990}, May 2020.

\bibitem{Huang&Zappone&Debbah&Yuen:18ICASSP}
C.~Huang, A.~Zappone \emph{et~al.}, ``{Efficient beam alignment in millimeter
  wave systems using contextual bandits},'' in \emph{Proc. IEEE Int. Conf.
  Acoust. Speech and Signal Process. (ICASSP)}, Calgary, AB, Apr. 2018.

\bibitem{Ye&Guo&Alouini:20TWC}
J.~Ye, S.~Guo, and M.-S. Alouini, ``{Joint reflecting and precoding designs for
  SER minimization in reconfigurable intelligent surfaces assisted MIMO
  systems},'' \emph{IEEE Trans. Wireless Commun.}, vol.~19, no.~8, pp. 5561 --
  5574, Aug. 2020.

\bibitem{Mishra&Johansson:ICASSP19}
D.~{Mishra} and H.~{Johansson}, ``{Channel estimation and low-complexity
  beamforming design for passive intelligent surface assisted {MISO} wireless
  energy transfer},'' in \emph{Proc. IEEE Int. Conf. Acoust. Speech and Signal
  Process. (ICASSP)}, Brighton, UK, May 2019.

\bibitem{He&Yuan:WCL20}
Z.~{He} and X.~{Yuan}, ``{Cascaded channel estimation for large intelligent
  metasurface Assisted massive MIMO},'' \emph{IEEE Wireless Commun. Lett.},
  vol.~9, no.~2, pp. 210 -- 214, Feb. 2020.

\bibitem{Jensen&Carvalho:ICASSP20}
T.~L. {Jensen} and E.~{De Carvalho}, ``{An optimal channel estimation scheme
  for intelligent reflecting surfaces based on a minimum variance unbiased
  estimator},'' in \emph{Proc. IEEE Int. Conf. Acoust. Speech and Signal
  Process. (ICASSP)}, Barcelona, Spain, May 2020.

\bibitem{Zheng&Zhang:WCL20}
B.~{Zheng} and R.~{Zhang}, ``Intelligent reflecting surface-enhanced {OFDM}:
  Channel estimation and reflection optimization,'' \emph{IEEE Wireless Commun.
  Lett.}, vol.~9, no.~4, pp. 518 -- 522, Apr. 2020.

\bibitem{Bertsekas:76AC}
D.~P. Bertsekas, ``{On the Goldstein - Levitin - Polyak Gradient Projection
  Method},'' \emph{IEEE Trans. Automat. Contr.}, vol. AC-21, no.~2, pp. 174 --
  184, Apr. 1983.

\bibitem{Bubeck:17book}
S.~Bubeck, \emph{Convex Optimization: Algorithms and Complexity}.\hskip 1em
  plus 0.5em minus 0.4em\relax now Publishers Inc., 2017.

\bibitem{Sayeed&Raghavan:07STSP}
A.~M. Sayeed and V.~Raghavan, ``{Maximizing MIMO capacity in sparse multi path
  with reconfigurable antenna arrays},'' \emph{IEEE J. Sel. Topics Signal
  Process.}, vol.~1, no.~1, pp. 156 -- 166, Jun. 2007.

\bibitem{Seo&Sung&Lee&Kim:15arXiv}
J.~Seo, Y.~Sung \emph{et~al.}, ``{Training beam sequence design for
  millimeter-wave MIMO systems: A POMDP framework},'' \emph{\emph{submitted for
  publication. [Online]. Available: \url{http://arxiv.org/abs/1410.3711}}}.

\bibitem{Zhang&Xu&Xu&Ng&Sun:20CL}
W.~Zhang, J.~Xu \emph{et~al.}, ``{Cascaded channel estimation for IRS-assisted
  mmwave multi-antenna with quantized beamforming},'' \emph{IEEE Commun. Lett.
  (Early Access)}, pp. 1 -- 1, Oct. 2020.

\bibitem{Hassibi&Hochwald:03IT}
B.~Hassibi and B.~M. Hochwald, ``{How much training is needed in
  multiple-antenna wireless links?}'' \emph{IEEE Trans. Inf. Theory}, vol.~49,
  no.~4, pp. 951 -- 963, Apr. 2003.

\bibitem{Carvalho&Slock:97SPAWC}
E.~de~Carvalho and D.~T.~M. Slock, ``{Cram$\acute{\mbox{e}}$r-Rao bounds for
  semi-blind, blind and training sequence based channel estimation},'' in
  \emph{Proc. IEEE Int. Workshop Signal Process. Adv. Wireless Commun.
  (SPAWC)}, Paris, France, Apr. 1997.

\bibitem{Kay:book}
S.~M. Kay, \emph{Fundamentals of Statistical Signal Processing: Detection
  Theory}.\hskip 1em plus 0.5em minus 0.4em\relax Englewood Cliffs, New Jersey:
  Prentice-Hall, 1998.

\bibitem{Ayach&Rajagopal&AduSurra&Pi&Heath:14WC}
O.~E. Ayach, S.~Rajagopal \emph{et~al.}, ``{Spatially sparse precoding in
  millimeter wave MIMO systems},'' \emph{IEEE Trans. Wireless Commun.},
  vol.~13, no.~3, pp. 1499 -- 1513, Mar. 2014.

\bibitem{Adhikaryetal:14JSAC}
A.~Adhikary, E.~A. Safadi \emph{et~al.}, ``{Joint spatial division and
  multiplexing for mm-Wave channels},'' \emph{IEEE J. Sel. Areas Commun.},
  vol.~32, no.~6, pp. 1239 -- 1255, Jun. 2014.

\bibitem{GarciaMorales&Femenias&Roy&Castor&Zorzi:20Access}
J.~Garc$\acute{\mbox{i}}$a-Morales, G.~Femenias, and F.~{Riera-Palou},
  ``{Energy-efficient access-point sleep-mode techniques for cell-free mmWave
  massive MIMO networks with non-uniform spatial traffic density},'' \emph{IEEE
  Access}, vol.~8, pp. 137\,587 -- 137\,605, Jul. 2020.

\bibitem{Omar&Slock&Bazzi:11PIMRC}
S.-M. Omar, D.~T. Slock, and O.~Bazzi, ``{Bayesian and deterministic CRBs for
  semi-blind channel estimation in SIMO single carrier cyclic prefix
  systems},'' in \emph{Proc. IEEE Int. Symp. Pers. Indoor Mobile Radio Commun.
  (PIMRC)}, Toronto, ON, Jan. 2011.

\bibitem{Brandwood:83IEEP}
D.~H. Brandwood, ``{A complex gradient operator and its application in adaptive
  array theory},'' \emph{IEE Proceedings}, vol. 130, no.~1, pp. 11 -- 16, Feb.
  1983.

\bibitem{Wang&Fang&Yuan&Chen&Duan&Li:20SPL}
P.~Wang, J.~Fang \emph{et~al.}, ``{Compressed channel estimation for
  intelligent reflecting surface-assisted millimeter wave systems},''
  \emph{IEEE Signal Process. Lett.}, vol.~27, pp. 905 -- 909, May 2020.

\bibitem{Lin&Yu&Zhu&Schober:20Arxiv}
T.~Lin, X.~Yu \emph{et~al.}, ``{Channel estimation for intelligent reflecting
  surface-assisted millimeter wave MIMO systems},'' \emph{\emph{submitted for
  publication. [Online]. Available: \url{https://arxiv.org/abs/2005.04720}}},
  May 2020.

\bibitem{Alkhateeb&Ayach&Leus&Heath:14STSP}
A.~Alkhateeb, O.~E. Ayach \emph{et~al.}, ``{Channel estimation and hybrid
  precoding for millimeter wave cellular systems},'' \emph{IEEE J. Sel. Topics
  Signal Process.}, vol.~8, no.~5, pp. 831 -- 846, Oct. 2014.

\bibitem{Xiaoetal:17JSAC}
M.~Xiao, S.~Mumtaz \emph{et~al.}, ``{Millimeter wave communications for future
  mobile networks},'' \emph{IEEE J. Sel. Areas Commun.}, vol.~35, no.~9, pp.
  1909 -- 1935, Sep. 2017.

\bibitem{Watt&Borhani&Katsaggelos:book}
R.~Borhani, J.~Watt, and A.~K. Katsaggelos, \emph{Machine Learning Refined:
  Foundations, Algorithms, and Applications}.\hskip 1em plus 0.5em minus
  0.4em\relax Cambridge, U.K.: Cambridge Univ. Press, 2016.

\end{thebibliography}
%\end{spacing}

\end{document}